\documentclass[aps,prd,reprint,superscriptaddress,nofootinbib,floatfix,showkeys]{revtex4-2}
\usepackage{amsmath,amssymb}
\usepackage{graphicx}
\usepackage{microtype}
\usepackage[colorlinks=true,linkcolor=blue,citecolor=blue,urlcolor=blue]{hyperref}
\hypersetup{pdftitle={},pdfauthor={},pdfsubject={},pdfkeywords={},pdfcreator={},pdfproducer={}}
\ifdefined\pdfinfoomitdate\pdfinfoomitdate=1\fi
\ifdefined\pdftrailerid\pdftrailerid{}\fi
\ifdefined\pdfsuppressptexinfo\pdfsuppressptexinfo=15\fi
\allowdisplaybreaks

\DeclareMathOperator{\Tr}{Tr}

\DeclareMathOperator{\Si}{Si}

\begin{document}
	\raggedbottom
	\title{Local subtractions and renormalization-group monotonicity:\\ exact defect and sphere flows}
	\author{Francesco Scardino}
	\email{scardino@uni-wuppertal.de}
	\affiliation{Department of Physics, School of Mathematics and Natural Sciences, Bergische Universit\"at Wuppertal, Gau\ss stra\ss e 20, 42119 Wuppertal, Germany}
	\begin{abstract}
We ask when local subtractions of the sphere free energy yield a function
that decreases monotonically along a renormalization-group flow. The
subtractions are polynomials in $D=R d/dR$ that remove the local
counterterms and leave fixed-point values unchanged, in a fixed
renormalization prescription. We prove that if the subtracted free energy
$C$ approaches its infrared value from below, with ordered endpoints and a
power-law approach to both fixed points, no such subtraction is monotone,
and that further factors with real zeros cannot remove a nonmonotonicity
already present. For a three-dimensional flow to a gapped phase,
$C-C_{\rm IR}=4\pi^2(M_2-3M_0)/R+\dots$, where $M_0$ and $M_2$ are
spectral moments of the flat-space stress-tensor two-point function, so
every subtraction fails when $M_2<3M_0$, as for a free conformally coupled massive scalar. In
two dimensions the leading term is always negative, so no unitary flow to
a gapped phase whose subtracted free energy starts at $c_{\rm UV}/3$
admits a monotone subtraction; the massive Dirac fermion on $S^2$ is an
exact example. In the conformally coupled massive $O(N)$ model, $M_2-3M_0$ is negative at
leading order in $N$ for every finite nonnegative quartic coupling. At
criticality it vanishes at that order and its first correction is negative.
At weak positive coupling it is negative for every curvature coupling
and every finite $N$. For the Gross--Neveu flow from an interacting fixed point, it
is negative already at leading order. In contrast, we prove monotonicity
for Robin boundary flows on every sphere $S^p$ with $p\ge2$ and for
massive Dirac fields on every odd sphere $S^d$ with $d\ge3$. For two
solvable double-trace families we find exact thresholds: a monotone
subtraction exists if and only if $\widehat\Delta\ge1/2$ on a spherical
surface defect, and $\Delta\ge(3-\sqrt{37/13})/2$ on $S^3$ at leading
large $N$.
	\end{abstract}
	
	\maketitle
	\section{Introduction}
	\label{intro:section}
	
	At a conformal fixed point, the finite part of the free energy of a
	three-dimensional quantum field theory on the round sphere $S^3$ is a
	universal number $F$.
	The $F$-theorem, conjectured in
	Refs.~\cite{MyersSinha2010,MyersSinha2011,JafferisKlebanovPufuSafdi2011,KlebanovPufuSafdi2011}
	and proved in Ref.~\cite{CasiniHuerta2012}, states that the ultraviolet
	(UV) value of $F$ exceeds its infrared (IR) value for every nontrivial
	unitary renormalization-group (RG) flow. The proof is entropic: it
	concerns the entanglement entropy of a disk, whose universal constant
	term is $-F$ at fixed points \cite{CasiniHuertaMyers2011}.
	
	Away from fixed points the two observables differ, and it is natural to
	ask whether the sphere free energy itself, as a function of the radius
	$R$, interpolates monotonically between the UV and IR values of $F$.
	The same question arises for defects, where the $b$-theorem orders the
	endpoint anomaly coefficients \cite{JensenOBannon2015} (see also
	Ref.~\cite{KobayashiNishiokaSatoWatanabe2019}), and where the
	endpoint ordering and the monotonicity along the flow have already been
	distinguished \cite{ShacharSinhaSmolkin2023}.
	
	Before this question can be posed, the scheme-dependent local terms
	of the free energy must be removed. On $S^3$ they are proportional to
	$R^3$ and $R$, and scale derivatives remove them: with $D=R\partial_R$,
	the operator $(1-D)(1-D/3)$ annihilates both powers and leaves constant
	fixed-point values unchanged. We denote the subtracted free energy by
	$C=(1-D)(1-D/3)F$, and its UV and IR values by $C_{\rm UV}$ and
	$C_{\rm IR}$.
	The same operator defines one of the sphere $F$-functions of
	Ref.~\cite{GhoshKiritsisNittiWitkowski2019}, in analogy with the
	differential subtraction of the entanglement entropy in
	Ref.~\cite{LiuMezei2013}, and it also appears in the replica-energy
	construction of Ref.~\cite{SantoniScardinoReplica2026}. That work
	distinguishes the sphere function from the entropic $F$-function
	obtained from the replica-defect free energy.
	
\begin{figure*}[t]
		\centering
		\includegraphics[width=0.92\textwidth]{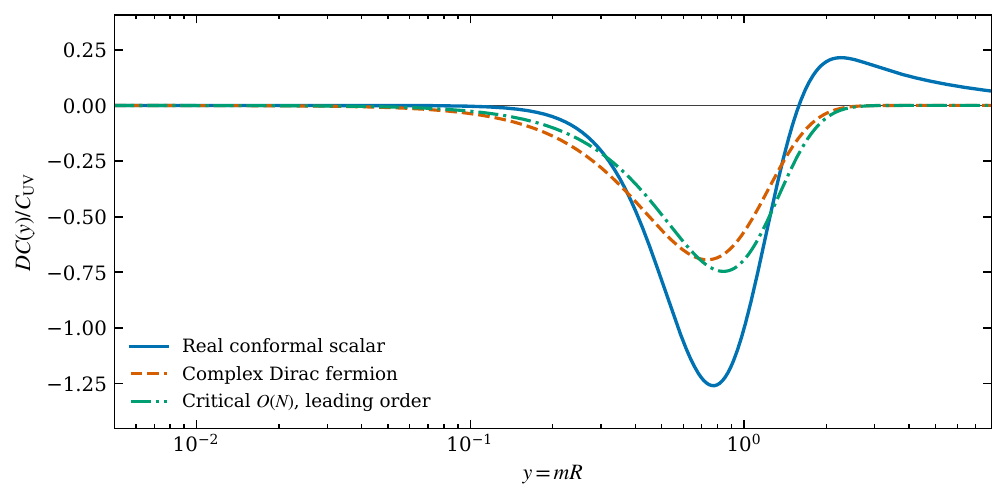}
		\caption{Scale derivatives of the subtracted free energies on $S^3$,
			with the common operator in Eq.~\eqref{st:filter}.
			Here $y=mR$, with $m$ the flat-space physical mass.
			The curves show a real conformally coupled scalar, a complex
			two-component Dirac fermion, and the critical $O(N)$ model at leading
			order per component, using Eq.~\eqref{on:leadingderivative}.
			The normalization is
			$C_{\rm UV}=\log2/8-3\zeta(3)/(16\pi^2)$ for the scalar and vector
			model, and $\log2/4+3\zeta(3)/(8\pi^2)$ for Dirac.
			The scalar derivative is positive in the infrared.
			The Dirac and leading vector-model derivatives remain negative.
			The curves illustrate the analytic sign results derived in the text.}
		\label{fig:running}
	\end{figure*}

	The outcome depends on the theory. For a massive Dirac
	fermion the subtracted free energy decreases monotonically
	\cite{GhoshKiritsisNittiWitkowski2019,Beneventano2017}, while for a
	massive conformally coupled scalar it falls below its IR value and then
	returns to it \cite{GhoshKiritsisNittiWitkowski2019,SantoniScardino2026},
	as shown in Fig.~\ref{fig:running}. A nonmonotone scalar free energy
	had also been observed after an additive subtraction
	\cite{KlebanovPufuSafdi2011}.
	For the scalar, Ref.~\cite{GhoshKiritsisNittiWitkowski2019} obtains a monotone function from the Legendre transform of the free
	energy with respect to the source, with the scale measured in units of the
	conjugate expectation value.
	
	In the present paper we explain the dependence on the theory and determine when a local subtraction may produce an RG monotone.
	Our starting point is that the counterterms fix only which powers of $R$
	must be removed, that is, the zeros of a polynomial in $D$. The
	polynomial then acts on the rest of the free energy, which is
	determined by the spectral data of the theory.
	
	In the following, we refer to the finite polynomials in $D$ with constant
	real coefficients that remove the local terms and preserve the
	fixed-point values as admissible subtractions (Sec.~\ref{ob:section}).
	Entropic monotones, Legendre transforms, and subtractions with
	scale-dependent coefficients lie outside this class.
	
	In the solvable families studied below, the scale derivative of the
	subtracted free energy is a Laplace transform of a derivative of a
	spectral function, and the order of the derivative grows with the number
	of subtracted powers. Unitarity fixes the sign of the spectral function
	but not the signs of its derivatives. Monotonicity therefore needs
	spectral information beyond positivity, and when this information has
	the wrong sign in the IR, no admissible polynomial subtraction can
	restore monotonicity.
	
	The previous Letter~\cite{SantoniScardino2026} established analytically
	that the canonical sphere subtraction fails for a massive free scalar.
	Here we determine when every admissible polynomial subtraction fails,
	and complement this obstruction with sufficient conditions for
	monotonicity, exact classifications, and interacting tests.
	
		\subsection{Main results}

	Our three main results are stated in the next three paragraphs: an
	obstruction that applies to every admissible subtraction at once, a
	flat-space stress-tensor formula for the leading IR coefficient that
	decides when it applies, and the sign of this coefficient in interacting
	vector models. The remaining paragraphs describe where monotonicity
	survives and how it is lost at the thresholds.

	If
	$C_{\rm UV}>C_{\rm IR}$, an approach to the IR endpoint from below
	excludes every admissible subtraction, under the decay assumptions of
	Sec.~\ref{ob:section}. The proof is a transform argument: the bilateral Laplace
	transform of the scale derivative has a real zero, and every finite
	polynomial in $D$ preserves it, so adding scale derivatives cannot
	repair the failure. Further subtractions with real zeros cannot remove a
	nonmonotonicity that is already present, whatever its origin.

	For a three-dimensional flow to a gapped phase, the leading IR
	coefficient is fixed by flat-space stress-tensor data
	(Sec.~\ref{st:sec}). With the contact prescription
	specified there,
	\begin{equation}
		C(R)-C_{\rm IR}
		=\frac{4\pi^2}{R}(M_2-3M_0)+O(R^{-3}).
		\label{intro:moment-criterion}
	\end{equation}
	The spectral moments $M_0$ and $M_2$ of the flat-space stress-tensor
	two-point function are separately nonnegative, but their difference need
	not be, and $M_2<3M_0$ excludes monotonicity throughout the subtraction
	class. For the conformally coupled massive scalar the difference is
	$-1/(384\pi m)$, so the
	dip in Fig.~\ref{fig:running} survives every admissible subtraction. In
	two dimensions the corresponding coefficient, $-8\pi M_0/R^2$, has a
	fixed sign. A unitary flow from a two-dimensional CFT to a gapped phase
	whose subtracted free energy reaches $c_{\rm UV}/3$ therefore admits no
	monotone subtraction. The massive Dirac fermion is an exact example,
	while the zero mode of the noncompact free scalar shows that the
	endpoint condition is needed.

	In the conformally coupled massive $O(N)$ model
	(Sec.~\ref{on:section}), at each fixed
	finite nonnegative $\lambda/m$, with $\lambda$ the
	quartic coupling, the moment difference $-N/[24(\lambda+16\pi m)]$ is
	negative already at order $N$, so no admissible subtraction of the
	leading free energy is monotone. In the strict critical limit this
	coefficient vanishes and the leading subtracted free energy is monotone.
	The order-$N^0$ coefficient is negative: we compute it exactly from a
	curvature expansion and confirm it with flat-space stress-tensor
	diagrams. Maximizing $M_2-3M_0$ over the curvature coupling $\xi$
	gives a negative leading weak-coupling coefficient for every finite $N$. At order $N^0$ this
	maximum is negative for every positive finite coupling and decreases
	to the critical value (Sec.~\ref{fin:section}). For the Gross--Neveu
	flow, a deformation of an interacting UV fixed point, we find
	$M_2-3M_0=-N/(96\pi m)$, confirmed by the exact sphere saddle and by
	flat-space correlators, so that no admissible subtraction of its leading
	free energy is monotone.

	Complete
	monotonicity of a rescaled integrated heat trace proves monotonicity for
	Robin boundary flows on every sphere $S^p$ with $p\geq2$, and a related
	identity proves it for massive Dirac fields on every odd sphere $S^d$
	with $d\geq3$ (Sec.~\ref{pos:section}). Subtractions of arbitrarily high
	order can therefore be monotone.

	For the two double-trace families of Secs.~\ref{dt:defect}
	and~\ref{dt:section}, we determine exactly when a monotone subtraction
	exists. For a generalized-free
	scalar of dimension $0<\widehat\Delta<1$ on a spherical surface defect,
	the threshold is $\widehat\Delta=1/2$, and the family is realized by a
	local four-dimensional monodromy defect. For a double-trace deformation
	of a three-dimensional large-$N$ CFT, with $1/2<\Delta<3/2$, the
	threshold on $S^3$ at leading large $N$ is
	$\Delta_*=(3-\sqrt{37/13})/2$. Below either threshold every admissible
	subtraction fails, although the spectra remain positive and the
	endpoints ordered.

	The depth of the dip vanishes faster than any power of the distance
	from the threshold for the defect. On $S^3$ it vanishes as the $3/2$
	power of the distance from $\Delta_*$. The same classifications apply to scalar double-trace boundary
	flows in AdS$_3$ and AdS$_4$, extending the endpoint analysis of
	Ref.~\cite{BasonCopettiDiPietroJiKomatsu2025} to monotonicity along
	these flows: a monotone subtraction exists if and only if
	$-1<M^2L^2\le-3/4$ and $-9/4<M^2L^2\le-20/13$, respectively.
	
	\subsection{Plan of the paper}
	
	Sections~\ref{ob:section} and~\ref{st:sec} contain the obstruction and the
	stress-tensor diagnostic. Section~\ref{pos:section} develops sufficient
	conditions, and Secs.~\ref{dt:defect}--\ref{on:section} treat the
	classifications and vector models. Section~\ref{discussion:section} discusses the results
	and the open problems. The appendices give the proofs in the same order:
	Appendix~\ref{st:app} derives the curvature matching behind
	Eq.~\eqref{intro:moment-criterion}, Appendix~\ref{pos:appendix} the
	sufficient conditions, Appendices~\ref{dt:defect-core}
	and~\ref{dt:appendix} the two classifications, and
	Appendix~\ref{on:appendix} the vector-model calculations.
	
	\section{Local subtractions and an infrared obstruction}
	\label{ob:section}
	
	In this section we define the admissible subtractions and derive a
	criterion under which none of them is monotone. Along each flow, the
	microscopic couplings to the metric and to the deformation sources
	are held fixed as the radius varies, including any scalar improvement
	term. Section~\ref{fin:section} compares the infrared coefficients for
	different fixed improvements. Renormalization
	is local, covariant, and preserves the UV scaling degree, so that
	independent higher-derivative gravitational interactions and source
	couplings singular at the undeformed fixed point are excluded, as
	detailed in Sec.~\ref{st:prescription}.
	
	For the real three-sphere free energy $F(R)=-\log|Z[S^3_R]|$ we define
	\begin{align}
		D&=R\partial_R,\qquad C=P_3(D)F,\nonumber\\
		P_3(D)&=(1-D)(1-D/3).
		\label{st:filter}
	\end{align}
	The zeros of $P_3$ remove the volume and Einstein--Hilbert terms,
	proportional to $R^3$ and $R$, while the normalization $P_3(0)=1$
	preserves the fixed-point constant. Every polynomial that removes the two local terms therefore has the form $QP_3$, and it preserves the endpoint
	values if $Q(0)=1$. More generally, on $S^p$ the local powers
	$R^p,R^{p-2},\ldots$ fix the zeros of the polynomial, and its normalization reproduces the fixed-point value, a signed
	universal constant for odd $p$ and a signed logarithmic coefficient for
	even $p$ (Sec.~\ref{pos:heat-theorem}).
	
	We call a polynomial in $D$ with constant real coefficients, the
	required zeros, and this normalization an admissible subtraction. The
	endpoint ordering, supplied on $S^3$ by the $F$-theorem, says nothing
	about the behavior between the endpoints.
	
	\subsection{An elementary example}
	\label{ob:example}
	
	The mechanism can already be seen in a simple case. Consider a gapped
	theory whose sphere free energy has, at large radius, the expansion
	\begin{equation}
		F(R)=k_3(mR)^3+k_1\,mR+F_{\rm IR}+\frac{c}{mR}+O(R^{-3}),
		\label{ob:toy}
	\end{equation}
	where the first two terms are the volume and Einstein--Hilbert terms,
	$F_{\rm IR}$ is the IR fixed-point value, and $c/(mR)$ is the leading
	term that survives the subtraction. In a gapped theory this term is
	induced by the curvature-squared terms of the IR effective action, which
	are not among the UV counterterms (Sec.~\ref{st:sec}). Since $DR^k=kR^k$, a polynomial in $D$ acts on each power
	by multiplication, $P(D)R^k=P(k)R^k$. Hence $P_3$ annihilates the two
	local terms, leaves the constant unchanged, and multiplies the inverse-radius term by
	$P_3(-1)=8/3$,
	\begin{equation}
		C(R)=F_{\rm IR}+\frac{8c}{3mR}+O(R^{-3}).
		\label{ob:toy-C}
	\end{equation}
	The subtraction thus removes the scheme-dependent terms, but it cannot change the sign of the leading inverse-radius term. If $c<0$, $C$ approaches
	its IR value from below. Since its UV value lies above the IR value, $C$
	must then fall below $C_{\rm IR}$ at some radius and rise again, so that
	it is not monotone. A different subtraction $QP_3$ multiplies the same
	term by $Q(-1)P_3(-1)$, which can cancel it or reverse its sign. The
	rest of this section shows that this does not help: under mild decay
	assumptions, an approach from below excludes every admissible
	subtraction at once. Section~\ref{st:sec} expresses $c$ through stress-tensor data.
	
	\subsection{A zero of the bilateral Laplace transform}
	\label{ob:theorem}
	
	The criterion rests on a simple observation. In the variable
	$t=\log(R/R_0)$ the scale derivative is $D=\partial_t$, so that a
	subtraction acts as a differential operator in $t$. Integration by parts
	against an exponential weight $e^{st}$ turns $\partial_t$ into
	multiplication by $-s$. Hence the weighted integral of the scale
	derivative is only multiplied by a number under any subtraction, and if
	it vanishes for some real $s$, it vanishes after every subtraction.
	Since the weight is positive, the subtracted derivative must then change
	sign.
	
	Let $g=DC$, with $C$ real and smooth, and
	assume that, for some $\alpha,\beta>0$ and every fixed integer $k\ge0$,
	\begin{align}
		D^k g(t)&=O(e^{\alpha t}) &&(t\to-\infty),\nonumber\\
		D^k g(t)&=O(e^{-\beta t}) &&(t\to+\infty).
		\label{ob:bounds}
	\end{align}
	These bounds express a power-law approach to both fixed points, as in conformal perturbation theory.
	The constants may depend on $k$, and for a polynomial of fixed degree
	only the derivatives that actually occur are needed. For $g\not\equiv0$ we define the bilateral Laplace transform, that is,
	a Laplace transform over the whole real line,
	\begin{equation}
		\widetilde g(s)=\int_{-\infty}^{\infty}e^{st}g(t)\,dt.
		\label{ob:transform}
	\end{equation}
	With these definitions the criterion reads: if $\widetilde g$ has a real
	zero $s_0\in(-\alpha,\beta)$, then $Q(D)C$ is nonmonotone for every
	nonzero real polynomial $Q$. The bounds allow repeated integrations by
	parts, which give
	\begin{equation}
		\int_{-\infty}^{\infty}e^{s_0t}Q(D)g(t)\,dt
		=Q(-s_0)\widetilde g(s_0)=0.
		\label{ob:ibp}
	\end{equation}
	Since the weight $e^{s_0t}$ is positive, the continuous function
	$Q(D)g$ takes both signs, unless it vanishes identically. In the latter
	case its Fourier transform obeys $Q(i\omega)\widehat g(\omega)=0$.
	Since $g$ is integrable, $\widehat g$ is continuous, and since $Q$ has
	finitely many real zeros, $\widehat g=0$, which contradicts
	$g\not\equiv0$. Hence $D[Q(D)C]=Q(D)g$ changes sign. In terms of the
	radius $\widetilde g$ is a Mellin transform, and the argument shows that its real
	zeros are preserved by every finite constant-coefficient differential
	operator. Related transform obstructions occur for monotonic responses
	\cite{Taghavian2025}.
	
	\subsection{Approach from below in the infrared}
	\label{ob:ir}
	
	A zero of $\widetilde g$ follows from two properties that are often available:
	ordered endpoints and an approach to the IR endpoint from below.
	Suppose that Eq.~\eqref{ob:bounds} holds,
	$C_{\rm UV}>C_{\rm IR}$, and
	\begin{equation}
		g(t)=b e^{-\beta t}+o(e^{-\beta t}),\qquad b>0.
		\label{ob:irtail}
	\end{equation}
	Equivalently, $C=C_{\rm IR}-(b/\beta)e^{-\beta t}+o(e^{-\beta t})$
	approaches its IR value from below. Then $\widetilde g(0)=C_{\rm IR}-C_{\rm UV}<0$,
	while the IR tail gives $\lim_{s\uparrow\beta}(\beta-s)\widetilde g(s)=b>0$. By
	continuity $\widetilde g$ vanishes in $(0,\beta)$, and every nonzero polynomial
	subtraction $Q(D)C$ is nonmonotone. No sign assumption on the UV derivative is needed. We shall see in Sec.~\ref{st:sec} that the conformally coupled massive scalar on
	$S^3$ approaches its IR value from below, so the nonmonotonicity of
	Fig.~\ref{fig:running} persists for every admissible subtraction.
	
	The criterion is sufficient, but not necessary: an approach from above does not guarantee that some subtraction is monotone. For
	$y=e^t$, the function
	\begin{equation}
		C(y)=\frac1{1+y}-\frac{16y^2}{(1+y)^4}
		\label{ob:counterexample}
	\end{equation}
	has endpoints $1$ and $0$ and approaches zero from above, but the
	beta-integral formula gives, for $0<s<1$,
	\begin{equation}
		\widetilde g(s)=-\frac{\pi s}{\sin(\pi s)}
		\left[1-\frac83s(1-s^2)\right].
		\label{ob:counterexample-transform}
	\end{equation}
	Hence $\widetilde g(1/2)=0$, and every polynomial subtraction is nonmonotone.
	
	\subsection{Further subtractions with real zeros}
	\label{ob:real}
	
	The transform criterion requires a real zero of $\widetilde g$. For
	subtractions whose zeros are all real, a weaker property suffices. A
	factor $1-D/r$ with real $r\neq0$ removes the power $R^r$, and $D$
	removes constants. The minimal subtractions, Eq.~\eqref{st:filter} and
	Eq.~\eqref{pos:filters} below, are products of such factors, and multiplying by further such factors
	removes further powers. Let $S(h)$ denote the
	number of sign changes of a function $h$. If $g=DC$ obeys
	Eq.~\eqref{ob:bounds} and $Q$ is a nonzero real polynomial with only
	real zeros, $n_Q$ of which lie in $(-\beta,\alpha)$, counted with
	multiplicity, then
	\begin{equation}
		S\bigl(Q(D)g\bigr)\geq S(g)+n_Q .
		\label{ob:real-count}
	\end{equation}
	In particular, if $C$ is not monotone, neither is $Q(D)C$.
	
	It suffices to consider a single factor $D-r$ applied to a function $f$
	that obeys the bounds~\eqref{ob:bounds}. Since
	$(D-r)f=e^{rt}(e^{-rt}f)'$, let $F=e^{-rt}f$ take nonzero alternating
	signs at points $t_0<\dots<t_n$, with $n=S(f)$. By the mean-value
	theorem, $F'$ has the sign of $F(t_i)$ at some point of
	$(t_{i-1},t_i)$. If $r>-\beta$, then $F\to0$ as $t\to+\infty$, and $F'$
	has the sign of $-F(t_n)$ at some point beyond $t_n$. If $r<\alpha$,
	then $F\to0$ as $t\to-\infty$, and $F'$ has the sign of $F(t_0)$ at some
	point before $t_0$. Every real $r$ satisfies at least one of these
	conditions. Hence $(D-r)f$ has at least $n$ sign changes, and at least
	$n+1$ if $r\in(-\beta,\alpha)$. Each application preserves the bounds~\eqref{ob:bounds}, so the
	estimate iterates over the roots of $Q$. This is the variation-diminishing
	property of P\'olya frequency functions \cite{Schoenberg1951,Karlin1968}.
	
	The nonmonotone examples of this paper already have a real zero of
	$\widetilde g$, so Eq.~\eqref{ob:real-count} does not change their
	classification. It extends the diagnostic to every nonmonotone running
	function: once a minimal subtraction fails, no product of further
	power-removing factors can repair it, whatever the sign of the infrared
	term. A repair
	would require complex zeros, which remove log-periodic terms
	$R^a\cos(b\log R)$ that are not local counterterms.
	
	\section{Stress-tensor spectral moments}
	\label{st:sec}
	\label{st:prescription}
	
	The approach-from-below criterion of Sec.~\ref{ob:ir} needs the sign of
	the leading IR coefficient. In this section we express this coefficient
	through flat-space stress-tensor data, without assuming a Gaussian
	determinant. We first treat a unitary three-dimensional flow from a UV
	conformal fixed point to a gapped phase, and specify the contact
	prescription used throughout the paper. In two dimensions the same
	construction fixes the sign of the coefficient (Sec.~\ref{st:2d}).
	
	\subsection{Three dimensions}
	\label{st:3d}
	
	For a gapped theory, the sphere free energy
	at large radius is given by a local effective action for the metric. Its
	volume and Einstein--Hilbert terms are removed by the subtraction, and
	the first surviving terms are quadratic in the curvature, which
	contribute at order $1/R$. Their coefficients are fixed by the
	flat-space stress-tensor two-point function. We fix the UV stress tensor, normalized by
	$\delta S=\frac12\int h_{\mu\nu}T^{\mu\nu}$, and assume a local,
	parity-even expansion of the metric effective action. Let $G(q)$ be the Fourier
	transform of the connected stress-tensor two-point function and $s_0>0$
	its spectral threshold. With the transverse projectors $\Pi_2$ and $\Pi_0$ onto spin two and
	spin zero, defined in Eq.~\eqref{st:projectors}, the spectral
	representation and its moments are
	\begin{align}
		G(q)&=q^4\int_{s_0}^{\infty}\frac{ds}{q^2+s}
		\bigl[\rho_2(s)\Pi_2+\rho_0(s)\Pi_0\bigr]
		+\text{contacts},
		\label{st:dispersion}\\
		M_j&=\int_{s_0}^{\infty}\frac{ds}{s}\,\rho_j(s)\geq0,
		\qquad j=0,2.
		\label{st:moments}
	\end{align}
	The spectral representation decomposes the two-point function into
	contributions of intermediate states of mass squared $s$. Their weights,
	the spin-two and spin-zero spectral densities $\rho_2$ and $\rho_0$, are
	nonnegative by reflection positivity, the Euclidean form of unitarity
	\cite{CappelliFriedanLatorre1991,Karateev2020}, while the UV CFT behavior
	and the mass gap make both moments convergent. The moments determine
	the curvature-squared terms of the effective action, an established
	construction discussed in four dimensions in
	Ref.~\cite{deRhamTolley2020}. Matching the spin-two and spin-zero parts of the
	quadratic metric response to $\int\sqrt g\,(A\mathcal R^2+B\mathcal
	R_{\mu\nu}^2)$ gives, as derived in Appendix~\ref{st:app},
	\begin{equation}
		A=\frac{3M_2-M_0}{16}+a_{\rm ct},
		\qquad B=-\frac{M_2}{2}+b_{\rm ct}.
		\label{st:AB}
	\end{equation}
	Here $a_{\rm ct}$ and $b_{\rm ct}$ collect the local four-derivative
	contributions. They include fixed contributions from the second metric
	variation of the action, usually called seagull terms, and they would be
	shifted by independent curvature-squared background couplings. On the round sphere these terms scale as $1/R$, and the subtraction
	multiplies a $1/R$ term by $P_3(-1)=8/3$, so that the subtracted sphere
	free energy behaves as
	\begin{align}
		C(R)-C_{\rm IR}
		={}&\frac{4\pi^2}{R}
		(M_2-3M_0+48a_{\rm ct}+16b_{\rm ct})
		\nonumber\\
		&+O(R^{-3}).
		\label{st:sphere-general}
	\end{align}
	
	We fix the contact terms by ultraviolet power counting: a local
	contact may not be more singular at coincidence than the UV
	stress-tensor two-point function. The source, metric couplings, and
	improvement term are held fixed during metric and scale variations.
	This defines the local covariant
	prescription used throughout the paper.
	
	In distributional language, the prescription preserves the UV scaling
	degree \cite{BrunettiFredenhagen2000,HollandsWald2001}. In three
	dimensions the stress-tensor two-point function has scaling degree six,
	whereas a four-derivative delta function has degree seven. Such a
	contact is therefore excluded from the full metric response, including
	the second variation of the action. In this prescription
	$a_{\rm ct}=b_{\rm ct}=0$, so
	\begin{equation}
		C(R)-C_{\rm IR}
		=\frac{4\pi^2}{R}(M_2-3M_0)+O(R^{-3}).
		\label{st:matching}
	\end{equation}
	For free fields this is the conventional $\zeta$-function prescription,
	while an independent curvature-squared coupling would reinstate
	Eq.~\eqref{st:sphere-general}. The restriction concerns UV couplings only, and it allows induced IR coefficients proportional to
	$1/m$.
	
	Equation~\eqref{st:matching} displays the limitation of spectral positivity. Each moment is nonnegative, but their difference can have
	either sign. A negative difference implies the approach from below
	required in Sec.~\ref{ob:ir}, while a positive difference implies only
	an approach from above, which does not guarantee monotonicity. For a
	real conformally coupled scalar, a complex two-component Dirac field,
	and the leading massive critical $O(N)$ flow, the moments are
	\begin{align}
		(M_0^\phi,M_2^\phi)
		&=\frac{1}{\pi m}\left(\frac1{640},\frac1{480}\right),
		\label{st:scalar-moments}\\
		(M_0^\psi,M_2^\psi)
		&=\frac{1}{\pi m}\left(\frac1{480},\frac1{160}\right),
		\label{st:dirac-moments}\\
		(M_0^{\rm crit},M_2^{\rm crit})
		&=\frac{N}{\pi m}\left(\frac1{1440},\frac1{480}\right).
		\label{st:critical-moments}
	\end{align}
	Thus the scalar difference is $-1/(384\pi m)$, which accounts for its
	approach from below, whereas the Dirac field and the leading vector
	model saturate $M_2=3M_0$. Appendix~\ref{st:app} derives the spectral
	densities and the convergent position-space sum rule
	\eqref{st:position-sum}. The two-point formula is specific to the
	curvature-squared terms: higher curvature invariants can involve higher
	stress-tensor correlators.
	
	\subsection{Two dimensions}
	\label{st:2d}
	
	In two dimensions the transverse spin-two projector vanishes
	identically, and
	$\mathcal R_{\mu\nu}=\tfrac12g_{\mu\nu}\mathcal R$ leaves
	$\int\sqrt g\,\mathcal R^2$ as the only curvature-squared invariant.
	Only the spin-zero density $\rho_0$ of Eq.~\eqref{st:dispersion}
	remains, now with $\Pi_0=\theta\theta$, and in these conventions the
	$c$-theorem sum rule reads
	$c_{\rm UV}-c_{\rm IR}=12\pi\int ds\,\rho_0(s)$
	\cite{CappelliFriedanLatorre1991}. The matching of
	Appendix~\ref{st:app}, with $h_{\mu\nu}=\theta_{\mu\nu}\varphi$, gives
	the coefficient $A=-M_0/8$ of $\int\sqrt g\,\mathcal R^2$, with $M_0$
	defined by Eq.~\eqref{st:moments}. The prescription again excludes a
	contact term, since the two-point function now has scaling degree four
	and a four-derivative delta function degree six. On $S^2_R$,
	$\int\sqrt g\,\mathcal R^2=16\pi/R^2$, and the minimal subtraction
	$C_2=-D(1-D/2)F_2$ of Eq.~\eqref{pos:filters} multiplies an $R^{-2}$
	term by $4$. Hence
	\begin{equation}
		C_2(R)-C_{\rm IR}=-\frac{8\pi M_0}{R^2}+O(R^{-4}),
		\qquad M_0>0 .
		\label{st:2d-tail}
	\end{equation}
	Unlike the three-dimensional coefficient~\eqref{st:matching}, this one
	has a fixed sign.
	
	This yields an obstruction that requires no expansion in a small
	parameter. Consider a unitary flow from a two-dimensional CFT with
	central charge $c_{\rm UV}>0$ to a gapped phase. Assume the contact
	prescription, the bounds~\eqref{ob:bounds} for $g=DC_2$, and that the
	sphere function reaches the conformal value, $C_2\to c_{\rm UV}/3$ as
	$R\to0$. The gapped phase has $C_{\rm IR}=0$, so the endpoints are
	ordered, and Eq.~\eqref{st:2d-tail} is an approach from below. By
	Sec.~\ref{ob:ir}, no admissible subtraction of $F_2=-\log Z[S^2_R]$ is
	then monotone. The obstruction concerns the partition function:
	Zamolodchikov's $c$-function \cite{Zamolodchikov1986} and the de Sitter
	functions of Ref.~\cite{Loparco2024} are built from correlation
	functions and are not affected.
	
	The hypothesis on the ultraviolet limit is not automatic. For a
	noncompact free scalar with $u=m^2R^2$, the $\zeta$-function
	prescription gives
	\begin{equation}
		C_2(u)=\frac13+u-u^2\sum_{\ell\geq0}
		\frac{2\ell+1}{[\ell(\ell+1)+u]^2}.
		\label{st:2d-scalar}
	\end{equation}
	Its tail $-2/(15u)$ agrees with Eq.~\eqref{st:2d-tail}, since
	$M_0=1/(60\pi m^2)$, but the constant mode contributes $-1$ at small
	radius, so that $C_2(0)=-2/3$ rather than $c_{\rm UV}/3=1/3$. The
	endpoints are then not ordered, and the obstruction does not apply. An
	analogous zero-mode effect in the ultraviolet limit of a
	correlator-based function is discussed in Ref.~\cite{Loparco2024}.
	
	The massive Dirac fermion satisfies all hypotheses. For a complex
	two-component field, $\Theta=m\bar\psi\psi$ gives
	$\rho_0(s)=m^2(1-4m^2/s)^{1/2}/(2\pi s^2)$ for $s\geq4m^2$, so that
	$c_{\rm UV}=1$ and $M_0=1/(120\pi m^2)$. With $x=mR$, the
	$\zeta$-function prescription gives the exact sphere function
	\begin{equation}
		C_2(x)=\frac13-2x^2-2x^3\,{\rm Im}\,\psi'(1+ix),
		\label{st:2d-dirac}
	\end{equation}
	which tends to $1/3$ at small radius and approaches $-1/(15x^2)$, as
	Eq.~\eqref{st:2d-tail} requires. In the variable $t=\log x$, the
	bilateral transform of $g=DC_2$ is
	\begin{equation}
		\widetilde g(s)=4\,\Gamma(2+s/2)\,\Gamma(1-s/2)\,\zeta(-1-s),
		\label{st:2d-transform}
	\end{equation}
	so that the zero required in Sec.~\ref{ob:theorem} is the trivial zero
	$\zeta(-2)=0$, at $s_0=1$. A Majorana fermion, which describes the
	thermal deformation of the Ising model, gives half of each of these
	functions.
	
	\section{Sufficient conditions for monotonicity}
	\label{pos:section}
	
	In this section we derive, for Gaussian determinants, the Laplace
	representation anticipated in the Introduction. Its sign provides
	sufficient conditions for monotonicity, which we verify for the Robin and
	Dirac flows in arbitrarily high dimension.
	
	\subsection{A condition on the integrated heat trace}
	\label{pos:heat-theorem}
	
	Let $\mathcal O$ be a scalar primary of dimension
	$\Delta=(p-\varepsilon)/2$ on $S^p$, with $p\ge2$ and
	$0<\varepsilon<2$. We take $\mathcal O$ to be a
	generalized free field, that is, an operator whose correlators follow
	from its two-point function by Wick's theorem, as for the boundary value
	of a free field. Its two-point function depends
	only on the geodesic distance, so it is diagonal in spherical harmonics. Its inverse has eigenvalue $a_\ell R^{-\varepsilon}$ on the $d_\ell$
	harmonics of angular momentum $\ell$, with the normalization absorbed
	into the coupling,
	\begin{align}
		a_\ell&=\frac{\Gamma(\ell+(p+\varepsilon)/2)}
		{\Gamma(\ell+(p-\varepsilon)/2)},\nonumber\\
		d_\ell&=\frac{(2\ell+p-1)\Gamma(\ell+p-1)}
		{\Gamma(p)\Gamma(\ell+1)}.
		\label{pos:spectrum}
	\end{align}
	All eigenvalues are positive. For $\nu$ real components and $x=\lambda R^\varepsilon>0$, the free
	energy $F_p=-\log Z[S^p_R]$ of the theory deformed by
	$\frac\lambda2\int\mathcal O^2$ is a Gaussian determinant,
	\begin{equation}
		F_p(x)-F_p(0)=\frac\nu2\sum_{\ell\geq0}{}_{\rm ren}
		d_\ell\log(1+x/a_\ell).
		\label{pos:determinant}
	\end{equation}
	The same formula gives the leading large-$N$ change of the free energy
	under such a double-trace deformation in a CFT with large-$N$
	factorization, where single-trace operators behave as generalized free
	fields at leading order
	\cite{GubserKlebanov2002,KlebanovPufuSafdi2011}. The renormalized sum
	includes the physical radius dependence of the eigenvalues
	$(a_\ell+x)/R^\varepsilon$ and retains the local logarithms at resonant
	dimensions, at fixed coupling and subtraction scale. For odd $p=2q-1$,
	we normalize the running function to the signed universal constant
	$(-1)^qF_p$, which is the quantity $\tilde F$ of
	Ref.~\cite{GiombiKlebanov2014} and reduces to $F$ for $p=3$. For even
	$p=2q$, the subtraction extracts $(-1)^q$ times the coefficient of
	$\log R$ in $F_p$, which equals $b/3$ for surface defects
	(Sec.~\ref{dt:defect}). With $D=\varepsilon x\partial_x$, the admissible
	subtraction of minimal order is
	\begin{equation}
		\mathcal P_p(D)=
		\begin{cases}
			(-1)^q\displaystyle\prod_{j=1}^{q}
			\left(1-\dfrac D{2j-1}\right),&p=2q-1,\\[6pt]
			(-1)^qD\displaystyle\prod_{j=1}^{q}
			\left(1-\dfrac D{2j}\right),&p=2q,
		\end{cases}
		\label{pos:filters}
	\end{equation}
	and we set $C_p=\mathcal P_p(D)F_p$. Its zeros at $D=p,p-2,\ldots$
	remove the local powers $R^p,R^{p-2},\ldots$, as in
	Sec.~\ref{ob:section}. In particular,
	$C_2=-D(1-D/2)F_2$ and $C_3=(1-D)(1-D/3)F_3$.
	
	The sign of the scale derivative is controlled by a single spectral
	function. Since $D\log(1+x/a_\ell)=\varepsilon x/(a_\ell+x)$ and
	$x/(a_\ell+x)=x/a_\ell-x^2\int_0^\infty ds\,e^{-s(x+a_\ell)}/a_\ell$,
	the scale derivative of Eq.~\eqref{pos:determinant} is, formally, the
	Laplace transform
	\begin{equation}
		-DF_p(x)=\frac{\nu\varepsilon}{2}\,x^2\int_0^\infty ds\,e^{-xs}K_p(s),
	\end{equation}
	up to a term linear in $x$. Here $K_p(s)=\Tr(A_p^{-1}e^{-sA_p})$,
	where $A_p$ is the inverse two-point operator with eigenvalues
	$a_\ell$. It is the heat trace $\Tr e^{-uA_p}$ \cite{Vassilevich2003}
	integrated over $u\ge s$, and it is positive because every $a_\ell$ and
	$d_\ell$ is positive. The identity is formal, because $K_p(s)$ grows like
	$s^{1-p/\varepsilon}$ as $s\to0$, so that both the integral and the
	linear term diverge in general. These divergences are local terms. The
	subtraction removes them, with the linear term cancelled by the
	spectral-zeta identities of Appendix~\ref{pos:appendix}, and gives the
	convergent representation \eqref{pos:response} below. On $K_p$ the
	subtraction acts as a differential operator in $s$, which becomes a
	single higher derivative after the rescaling
	\begin{align}
		K_p(s)&=\sum_{\ell\geq0}\frac{d_\ell}{a_\ell}e^{-sa_\ell},
		\nonumber\\*
		v&=s^{2/\varepsilon},\qquad
		\Phi_p(v)=s^{p/\varepsilon-1}K_p(s).
		\label{pos:heat}
	\end{align}
	Primes on $\Phi_p$ denote derivatives with respect to $v$. Setting $n_p=\lfloor p/2\rfloor+1$, which is the degree of $\mathcal P_p$,
	we prove in Appendix~\ref{pos:appendix} the convergent representation
	\begin{align}
		-DC_p(x)={}&\nu\varepsilon\kappa_p x^2
		\int_0^\infty ds\,e^{-xs}s^{1+(2n_p-p)/\varepsilon}
		\nonumber\\[-2pt]
		&\hspace{20mm}\times(-1)^{n_p}\Phi_p^{(n_p)}(s^{2/\varepsilon}),
		\label{pos:response}\\
		\kappa_{2q-1}&=\frac{2^{q-1}}{(2q-1)!!},\qquad
		\kappa_{2q}=\frac1{q!}.
		\nonumber
	\end{align}
	The representation includes the local logarithms. Since the remaining
	factors in the integrand are positive for $x,s>0$, the condition
	\begin{equation}
		(-1)^{n_p}\Phi_p^{(n_p)}(v)\geq0\qquad(v>0),
		\label{pos:criterion}
	\end{equation}
	implies $DC_p\le0$ for every positive coupling, and the decrease is
	strict if the inequality is strict on a set of positive measure.
	A function $f$ is called completely monotone if $(-1)^nf^{(n)}\ge0$ for
	every $n\ge0$, as for $e^{-x}$ or $1/(1+x)$. By Bernstein's theorem,
	such functions are exactly the Laplace transforms of nonnegative
	measures \cite{SchillingSongVondracek2012}. Hence, by the uniqueness of
	the Laplace transform, Eq.~\eqref{pos:criterion} is equivalent to the
	complete monotonicity of $f=-DC_p/x^2$ in $x$.
	Ordinary monotonicity, instead, only requires the integral in
	Eq.~\eqref{pos:response} to be nonnegative.
	
	In words, each factor of the subtraction becomes one derivative of
	$\Phi_p$, and $C_p$ decreases whenever this derivative has the
	alternating sign $(-1)^{n_p}$. In particular, it suffices that $\Phi_p$
	be completely monotone in $v$, since all its derivatives then alternate
	in sign. For $p=2$ and $p=3$ the condition is the same convexity, $\Phi_p''\ge0$,
	which the positivity of $K_p$ alone does not supply.
	
	\subsection{Robin boundary flows}
	\label{pos:robin}
	
	We now apply the criterion to the Robin flow. At $\varepsilon=1$ the
	operator $\mathcal O$ has dimension $(p-1)/2$, which is the dimension of
	the boundary value of a free scalar in $p+1$ dimensions with Neumann
	boundary conditions. Correspondingly, the spectrum becomes
	$a_\ell=\ell+(p-1)/2$, the Dirichlet-to-Neumann spectrum of a conformal
	scalar on a $(p+1)$-dimensional hemisphere. A positive boundary mass $h$
	then imposes the Robin condition $\partial_n\phi+h\phi=0$, which
	interpolates between Neumann and Dirichlet boundary conditions, and drives the flow from the former to the latter
	\cite{GiombiKhanchandani2020,Chalabi2023Boundary}. The endpoint values of this flow were
	studied in arbitrary dimensions in Ref.~\cite{NishiokaSato2021}. The generating
	function of the spherical multiplicities and Euler's product for
	$\sinh$ \cite[Eqs.~4.36.1 and 4.36.2]{DLMFHyperbolicProducts} give
	\begin{align}
		K_p(s)&=\frac2{p-1}[2\sinh(s/2)]^{-(p-1)},\nonumber\\
		\Phi_p(v)&=\frac2{p-1}
		\prod_{k=1}^\infty\left(1+\frac{v}{4\pi^2k^2}\right)^{-(p-1)}.
		\label{pos:robin-product}
	\end{align}
	The product is completely monotone, since it is the Laplace transform of
	a positive density (Appendix~\ref{pos:appendix}), and its derivatives
	alternate strictly in sign. Therefore, by Eq.~\eqref{pos:response}, the subtracted
	Robin free energy $C_p$ decreases strictly for every positive boundary
	mass in every sphere dimension $p\ge2$. The difference between its UV
	and IR values is $\nu/24$ for $p=2$ and $\nu\zeta(3)/(8\pi^2)$ for
	$p=3$. For $p=2$ the lowest eigenvalue is $a_0=1/2$, so that the
	coupling is half the one of the surface defect of Sec.~\ref{dt:defect},
	whose lowest eigenvalue is normalized to one. The case $p=1$ is
	excluded, because the Neumann operator has a zero mode.
	
	\subsection{Massive Dirac fields on odd spheres}
	\label{pos:dirac}
	
	A free massive Dirac field needs a separate argument. Its free energy is
	a fermionic determinant, with the Dirac spectrum and multiplicities in
	place of Eq.~\eqref{pos:spectrum}, so the criterion of
	Sec.~\ref{pos:heat-theorem} does not apply directly. However, the mass
	derivative of the Dirac free energy has a closed form, a polynomial in
	$x^2$ times $\tanh(\pi x)$, and the required sign follows from a
	differential identity for such products. For $d=3$ the monotonicity of
	the subtracted free energy is known
	\cite{Beneventano2017,GhoshKiritsisNittiWitkowski2019}, and the identity
	extends it to every odd sphere.
	
	For a complex irreducible Dirac field on $S^d$, with $d=2r+1\ge3$, the
	mass enters through $x=|m|R$, so that $D=x\partial_x$, and we use the
	subtraction $\mathcal P_d$ of Eq.~\eqref{pos:filters}, whose zeros
	remove the local powers $x,x^3,\ldots,x^{d}$. We set $w=x^2$ and
	\begin{equation}
		Q_r(w)=\prod_{j=0}^{r-1}\left[w+(j+1/2)^2\right].
		\label{pos:dirac-polynomial}
	\end{equation}
	In the covariant zeta prescription \cite{KlebanovPufuSafdi2011} the
	real free energy $F_d=-\log|Z|$ satisfies
	\begin{equation}
		\frac{d}{dx}\left[(-1)^{r+1}F_d(x)\right]
		=A_rQ_r(x^2)\tanh(\pi x),
		\label{pos:dirac-resolvent}
	\end{equation}
	where $A_r=2^r\pi/(2r)!$, and we set $C_d=\mathcal P_d(D)F_d$.
	Applying the subtraction to Eq.~\eqref{pos:dirac-resolvent} gives
	\begin{align}
		-DC_d={}&\frac{2^{r+1}A_r}{(2r+1)!!}w^{r+3/2}
		\nonumber\\[-2pt]
		&\times(-1)^r\frac{d^{r+1}}{dw^{r+1}}
		[Q_r(w)\tanh(\pi\sqrt w)].
		\label{pos:dirac-response}
	\end{align}
	The sign of the right-hand side follows from an identity proved in
	Appendix~\ref{pos:dirac-proof}: if $Q(w)=\prod_{i=1}^r(w+c_i)$, with
	$c_i>0$, and $Q^2f'$ is completely monotone, then $(-1)^r(Qf)^{(r+1)}$
	is completely monotone. For $f(w)=\tanh(\pi\sqrt w)$ the product formula
	for $\cosh$ verifies the hypothesis, because $Q_r^2$ cancels its first
	$r$ reciprocal squared factors. Therefore, on every odd sphere $d\ge3$,
	$C_d$ decreases strictly from its positive massless value to zero, and
	\begin{equation}
		\frac{-DC_d(\sqrt w)}{w^{r+3/2}}
		\quad\hbox{is completely monotone in }w.
		\label{pos:dirac-normalized}
	\end{equation}
	For $d=3$, $C_d$ coincides with the function $S_{\rm ren}$ of
	Ref.~\cite{Beneventano2017}. Indeed, their relations
	$-S=(1-D/3)S_{\rm eff}$ and $S_{\rm ren}=(D-1)S$ give
	$S_{\rm ren}=P_3(D)S_{\rm eff}$, and the local linear and cubic terms
	that distinguish their determinant convention are annihilated by $P_3$.

	\section{Surface-defect flows}
	\label{dt:defect}
	
	In the surface-defect family monotonicity holds only in part of the
	parameter space, and we determine exactly where. A generalized-free
	scalar $\widehat{\mathcal O}$ of dimension $0<\widehat\Delta<1$ lives on
	a spherical two-dimensional defect of radius $R$, and it is deformed by
	a positive double-trace coupling $f$. Its two-point function on the
	defect is diagonal in spherical harmonics, with eigenvalue $\lambda_n$ on
	the $2n+1$ harmonics of angular momentum $n$, and the $n=0$ eigenvalue
	scales as $\lambda_0\propto R^\varepsilon$, with
	$\varepsilon=2-2\widehat\Delta$. We write $\lambda_n=\lambda_0/p_n$, so
	that $p_n$ is the spectrum of the inverse two-point function in units of
	its smallest eigenvalue, and we set $x=f\lambda_0>0$ and
	$D=\varepsilon x\partial_x$. The ratios $p_n$ and the mode variables
	$u_n$ are then
	\begin{equation}
		p_0=1,\qquad
		\frac{p_n}{p_{n-1}}=\frac{n+\varepsilon/2}{n-\varepsilon/2},
		\qquad u_n=\frac{x}{x+p_n}.
		\label{dt:defect-spectrum}
	\end{equation}
	Here $u_n=f\lambda_n/(1+f\lambda_n)$ lies between $0$, where the mode $n$
	is unaffected by the coupling, and $1$, where it is fully suppressed.
	This is the $p=2$ spectrum of Eq.~\eqref{pos:spectrum}, with
	$p_n=a_n/a_0$ and the coupling rescaled accordingly.
	
	On a two-dimensional defect the local terms are the area term and a
	constant, so that the admissible subtraction of minimal order is $C_2$,
	normalized to give $b/3$ at a fixed point. The resulting running
	function ${\cal S}_b$ coincides with the renormalized defect entropy of
	Ref.~\cite{ShacharSinhaSmolkin2023}. For $\nu$ real components it obeys
	the exact identity
	\begin{align}
		{\cal S}_b&=-D(1-D/2)F_{\rm def}\nonumber\\
		&=\frac{b_{\rm UV}}3-\frac{\nu\varepsilon^2}{8}J_\varepsilon^{(2)},
		\nonumber\\
		J_\varepsilon^{(2)}&=2\sum_{n\ge1}n(u_{n-1}-u_n)^2,
		\label{dt:defect-identity}\\
		b_{\rm IR}-b_{\rm UV}&=-\frac{\nu\varepsilon^3}{8}.
		\label{dt:b-endpoints}
	\end{align}
	Here $F_{\rm def}=-\log(Z_{\rm defect}/Z_{\rm bulk})$, and $b_{\rm UV}$
	and $b_{\rm IR}$ are the defect anomaly coefficients at the two
	endpoints. The endpoint values agree with the $b$-theorem throughout the
	family \cite{JensenOBannon2015}. Since $J_\varepsilon^{(2)}$ is a sum of
	squares, ${\cal S}_b\le b_{\rm UV}/3$ at every scale. Monotonicity,
	however, requires the stronger inequality $(J_\varepsilon^{(2)})'\ge0$,
	where the prime denotes differentiation with respect to $x$.
	
\begin{figure}[t]
		\centering
		\includegraphics[width=\columnwidth]{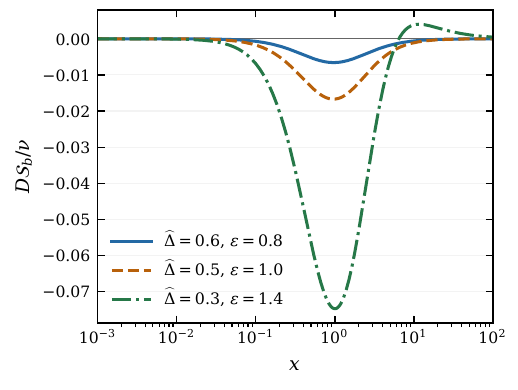}
		\caption{Scale derivative $D{\cal S}_b$ of the subtracted defect free
			energy, Eq.~\eqref{dt:defect-identity}, as a function of the coupling
			$x=f\lambda_0$, with $p_0=1$ and $\varepsilon=2-2\widehat\Delta$, for
			values of $\widehat\Delta$ on both sides of the threshold
			$\widehat\Delta=1/2$. For $\widehat\Delta\ge1/2$ the derivative is
			negative at every coupling, as guaranteed by the bound
			\eqref{dt:defect-global-bound}. For $\widehat\Delta<1/2$ it becomes
			positive at large coupling, where ${\cal S}_b$ approaches its IR value
			from below, Eq.~\eqref{dt:defect-IR}. As $\widehat\Delta\to1/2$ from
			below, the sign change moves to exponentially large coupling,
			Eq.~\eqref{bmin:essential}. The signs are proved in
			Appendices~\ref{dt:defect-monotonicity-proof} and
			\ref{dt:defect-ir-proof}.}
		\label{fig:defect}
	\end{figure}

	The recurrence \eqref{dt:defect-spectrum} proves
	$(J_\varepsilon^{(2)})'>0$ for $0<\varepsilon\le1$
	(Appendix~\ref{dt:defect-monotonicity-proof}). For $1<\varepsilon<2$,
	instead, the IR expansion is
	\begin{align}
		J_\varepsilon^{(2)}-\frac{\varepsilon}{3}
		&=A_2(\varepsilon)x^{-2/\varepsilon}+o(x^{-2/\varepsilon}),
		\nonumber\\
		A_2(\varepsilon)&=\frac{2\varepsilon^3}{3}
		\mathrm B(3-2/\varepsilon,3+2/\varepsilon)
		c_\varepsilon^{-2/\varepsilon}>0,
		\label{dt:defect-IR}
	\end{align}
	where $c_\varepsilon=\Gamma(1+\varepsilon/2)/\Gamma(1-\varepsilon/2)$
	(Appendix~\ref{dt:defect-ir-proof}). Since $J_\varepsilon^{(2)}$ exceeds
	its limit $\varepsilon/3$, ${\cal S}_b$ approaches its IR value
	$b_{\rm IR}/3$ from below, and the expansion survives any fixed number of
	scale derivatives. Every subtraction that removes the area and constant
	terms contains $-D(1-D/2)$, so that the admissible subtractions are
	$Q(D){\cal S}_b$ with $Q(0)=1$, and the criterion of Sec.~\ref{ob:ir}
	excludes all of them. We conclude that a monotone admissible subtraction
	exists if and only if $1/2\le\widehat\Delta<1$, in which case
	${\cal S}_b$ itself decreases strictly. Figure~\ref{fig:defect} displays
	the change of sign of its derivative, and
	Appendix~\ref{dt:defect-trace-proof} derives the trace identity,
	including the anomaly contact term.
	
	Nothing in the spectral data changes sign at $\widehat\Delta=1/2$.
	Instead, near the threshold,
	$J_\varepsilon^{(2)}-\varepsilon/3$ contains two competing IR terms,
	$A_2(\varepsilon)x^{-2/\varepsilon}>0$ from the cubic sum and
	$-\varepsilon x^{-2}$ from the explicit dependence on $u_0$, by
	Eq.~\eqref{dt:defect-cubic} and Appendix~\ref{dt:defect-ir-proof}. For $\varepsilon<1$ the second term decays
	more slowly, so that ${\cal S}_b$ approaches its IR value from above,
	consistently with the monotonicity proved in
	Appendix~\ref{dt:defect-monotonicity-proof}. For $\varepsilon>1$ the
	first term dominates, and the approach is from below. At $\varepsilon=1$
	the two exponents coincide, and the coefficient $A_2(1)-1=-7/15$ is
	negative, so that the threshold itself is monotone. The loss of
	monotonicity is therefore driven by the crossing of two IR exponents.
	
	This mechanism also fixes how monotonicity is lost. Just above the
	threshold, with $\delta=\varepsilon-1\downarrow0$, the minimum of
	${\cal S}_b$ sits at $x_{\min}\sim X_*e^{\log(15/8)/(2\delta)}$, with
	$X_*$ given in Eq.~\eqref{bmin:essential}, exponentially far in the IR.
	The depth of the dip below the IR value vanishes as
	$\delta\,e^{-\log(15/8)/\delta}$, faster than any power of $\delta$.
	The loss of monotonicity is thus invisible at every finite order in
	$\delta$, while the free energy remains analytic at every finite
	coupling. Appendix~\ref{bmin:onset} gives the precise asymptotics.
	
	The entire family is realized in a local four-dimensional
	free complex scalar theory \cite{BianchiMonodromy2021,GiombiMonodromy2021}.
	With monodromy $e^{-2\pi i\alpha}$, $0<\alpha<1$, the bulk field near the
	defect contains a mode $r_\perp^{-\alpha}$, whose coefficient is a defect
	operator of dimension $1-\alpha$. A positive quadratic deformation in
	this operator drives the flow to the boundary condition with
	$r_\perp^{\alpha}$. Thus $\varepsilon=2\alpha$, while $\nu=2$ for the
	complex field, and the classification holds without any large-$N$
	approximation.
	
	\section{Sphere flows in three dimensions}
	\label{dt:section}
	\label{dt:classification}
	
	The second family consists of the leading large-$N$ double-trace flows
	on $S^3$. They are the case $p=3$ of the determinants of
	Sec.~\ref{pos:section}, but here we decide monotonicity directly,
	without the heat-trace condition, and find an exact threshold. Let
	$\mathcal O$ be a scalar primary of dimension $\Delta$ in a
	three-dimensional CFT with large-$N$ factorization. A positive deformation
	$\lambda\int\mathcal O^2/2$, with the other relevant sources tuned to the
	chosen trajectory, gives at leading order the determinant
	\cite{GubserKlebanov2002,DiazDorn2007}
	\begin{align}
		\varepsilon&=3-2\Delta,\qquad0<\varepsilon<2,\qquad
		x=\lambda R^\varepsilon,
		\label{dt:parameters}\\
		p_\varepsilon(n)&=
		\frac{\Gamma(n+(1+\varepsilon)/2)}
		{\Gamma(n+(1-\varepsilon)/2)},\qquad n\ge1,
		\label{dt:spectrum}\\
		F_\varepsilon(x)-F_\varepsilon(0)
		&=\frac\nu2\sum_{n\ge1}{}_{\rm ren}
		n^2\log(1+x/p_\varepsilon(n)).
		\label{dt:determinant}
	\end{align}
	Here $\nu$ counts real components and the two-point normalization is
	absorbed into $\lambda$. In the notation of Sec.~\ref{pos:section},
	$n=\ell+1$, the multiplicity is $n^2=d_\ell$, and
	$p_\varepsilon(n)=a_{n-1}$. For a complex operator the deformation is
	$\lambda\int\mathcal O^\dagger\mathcal O$. The admissible subtraction of
	minimal order is
	\begin{equation}
		D=\varepsilon x\partial_x,\qquad
		C_\varepsilon=(1-D)(1-D/3)F_\varepsilon.
		\label{dt:filtered}
	\end{equation}
	The endpoint ordering is the standard double-trace result
	\cite{GiombiKlebanov2014}.
	
	The main result of this section is that the monotone range is exactly
	\begin{equation}
		DC_\varepsilon(x)<0\quad\hbox{for all }x>0
		\quad\Longleftrightarrow\quad
		0<\varepsilon\le\sqrt{37/13}.
		\label{dt:sharp}
	\end{equation}
	Above the threshold, moreover, every admissible subtraction is
	nonmonotone. Equivalently, a monotone admissible subtraction exists if
	and only if
	\begin{equation}
		\Delta\ge\Delta_*,\qquad
		\Delta_*=\frac{3-\sqrt{37/13}}2.
		\label{dt:dimension}
	\end{equation}
	
	We first prove sufficiency, which rests on summation identities. As in
	Sec.~\ref{dt:defect}, the variables $u_n=x/[x+p_\varepsilon(n)]$ measure
	how strongly the mode $n$ is suppressed by the coupling. The recurrence
	$p_\varepsilon(n+1)/p_\varepsilon(n)=(2n+1+\varepsilon)/(2n+1-\varepsilon)$
	turns $-DC_\varepsilon$ into an absolutely convergent sum over adjacent
	modes, Eq.~\eqref{dt:convergent}, whose individual summands need not be
	positive. A summation by parts with a quadratic polynomial moves part of
	each summand into a telescoping boundary term, chosen so that the
	remainder has a definite sign below the threshold. The constant left
	over in this identity is $(37-13\varepsilon^2)/15$, so the factor that
	fixes the threshold appears explicitly. Elementary bounds then prove
	monotonicity for $0<\varepsilon\le13/10$, while one further summation by
	parts and explicit polynomial coefficient inequalities cover the
	remaining interval, up to and including $\varepsilon=\sqrt{37/13}$
	(Appendix~\ref{dt:appendix}). The argument establishes ordinary
	monotonicity. It neither requires nor proves the convexity condition
	$\Phi_3''\ge0$ of Sec.~\ref{pos:section}.
	
	Necessity, instead, follows from the IR expansion, which is controlled
	by the large-eigenvalue expansion of the spectrum. The terms of this expansion
	are governed by the poles of the spectral zeta function
	$\mathcal Z_\varepsilon(w)=\sum n^2p_\varepsilon(n)^{-w}$, and the pole
	at $w=-1/\varepsilon$ produces the term $x^{-1/\varepsilon}=1/y$, with a
	residue proportional to $(\varepsilon^2-1)(37-13\varepsilon^2)$
	(Appendix~\ref{dt:zeta-proof}). Consequently, near and above the threshold, with $y=x^{1/\varepsilon}$,
	\begin{align}
		C_\varepsilon-C_{\rm IR}&=\frac{\nu A(\varepsilon)}y+o(y^{-1}),
		\nonumber\\
		A(\varepsilon)&=
		\frac{\pi(\varepsilon^2-1)(37-13\varepsilon^2)}
		{4320\sin(\pi/\varepsilon)}.
		\label{dt:IR}
	\end{align}
	The same factor $37-13\varepsilon^2$ thus fixes the sign of the IR
	coefficient. For $\varepsilon>\sqrt{37/13}$ the approach is from below,
	and the criterion of Sec.~\ref{ob:ir} excludes every admissible
	subtraction. As in the defect case, the minimum of the running function
	moves to infinite scale near the threshold, but here as a power law:
	$y_{\min}\propto(\varepsilon-\sqrt{37/13})^{-1/2}$, with a depth
	proportional to $(\varepsilon-\sqrt{37/13})^{3/2}$
	(Appendix~\ref{dt:onset-proof}).
	
	At $\varepsilon=1$, $p_1(n)=n$, and the determinant describes the flow
	from the free to the critical $O(N)$ model at order $N^0$. The massive
	critical $O(N)$ flow studied in Sec.~\ref{on:section} starts at that
	critical fixed point.
	
	One of the holographic $F$-functions of
	Ref.~\cite{GhoshKiritsisNittiWitkowski2019} is built with the same
	operator $(1-D)(1-D/3)$, so a direct comparison is possible. That
	reference studies holographic RG flows on $S^3$ driven by a relevant
	operator. Its numerical tests support monotone functions built from the
	free energy when the perturbing operator has dimension above $3/2$, and from the Legendre transform of the free energy below $3/2$. The
	distinction follows the choice of quantization in those models. In
	the double-trace family of this section the perturbing operator is
	$\mathcal O^2$, of dimension $2\Delta$ at leading order, and the
	dividing line lies instead at $2\Delta_*=3-\sqrt{37/13}$. Thus, for $2\Delta_*\le2\Delta<3/2$, the subtraction of the free energy
	itself remains monotone, although the perturbing operator lies below $3/2$.
	The dividing line is therefore not universal: in this family it is
	fixed by the sign of the IR coefficient, through the factor
	$37-13\varepsilon^2$ in Eq.~\eqref{dt:IR}. The free massive scalar treated in that reference, whose mass term has
	dimension $1$, approaches its IR value from below (Sec.~\ref{st:sec}),
	and its nonmonotonicity agrees with the obstruction of
	Sec.~\ref{ob:ir}.
	
	\subsection{Boundary flows in anti-de Sitter space}
	\label{dt:ads}
	
	The determinants of Secs.~\ref{dt:defect} and \ref{dt:section} also
	describe a boundary RG flow in anti-de Sitter space. Consider a free
	scalar of mass $M$ in Euclidean AdS$_{p+1}$ of radius $L$, whose
	conformal boundary is a sphere $S^p$ of radius $R$, and set
	$\alpha=\sqrt{p^2/4+M^2L^2}$. For $0<\alpha<1$ two boundary conditions
	are allowed \cite{BreitenlohnerFreedman1982,KlebanovWitten1999}. The corresponding boundary operators
	$\widehat{\mathcal O}_\pm$ have dimensions $\Delta_\pm=p/2\pm\alpha$
	\cite{GubserMitra2003,HartmanRastelli2008}, and since the bulk field is
	free, $\widehat{\mathcal O}_-$ is a generalized free field.
	
	A positive double-trace boundary deformation
	$\frac f2\int\widehat{\mathcal O}_-^2$ is relevant, because
	$2\Delta_-<p$. It drives the flow from the $\Delta_-$ boundary condition
	in the UV to the $\Delta_+$ boundary condition in the IR. As in
	Secs.~\ref{dt:defect} and \ref{dt:section}, integrating out
	$\widehat{\mathcal O}_-$ gives the determinant \eqref{pos:determinant},
	now with $\Delta=\Delta_-$, that is, $\varepsilon=p-2\Delta_-=2\alpha$,
	and with one real component \cite{DiazDorn2007}. The coupling $f$ has
	dimension $2\alpha$, so that the flow is controlled by $fR^{2\alpha}$.
	Accordingly, $D$ varies the boundary radius $R$ at fixed coupling, bulk
	mass, and bulk radius, and the local boundary counterterms are matched
	as before.
	
	The classifications of Secs.~\ref{dt:defect} and \ref{dt:section}
	therefore apply directly. For AdS$_3$, with $p=2$, the boundary flow is
	the surface-defect family with $\widehat\Delta=\Delta_-=1-\alpha$. For
	AdS$_4$, with $p=3$, it is the three-sphere family with
	$\Delta=\Delta_-=3/2-\alpha$. A monotone admissible subtraction of the
	boundary free energy thus exists if and only if $\alpha\le1/2$ in
	AdS$_3$, that is, $-1<M^2L^2\le-3/4$, and if and only if
	$\alpha\le\sqrt{37/13}/2$ in AdS$_4$, that is,
	$-9/4<M^2L^2\le-20/13$. These
	ranges lie within the windows $-1<M^2L^2<0$ and $-9/4<M^2L^2<-5/4$, in
	which both boundary conditions exist. The endpoint inequality, in
	contrast, holds for every $\alpha$. In AdS$_3$ the endpoint values are $b/3$, with $b$ the boundary anomaly
	coefficient, and their difference is $\alpha^3/3$ for one real
	scalar.
	
	These results bear directly on the AdS $F$-theorem proposed in
	Refs.~\cite{GiombiKhanchandani2020,BasonCopettiDiPietroJiKomatsu2025},
	according to which a regularized AdS free energy decreases under
	boundary RG flows. Reference~\cite{BasonCopettiDiPietroJiKomatsu2025}
	proves monotonicity along the flow in AdS$_2$, where a single
	subtraction suffices, and tests the endpoint inequality in higher
	dimensions. For the flow above, however, a running monotone obtained by a local
	polynomial subtraction of the boundary determinant exists in AdS$_3$ and AdS$_4$
	only within the ranges just stated. Reference~\cite{BasonCopettiDiPietroJiKomatsu2025}
	removes the bulk UV terms by means of an analytically continued sphere
	free energy. Applying our classification to a running version of that
	AdS free energy requires matching its boundary subtraction to the
	polynomial prescription used here.
	
	\section{Interacting vector models}
	\label{on:section}
	
	Among the free fields of Sec.~\ref{st:sec}, the conformally coupled
	scalar has
	$M_2-3M_0<0$, while the Dirac field has $M_2=3M_0$. Interacting vector
	models test how dynamics changes this sign. We first treat the strict
	critical limit, then compare it with the family at finite quartic
	coupling in Sec.~\ref{on:finite-coupling}. A general response formula
	also applies to the Gross--Neveu model (Sec.~\ref{on:response}). We
	then study the scalar improvement and derive a weak-coupling result
	for every finite $N$ in Sec.~\ref{fin:section}.
	
	We consider the three-dimensional critical $O(N)$ model deformed by its
	singlet scalar, with the sign that gives all $N$ fields a mass gap, in
	the UV prescription of Sec.~\ref{st:prescription}. In the
	auxiliary-field formulation \cite{MosheZinnJustin2003}, its action on a
	curved background is
	\begin{align}
		S[\phi,\sigma]
		={}&\int d^3x\sqrt g\,\Bigl[\frac12\,g^{\mu\nu}\partial_\mu\phi\cdot\partial_\nu\phi
		+\frac{\mathcal R}{16}\,\phi\cdot\phi\nonumber\\
		&\qquad+\frac12\,\sigma\,\phi\cdot\phi\Bigr]
		+NJ\int d^3x\sqrt g\,\sigma,
		\label{on:microaction}
	\end{align}
	where $\phi=(\phi_1,\ldots,\phi_N)$, the auxiliary field $\sigma$ acts as
	a mass squared for the scalars, and $J>0$ is the source of the singlet
	deformation. Integrating out $\phi$ gives $N$ times the effective action
	\begin{align*}
		S_0[g,\sigma,J]={}&\frac12\Tr\log\bigl(-\nabla_g^2+\mathcal R/8+\sigma\bigr)\\
		&+J\int d^3x\sqrt g\,\sigma,
	\end{align*}
	as in Eq.~\eqref{on:action}, so that at large $N$ the
	integral over $\sigma$ is dominated by a saddle point. At leading order
	the flat-space physical mass is $m=8\pi J$. We set $y=mR$, with
	the flat-space physical mass $m$ held fixed, and expand the free energy and its subtraction in $1/N$ at fixed $y$,
	\begin{align}
		F_N&=Nf_0+f_1+O(N^{-1}),\nonumber\\
		C_N&=NC_0+C_1+O(N^{-1}),\qquad C_j=P_3(D)f_j.
		\label{on:expansion}
	\end{align}
	On the round sphere, $\sigma$ is constant at the leading saddle. We
	write $a^2=R^2\sigma-1/4$, so that the scalar
	eigenvalues are $(n^2+a^2)/R^2$, with multiplicity $n^2$. The saddle
	equation is $y=a\coth(\pi a)$ \cite{AnninosDenefHarlow2012}. For
	$0<y<1/\pi$, $a$ is imaginary, while the scalar spectrum remains
	positive. For both real and imaginary $a$, the leading contribution $C_0$ is
	positive and strictly decreasing, with
	\begin{equation}
		C_0(y)=\frac{2\pi^2}{3}y^4e^{-2\pi y}
		[1+O(y^{-1})].
		\label{on:leading-short}
	\end{equation}
	Hence every algebraic IR term cancels (Appendix~\ref{on:leading}).
	
	\subsection{The first subleading infrared coefficient}
	\label{on:coefficient}
	
	At the first subleading order the free energy receives the one-loop determinant of the fluctuations of $\sigma$ around
	the saddle, whose quadratic action is given by the bubble $B$. This
	gives
	\begin{equation}
		f_1=\frac12\operatorname{Tr}\log B
		+S_{\mathrm{ct},1}+F_{\mathrm{norm}},\qquad B(x,x')=G(x,x')^2.
		\label{on:determinant}
	\end{equation}
	Here $G=(-\nabla^2+\mathcal R/8+\sigma)^{-1}$ is the scalar propagator at the leading saddle, $S_{\mathrm{ct},1}$
	collects the order-$N^0$ local counterterms, and $F_{\mathrm{norm}}$ is
	a source-independent normalization. The
	normalized auxiliary-field integral fixes the contour and the sign of the
	determinant (Appendix~\ref{on:measure}). A consistent regulation of the
	auxiliary field is also essential for sphere free energies at conformal
	fixed points \cite{FraserTaliente2025}. With the physical mass fixed and
	no independent four-derivative background coupling, we find
	\begin{align}
		f_1(y)&=f_{1,\mathrm{local}}(y)+\frac{c_{-1}}{y}+O(y^{-3}),
		\nonumber\\*
		C_1(y)&=-\frac{b_1}{y}+O(y^{-3}),
		\label{on:subleadingIR}\\
		c_{-1}&=\frac{-47\Si(\pi)+64\Si(2\pi)-27\Si(3\pi)}{5760},\nonumber\\*
		b_1&=-\frac83c_{-1}>0,
		\label{on:b}
	\end{align}
	where $\Si(x)=\int_0^xdt\,\sin t/t$ is the sine integral. The expansions remain valid after
	scale differentiation.
	
	We obtain this coefficient from the curvature expansion on hyperbolic
	space. Since $c_{-1}$ multiplies a local curvature-squared term, it can be computed on any space of constant curvature. Hyperbolic space
	is convenient, because its spectral decomposition is continuous and the
	bubble reduces to a single radial integral. For sectional curvature $\kappa=-k^2$, the gap equation gives
	$G(r)=e^{-mr}/[4\pi S_\kappa(r)]$, with $S_\kappa(r)=\sinh(kr)/k$, and
	the bubble eigenvalue is
	\begin{equation}
		B_\kappa(p)=\frac1{4\pi p}\int_0^\infty
		e^{-2mr}\frac{\sin(pr)}{S_\kappa(r)}\,dr.
		\label{on:hyperbolicB}
	\end{equation}
	Expanding $S_\kappa^{-1}=r^{-1}+\kappa r/6+7\kappa^2r^3/360+\cdots$
	gives the coefficient $Q=c_{-1}/(2\pi^2m)$ of $\kappa^2$ in the
	free-energy density. The spectral integral converges and gives
	Eq.~\eqref{on:b}; its positive representation
	\eqref{on:bpositiveintegral} establishes $b_1>0$
	(Appendix~\ref{lean:curvaturedetails}). Local covariance carries the
	coefficient to the sphere, where it gives $2\pi^2Q/R=c_{-1}/y$, and
	$P_3(-1)=8/3$ yields $C_1$. The uniform spherical expansion of
	Appendix~\ref{on:symbol}, based on the discrete sphere spectrum
	\cite{DiPietroGorbenkoKomatsu2023}, evaluates the same coefficient and
	controls the remainder and its derivatives.
	
	The coefficient is insensitive to the remaining ambiguities.
	A
	radius-independent order-$1/N$ change of mass normalization adds a multiple of $yf_0'$ to $f_1$. This leaves $c_{-1}$ unchanged,
	because the leading term has no inverse-power corrections, and the
	allowed source counterterms leave it unchanged as well (Appendix~\ref{on:source}).
	
	This fixes the sign at order $N^0$; carrying it to sufficiently large
	finite $N$ requires control of the $1/N$ remainder as $y\to\infty$,
	which we leave open.
	
	\subsection{Flat-space stress-tensor calculation}
	\label{on:flat-summary}
	
	Through Eq.~\eqref{st:matching}, the sphere coefficient predicts the
	order-$N^0$ part of the flat-space moment difference, which we denote by
	the superscript $(1)$,
	\begin{equation}
 (M_2-3M_0)^{(1)}=-\frac{b_1}{4\pi^2m}.
		\label{on:moment}
	\end{equation}
	We verify this prediction independently from the flat-space metric
	response (Appendix~\ref{flat:check}), where the moment difference is a
	combination of fourth-order momentum coefficients of the Weyl and
	transverse-traceless metric Hessians. The determinant
	$\frac12\operatorname{Tr}\log B$ needs an analytic regulator. Setting it
	to zero in the combined integrand gives a dimensionless loop integral
	$I_0$, whose
	integrand differs from the curvature integrand of
	Sec.~\ref{on:coefficient} by a total derivative with boundary
	contribution $4/(75\pi^3)$. Keeping the analytic regulator through the
	metric variations supplies $-4/(75\pi^3)$ and cancels this contribution,
	so that
	\begin{equation}
		m(M_2-3M_0)^{(1)}
		=I_0-\frac4{75\pi^3}=-\frac{b_1}{4\pi^2}.
		\label{on:flat-cancellation}
	\end{equation}
	Thus, within the covariant prescription, the equality between the moments
	at leading order is lifted with a negative sign at order $N^0$.
	
	\subsection{Finite quartic coupling}
	\label{on:finite-coupling}
	
	At the conformal improvement $\xi=1/8$, the cancellation at the
	leading critical saddle is special to the strict critical limit. To see
	this, consider the massive vector model
	with action
	\begin{align}
		S_\lambda[\phi]={}&\int d^3x\sqrt g\,\Bigl[
		\frac12(\nabla\phi)^2
		+\frac12\left(r_0+\frac{\mathcal R}{8}\right)\phi^2
		\nonumber\\
		&\hspace{33mm}+\frac{\lambda}{8N}(\phi^2)^2\Bigr].
		\label{on:finite-coupling-action}
	\end{align}
	Here $\lambda\geq0$ and $\phi^2=\phi\cdot\phi$. We tune $r_0$ to
	fix the flat-space physical mass at $m>0$, and keep the improvement
	coupling at $\xi=1/8$.
	The normalization is $\lambda=2u$ in Eq.~\eqref{on:regulatedaction}.
	Both $m$ and $\lambda$ are held fixed when the radius is varied.
	Writing $\bar\lambda=\lambda/(8\pi m)$ and
	$F_{N,\lambda}=Nf_{0,\lambda}+O(N^0)$, the sphere saddle gives
	\begin{equation}
		C_{0,\lambda}(y)=P_3(D)f_{0,\lambda}(y)
		=-\frac{\pi}{48(2+\bar\lambda)y}+O(y^{-3}).
		\label{on:finite-coupling-tail}
	\end{equation}
	Appendix~\ref{on:finite-coupling-proof} derives this expansion, including
	its validity after scale differentiation. Through
	Eq.~\eqref{st:matching}, it fixes the leading spectral moment difference,
	\begin{equation}
		\left.(M_2-3M_0)\right|_{\text{leading in }N}
		=-\frac{N}{24(\lambda+16\pi m)}.
		\label{on:finite-coupling-moments}
	\end{equation}
	At $\lambda=0$ this is $-N/(384\pi m)$, the result for $N$ free
	conformally coupled scalars. It is strictly negative for every fixed
	finite $\lambda/m$, and tends to zero as $\lambda/m\to\infty$.
	
	The critical cancellation has a curvature interpretation. To all
	algebraic orders in the sphere IR expansion, the saddle has
	$\sigma=m^2+\mathcal R/24$, making the scalar operator
	$-\nabla^2+m^2+\mathcal R/6$. For a free scalar,
	Eq.~\eqref{st:scalar-integral} gives
	$(M_2-3M_0)_\phi=-(6\xi-1)^2/(24\pi m)$, which vanishes at this
	effective curvature coupling. The critical linear source contributes
	only volume and Einstein--Hilbert terms in the same expansion.
	The exact sphere saddle retains exponentially small corrections
	(Appendix~\ref{on:leading}); the microscopic improvement remains
	$\xi=1/8$.
	
	For finite coupling the UV endpoint is Gaussian and the IR theory is
	gapped. Thus $C_{0,\lambda}(0)$ is the positive free-scalar value in
	Eq.~\eqref{on:leadingendpoints}, while $C_{0,\lambda}(\infty)=0$.
	By Sec.~\ref{ob:ir}, the negative tail excludes every admissible
	polynomial subtraction of the leading large-$N$ free energy.
	The obstruction therefore occurs throughout an interacting family,
	already at order $N$. If the exact moment difference has the usual
	large-$N$ expansion with an $O(N^0)$ remainder at fixed $\lambda/m$,
	the same sign follows at sufficiently large finite $N$. This does not
	follow from the saddle alone, and no bound uniform in $\lambda/m$ is
	asserted.
	In the strict critical theory the leading coefficient vanishes and
	the order-$N^0$ result of Sec.~\ref{on:coefficient} is needed.
	Section~\ref{fin:section} studies the improvement dependence and gives
	the weak-coupling coefficient at finite $N$.
	
	\subsection{Auxiliary-field response and the Gross--Neveu flow}
	\label{on:response}
	
	The finite-coupling coefficient follows from a general formula for
	auxiliary-field flows at leading order in $N$. Suppose that the theory
	at fixed constant $\sigma$ is gapped, with the local expansion
	\begin{align}
		\frac{F[g,\sigma]}{N}={}&\int d^3x\sqrt g\,\bigl[w_0+w_1\mathcal R
		\nonumber\\
		&\qquad+A\mathcal R^2+B\mathcal R_{\mu\nu}^2+\cdots\bigr],
		\label{on:response-expansion}
	\end{align}
	whose coefficients depend on $\sigma$, and let $V(\sigma)$ be $w_0$ plus
	the source and coupling terms of $\sigma$, with a nondegenerate
	flat-space saddle, $V'(\sigma_0)=0$ and $V''(\sigma_0)\ne0$. On a
	curved background the saddle moves by
	$\delta\sigma=-w_1'\mathcal R/V''$. Eliminating it changes only the
	coefficient of $\mathcal R^2$, by $-w_1'^2/(2V'')$, since gradients of
	$\sigma$ first enter at six derivatives. By Eq.~\eqref{st:AB},
	\begin{equation}
		\Delta(M_2-3M_0)=-\frac{24N\,w_1'(\sigma_0)^2}{V''(\sigma_0)} .
		\label{on:response-formula}
	\end{equation}
	The shift has the sign of $-V''(\sigma_0)$. In the examples below,
	stability along the auxiliary integration contour fixes the sign of
	$V''$. The shift itself, a squared cross response divided by the
	curvature of a potential, resembles thermodynamic relations such as
	$C_P-C_V=TV\alpha^2/\kappa_T$.
	
	For the scalar determinant, $w_0=-\sigma^{3/2}/(12\pi)$ and
	$w_1=(1-6\xi)\sigma^{1/2}/(48\pi)$, with $\xi$ the curvature coupling of
	Eq.~\eqref{st:scalar-integral}. The auxiliary contour is imaginary and
	$V''<0$. In the critical model $V''=w_0''=-1/(16\pi m)$, and the shift
	$(1-6\xi)^2N/(24\pi m)$ cancels the free value
	$-(6\xi-1)^2N/(24\pi m)$ for every $\xi$, as the $\xi$-independence of
	the critical model requires: there $\sigma$ absorbs $\xi\mathcal R$ up to
	a local term. At finite coupling $V''=w_0''-1/\lambda$, giving
	\begin{equation}
	    M_2-3M_0=-\frac{2N(1-6\xi)^2}{3(\lambda+16\pi m)}+O(N^0).
	    \label{fin:leading}
	\end{equation}
	With $\xi=1/8$, this reproduces
	Eq.~\eqref{on:finite-coupling-moments} without the sphere saddle.
	
	For a fermion bilinear the contour is real and the shift has the
	opposite sign. Consider $N$ complex two-component Dirac fermions with the
	renormalized action
	\begin{equation}
		S=\int d^3x\sqrt g\,\bigl[\bar\psi_i(\gamma^\mu\nabla_\mu+\sigma)\psi_i
		+NJ\sigma\bigr],
		\label{on:gn-action}
	\end{equation}
	where the quadratic auxiliary coupling has been tuned to criticality.
	At $J=0$ this is the critical Gross--Neveu model at leading order
	\cite{GrossNeveu1974,MosheZinnJustin2003}. The source $J$ couples to its
	relevant singlet $\sigma$, of dimension $1+O(1/N)$, and drives the flow
	to a gapped phase. Now $w_0=|\sigma|^3/(6\pi)$ is convex, and the
	flat-space saddle $\sigma_0=m$ has $J=-m^2/(2\pi)$ and $V''=m/\pi>0$.
	The Dirac determinant gives $w_1=|\sigma|/(48\pi)$, and the free Dirac
	field saturates $M_2=3M_0$, so that Eq.~\eqref{on:response-formula} yields
	\begin{equation}
		(M_2-3M_0)_{\rm GN}=-\frac{N}{96\pi m}
		\label{on:gn-moments}
	\end{equation}
	at leading order. Appendix~\ref{on:gn-flat} confirms this value from the
	flat-space auxiliary propagator.
	
	The sphere saddle gives the same result. With $F_3$ the Dirac free
	energy of Eq.~\eqref{pos:dirac-primitive} for $r=1$ and $y=mR$, the
	leading free energy per flavor is $f_0^{\rm GN}=F_3(x_*)-\pi y^2x_*$,
	where $(x_*^2+1/4)\tanh(\pi x_*)=y^2$. Up to exponentially small terms
	$x_*=\sqrt{y^2-1/4}$, and
	\begin{align}
		C_0^{\rm GN}(y)&=P_3(D)f_0^{\rm GN}\nonumber\\
		&=-\frac{\pi}{24\sqrt{y^2-1/4}}+O(y^4e^{-2\pi y}),
		\label{on:gn-sphere}
	\end{align}
	whose $1/y$ term reproduces Eq.~\eqref{on:gn-moments} through
	Eq.~\eqref{st:matching}. The endpoints are
	$C_0^{\rm GN}(0)=\log2/4+3\zeta(3)/(8\pi^2)$, the free Dirac value, and
	$C_0^{\rm GN}(\infty)=0$. By Sec.~\ref{ob:ir}, no admissible subtraction
	of the leading Gross--Neveu free energy is monotone.
	
	The three families thus differ in where the sign is decided. The
	finite-coupling scalar family has a Gaussian UV endpoint, and the
	critical $O(N)$ point requires the order-$N^0$ calculation of
	Sec.~\ref{on:coefficient}. The Gross--Neveu flow deforms an interacting
	UV fixed point and is obstructed already at order $N$.
	
	\subsection{Dependence on the scalar improvement}
	\label{fin:section}

	We now compare the infrared coefficient for different scalar
	improvements. We use the finite-coupling family
	\eqref{on:finite-coupling-action}, replacing $1/8$ by $\xi$, at fixed
	$N$, $\lambda\geq0$ and flat-space physical mass $m>0$.
	The improvement coupling is held fixed as $R$ varies; changing it
	leaves the flat-space theory unchanged.

	The improvement leaves the spin-two moment unchanged and shifts the
	trace by $2\xi\partial^2\phi^2$. The moment difference is therefore a
	concave quadratic function of $\xi$:
	\begin{align}
	    G(\xi)&\equiv M_2-3M_0
	    =G_*-6\chi(\xi-\xi_*)^2,\label{fin:quadratic}\\
	    \chi&=\int d^3x\,\langle\phi^2(x)\phi^2(0)\rangle_c>0.
	    \label{fin:chi}
	\end{align}
	Here $G_*$ is the maximum over improvements, attained at $\xi_*$.
	The susceptibility $\chi$ is finite in the prescription of Sec.~\ref{st:prescription}; its
	normalization and the quadratic identity are derived in
	Appendix~\ref{fin:app}. For free fields the maximum is zero and occurs
	at $\xi=1/6$, as Eq.~\eqref{st:scalar-integral} shows.

	At weak coupling, interactions lower this maximum. Writing
	$\ell=\lambda/m=8\pi\bar\lambda$, the weak-coupling expansion at each
	finite $N$ gives
	\begin{equation}
	    mG_*=-\frac{N+2}{368640\pi^3N}\,\ell^2+O(\ell^3).
	    \label{fin:weak}
	\end{equation}
	Thus the negative infrared coefficient survives the most favorable
	choice of improvement: the sign found at $\xi=1/8$ persists for every
	curvature coupling.

	At large $N$, Eq.~\eqref{fin:leading} has its maximum at $\xi=1/6$ and
	vanishes there. Its first correction gives, at fixed positive finite
	$\ell$,
	\begin{equation}
	    mG_*=g_1(\ell)+O(N^{-1}).
	    \label{fin:largeN}
	\end{equation}
	The determinant on hyperbolic space yields a convergent integral whose
	sign is explicit (Appendix~\ref{lean:curvaturedetails}):
	\begin{equation}
	    g_1(\ell)<0,\qquad g_1'(\ell)<0.
	    \label{fin:g1sign}
	\end{equation}
	Its weak-coupling limit agrees with Eq.~\eqref{fin:weak} at large $N$,
	and its strong-coupling limit is the critical coefficient:
	\begin{align}
	    g_1(\ell)&\sim-\frac{\ell^2}{368640\pi^3}\quad(\ell\to0),\nonumber\\
	    g_1(\infty)&=-\frac{b_1}{4\pi^2}.
	    \label{fin:g1endpoints}
	\end{align}
	This connects the weak-coupling and critical results through an
	analytic function of the quartic coupling.
	
	\section{Discussion}
	\label{discussion:section}
	
	Local subtraction can produce an RG monotone, but only when the spectrum
	supplies sign information that positivity alone does not. Robin boundary
	flows on every sphere $S^p$ with $p\ge2$ and massive Dirac fields on
	every odd sphere $S^d$ with $d\ge3$ are examples, and they show that
	subtractions of arbitrarily high order can work. The same holds for
	surface defects with $\widehat\Delta\ge1/2$, for double-trace flows on
	$S^3$ with $\Delta\ge\Delta_*$, and for the massive critical $O(N)$
	model at leading order. The heat-trace condition of
	Sec.~\ref{pos:section} is sufficient but not necessary, since a
	derivative of the rescaled heat trace can change sign while its Laplace
	integral remains positive. The defect and three-sphere classifications
	are sharp.
	
	Where the infrared coefficient has the wrong sign, no admissible
	subtraction works, although the endpoint inequalities continue to hold.
	In three dimensions this sign is a difference of positive moments and
	depends on the dynamics. With conformal coupling, the
	finite-quartic-coupling vector model has a negative coefficient already
	at leading order in $N$, so the cancellation at criticality is a limiting
	property. The improvement $\xi=1/6$ removes the leading coefficient at
	every coupling. The maximum over improvements is negative at weak
	positive coupling for every finite $N$, and at order $N^0$ for every
	positive finite coupling, decreasing to the critical coefficient.
	The Gross--Neveu flow, a
	deformation of an interacting UV fixed point, has a negative coefficient
	already at order $N$. In two dimensions the sign is fixed for every unitary
	gapped flow whose sphere function reaches its conformal value. Where
	running monotones are known, the missing sign comes from additional
	structure. For the entropic $F$-function it is the strong subadditivity
	of the entanglement entropy of a disk \cite{CasiniHuerta2012}, and for
	boundary flows in AdS$_2$ it is a positive two-point sum rule that
	involves only the first scale derivative
	\cite{BasonCopettiDiPietroJiKomatsu2025}.
	
	The loss of monotonicity does not require a phase transition. Near both thresholds the minimum escapes to infinite radius, while the
	free energy remains analytic at every finite radius. The onset nevertheless differs between the two families: the dip depth
	vanishes faster than any power of the distance from the threshold for
	the defect, and algebraically on $S^3$. Holographic flavored ABJM
	theory on $S^3$ behaves differently. There, the four candidates built
	from the free energy are not monotone, and the flow has a first-order
	quantum phase transition at finite radius, across which three of them
	are discontinuous
	\cite[Secs.~5.2 and 6.2, Figs.~4 and 6]{JokelaKastikainenKiritsisNitti2022}.
	The authors suggest that the transition causes the loss of monotonicity.
	Their candidates also approach the IR endpoint from below, with a
	leading inverse-radius coefficient whose sign their differential
	operators preserve (Eq.~(6.19) of
	Ref.~\cite{JokelaKastikainenKiritsisNitti2022}). For smooth flows, the
	transform criterion turns this observation into an exclusion of every
	admissible subtraction, while the discontinuous candidates lie outside
	its assumptions. The same authors suggested relating the sign of this
	coefficient to IR CFT data. For a gapped flow the IR CFT is trivial, and
	Eq.~\eqref{st:matching} shows that the sign is a property of the massive
	theory, encoded in its stress-tensor moments.
	
	Related constructions on curved spacetimes differ in the observable and
	in the scale that is varied. In two-dimensional de Sitter space,
	stress-tensor sum rules establish the endpoint ordering, while the
	monotonicity in the curvature radius remains open in general
	\cite{Loparco2024}. Entropic inequalities in de Sitter and anti-de
	Sitter space constrain the size of a bulk entangling region at fixed
	curvature \cite{AbateTorroba2025,AbateTorroba2024,AbateSalazarTorroba2026}.
	Their application to a subtracted sphere partition function would require
	an additional identity relating the two observables.
	
	Two problems remain open. First, extending the negative order-$N^0$
	coefficient of the critical $O(N)$ model to a conclusion at finite $N$
	requires control of the large-$N$ remainder in the infrared limit.
	The improvement identity and the weak-coupling expansion constrain
	the finite-coupling family. The next $1/N$ coefficient would improve estimates at
	smaller $N$; establishing the sign there requires additional control or
	nonperturbative input. Second, in three dimensions it is not known whether dynamics supplies the
	required spectral inequalities in broader classes of flows. Ward
	identities, dispersion relations, and entropy inequalities could address
	this question, once their relation to the chosen subtracted partition
	function is established.
	
	\begin{samepage}
		\begin{acknowledgments}
			I thank Zohar Komargodski for posing the question that motivated this work and for insightful comments that clarified the distinction between the entropic and sphere realizations of $F$, and between running monotones and endpoint inequalities.
		\end{acknowledgments}
	\end{samepage}
	
	\appendix 
	\section{Stress-tensor moments and curvature matching}
	\label{lean:stressdetails}
	\label{st:app}
	
	In this appendix we define the spin projectors, derive the curvature
	matching \eqref{st:AB}, normalize the spectral densities of the free
	fields and of the leading vector model, and prove the position-space sum
	rule for the fourth moment.
	
	\subsection{Metric response}
	
	We first match the quadratic metric response to the curvature-squared
	terms. For the spectral convention in Eq.~\eqref{st:dispersion}, the
	transverse projectors are
	\begin{align}
		\theta_{\mu\nu}&=\delta_{\mu\nu}-q_\mu q_\nu/q^2,\nonumber\\
		(\Pi_0)_{\mu\nu,\rho\sigma}
		&=\tfrac12\theta_{\mu\nu}\theta_{\rho\sigma},\nonumber\\
		(\Pi_2)_{\mu\nu,\rho\sigma}
		&=\tfrac12(\theta_{\mu\rho}\theta_{\nu\sigma}
		+\theta_{\mu\sigma}\theta_{\nu\rho})
		-(\Pi_0)_{\mu\nu,\rho\sigma}.
		\label{st:projectors}
	\end{align}
	Since $\delta^2F=\langle\delta^2S\rangle-
	\langle\delta S\delta S\rangle_c$, the quadratic four-derivative action, written with the moments $M_j$ of
	Eq.~\eqref{st:moments}, is
	\begin{equation}
		F_{(4)}^{(2)}=-\frac18\int_q h(-q)q^4
		(M_2\Pi_2+M_0\Pi_0)h(q)+F_{\rm ct}^{(2)},
		\label{st:contact-matching}
	\end{equation}
	where $\int_q\equiv\int d^3q/(2\pi)^3$, and
	$F_{\rm ct}=\int\sqrt g\,(a_{\rm ct}\mathcal R^2+
	b_{\rm ct}\mathcal R_{\mu\nu}^2)$ includes the local contributions from
	the second metric variation of the action and the independent contact
	terms. For a transverse traceless perturbation, $\mathcal R^{(1)}=0$ and
	$(\mathcal R_{\mu\nu}^{(1)})^2=q^4h^2/4$, while for
	$h_{\mu\nu}=\theta_{\mu\nu}\varphi/\sqrt2$ the corresponding squares are
	$2q^4\varphi^2$ and $3q^4\varphi^2/4$. Hence
	\begin{align}
		(B-b_{\rm ct})/4&=-M_2/8,\nonumber\\
		2(A-a_{\rm ct})+3(B-b_{\rm ct})/4&=-M_0/8,
		\label{st:polarization-matching}
	\end{align}
	which proves Eq.~\eqref{st:AB}. The round sphere has
	$\mathcal R^2=36/R^4$, $\mathcal R_{\mu\nu}^2=12/R^4$, and volume
	$2\pi^2R^3$, and applying $P_3(-1)=8/3$, with $P_3$ from Eq.~\eqref{st:filter},
	gives Eq.~\eqref{st:sphere-general}. Although the spin-two quadratic form vanishes for a conformal perturbation, its nonlinear
	invariant $\mathcal R_{\mu\nu}^2-3\mathcal R^2/8$ does not vanish on the
	sphere.
	
	\subsection{Normalization of the spectral densities}
	
	We now compute the spectral densities of the free fields and of the
	leading vector model. The scalar bubble is
	\begin{equation}
		B(q)=\frac1{8\pi}\int_0^1\frac{dx}{\sqrt{m^2+x(1-x)q^2}}
		=\frac{\arctan(q/(2m))}{4\pi q},
		\label{st:bubble}
	\end{equation}
	so that $\langle\phi^2\phi^2\rangle_c=2B$ has spectral density
	$\mathbf1_{s\geq4m^2}/(4\pi\sqrt s)$. Let $\xi$ be the scalar curvature
	coupling, normalized by the action term
	$\tfrac12\xi\int\sqrt g\,\mathcal R\phi^2$. At noncoincident points, the
	trace of the improved stress tensor is
	$\Theta=(2\xi-1/4)\Box\phi^2-m^2\phi^2$, and the normalization of $\rho_0$ in Eq.~\eqref{st:dispersion} gives
	\begin{equation}
		\rho_{\Theta\Theta}(s)=2s^2\rho_0(s).
		\label{st:trace-density}
	\end{equation}
	For a normalized spatial traceless polarization $e_{ij}$, the scalar
	spin-two amplitude is $2e_{ij}k_i k_j$, and its angular average uses
	$\langle(e_{ij}\widehat k_i\widehat k_j)^2\rangle=1/4$ and
	$|\mathbf k|^2=(s-4m^2)/4$. These give
	\begin{align}
		\rho_0^\phi(s)&=\frac{\mathbf1_{s\geq4m^2}}{8\pi\sqrt s}
		\left(2\xi-\frac14-\frac{m^2}{s}\right)^2,
		\nonumber\\
		\rho_2^\phi(s)&=\frac{\mathbf1_{s\geq4m^2}}{256\pi\sqrt s}
		\left(1-\frac{4m^2}{s}\right)^2.
		\label{st:scalar-densities}
	\end{align}
	The complex two-component Dirac bilinear has
	$G_{\bar\psi\psi}(q)=-(q^2+4m^2)B(q)$, up to local terms, and we use
	$\Theta=m\bar\psi\psi$ for its spin-zero density. In the spin-two
	channel, the spin sum in the two-particle rest frame gives a
	squared-amplitude ratio $(s+4m^2)/16$ relative to the scalar bilinear.
	Consequently,
	\begin{align}
		\rho_0^\psi(s)&=\frac{m^2\mathbf1_{s\geq4m^2}}{16\pi s^{3/2}}
		\left(1-\frac{4m^2}{s}\right),\nonumber\\
		\rho_2^\psi(s)&=\frac{\mathbf1_{s\geq4m^2}}{128\pi\sqrt s}
		\left(1-\frac{16m^4}{s^2}\right).
		\label{st:dirac-densities}
	\end{align}
	Direct integration with weight $1/s$ proves Eqs.~\eqref{st:scalar-moments}
	and \eqref{st:dirac-moments}, using $\xi=1/8$ for the scalar. More
	generally, writing $a=2\xi-1/4$, we find
	\begin{equation}
		M_0^\phi=\frac{a^2-a/6+1/80}{8\pi m},\qquad
		M_2^\phi=\frac1{480\pi m}.
		\label{st:scalar-integral}
	\end{equation}
	
	For the leading critical vector model, $\sigma$ is the auxiliary field
	that enters the scalar operator as a mass squared, and its source is
	normalized by the action term $NJ\int\sqrt g\,\sigma$, as in
	Eq.~\eqref{on:action}. The flat-space saddle gives $J=m/(8\pi)$, and the
	quadratic variation of the scalar determinant gives
	$G_{\sigma\sigma}=-2/(NB)$. Together with the separated trace identity
	$\Theta=NJ\sigma$, these relations imply
	\begin{align}
		G_{\Theta\Theta}(q)
		&=-\frac{Nm^2}{32\pi^2B(q)}+\text{contacts},
		\label{st:critical-trace}\\
		B(q)^{-1}
		&=8\pi m\left(1+\frac{q^2}{12m^2}
		-\frac{q^4}{180m^4}+O(q^6)\right).
		\label{st:inverse-bubble}
	\end{align}
	The $q^4$ coefficient is $2M_0$, which gives
	$M_0^{\rm crit}=N/(1440\pi m)$. Since the auxiliary-scalar exchange
	affects only spin zero, $\rho_2^{\rm crit}=N\rho_2^\phi$, and this proves
	Eq.~\eqref{st:critical-moments}.
	
	\subsection{The fourth-moment sum rule}
	
	Finally, we express the moment difference as a position-space sum rule.
	We define $T_\perp=T_{11}+T_{22}$ and
	\begin{equation}
		\mathcal K(x)=2\langle T_{12}(x)T_{12}(0)\rangle_c
		-\tfrac32\langle T_\perp(x)T_\perp(0)\rangle_c.
		\label{st:K}
	\end{equation}
	For momentum $qe_3$, the nonzero contractions are
	$(\Pi_2)_{12,12}=1/2$ and $\sum_{a,b=1}^2(\Pi_0)_{aa,bb}=2$, while the
	corresponding trace contraction of $\Pi_2$ and the $12,12$ component of
	$\Pi_0$ vanish. Hence
	$\widetilde{\mathcal K}(qe_3)=q^4(M_2-3M_0)+O(q^6)$, up to contact terms
	of lower order in momentum, and four momentum derivatives give
	\begin{equation}
		M_2-3M_0=\frac1{24}\int_{\mathbb R^3}d^3x\,x_3^4\mathcal K(x).
		\label{st:position-sum}
	\end{equation}
	The CFT short-distance behavior $|x|^{-6}$ makes this weighted integral
	locally convergent, including perturbative logarithms, while massive
	clustering controls infinity. Contacts with at most three derivatives
	integrate to zero, and the ultraviolet prescription excludes the
	four-derivative contacts, which would otherwise change the sum rule.
	
	\section{Derivation of the sufficient conditions}
	\label{pos:appendix}
	
	In this appendix we prove the Laplace representation
	\eqref{pos:response}, the Robin product \eqref{pos:robin-product}, and
	the differential identity behind the Dirac result of
	Sec.~\ref{pos:dirac}.
	
	\subsection{The heat-trace identity}
	
	We start from the elementary form of the identity, with the eigenvalues
	$a_\ell$ and multiplicities $d_\ell$ of Eq.~\eqref{pos:spectrum}, and
	ignore convergence for the moment. Since
	$D\log(1+x/a_\ell)=\varepsilon x/(a_\ell+x)$ and
	$x/(a_\ell+x)=x/a_\ell-x^2\int_0^\infty ds\,e^{-s(x+a_\ell)}/a_\ell$,
	the sum in Eq.~\eqref{pos:determinant} gives
	\begin{equation}
		-DF_p(x)=\frac{\nu\varepsilon}{2}\,x^2\int_0^\infty ds\,e^{-xs}K_p(s)
		-\frac{\nu\varepsilon}{2}\,x\sum_{\ell\geq0}\frac{d_\ell}{a_\ell}.
	\end{equation}
	Thus the scale derivative of the free energy is, up to a term linear in
	$x$, a Laplace transform of the integrated heat trace $K_p$ of
	Eq.~\eqref{pos:heat}.
	
	As noted in Sec.~\ref{pos:heat-theorem}, this form is only formal:
	since $a_\ell\sim\ell^\varepsilon$ and $d_\ell\sim\ell^{p-1}$, the sum
	$\sum_\ell d_\ell/a_\ell$ diverges, and the Laplace integral converges at
	small $s$ only for $\varepsilon>p/2$. The rest of this subsection shows
	that the subtraction polynomial removes these local divergences exactly,
	which turns the formal identity into the convergent representation
	\eqref{pos:laplace-proof}.
	
	We regularize with spectral zeta functions. Let
	$Z(z)=\sum_\ell d_\ell a_\ell^{-z}$ and
	$Z_x(z)=\sum_\ell d_\ell(a_\ell+x)^{-z}$, defined first where the sums
	converge. Since the eigenvalues $(a_\ell+x)/R^\varepsilon$ carry
	dimensions, the physical determinant contains
	\begin{equation}
		-\frac\nu2\left[Z_x'(0)+\varepsilon Z_x(0)\log(\mu R)\right],
		\label{pos:physical-determinant}
	\end{equation}
	where $\mu$ is the subtraction scale, and subtracting the expression at
	$x=0$ gives Eq.~\eqref{pos:determinant}. The integrated heat trace has
	the Mellin representation
	\begin{equation}
		K_p(s)=\frac1{2\pi i}\int_{c_0-i\infty}^{c_0+i\infty}
		\Gamma(z-1)Z(z)s^{1-z}\,dz,
		\label{pos:mellin}
	\end{equation}
	where $c_0>\max(p/\varepsilon,1)$, so that the contour lies to the right
	of all the poles of $Z$.
	
	The location of these poles explains why the subtraction works. In terms
	of the shifted angular-momentum index $n=\ell+(p-1)/2$, the ratio
	$a_\ell/n^\varepsilon$ admits an expansion in even inverse powers of
	$n$, so that the positive poles of $Z$ lie at $z=(p-2j)/\varepsilon$. A
	pole at $\varepsilon z=p-2j$ corresponds to the power $R^{p-2j}$ of the
	free energy, which is precisely one of the local terms. Hence the zeros
	of the subtraction polynomial, which remove those powers, also cancel
	the poles.
	
	To make this explicit, we write $Q_p=(-1)^q\mathcal P_p$, with $p=2q-1$
	or $p=2q$ as in Eq.~\eqref{pos:filters}, which removes the overall sign
	of the subtraction, and we put $E=s\partial_s$. Since
	$(1-E)s^{1-z}=z\,s^{1-z}$, the operator $Q_p(\varepsilon(1-E))$
	multiplies the integrand of Eq.~\eqref{pos:mellin} by
	$Q_p(\varepsilon z)$, which cancels every positive pole of $Z$. The
	contour may then be moved to any $1<c<2$. Subtracting finitely many terms
	of the gamma-ratio expansion gives polynomial bounds on each closed
	vertical strip, while the gamma factor gives exponential decay on the
	contour. The resulting Laplace density is $O(s^{1-c})$ at the origin and
	decays exponentially at infinity by the spectral gap, so that its
	Laplace integral converges absolutely. Using
	$\Gamma(z-1)\Gamma(2-z)=-\pi/\sin(\pi z)$, we obtain
	\begin{align}
		&\frac{\nu\varepsilon x^2}{2}
		\int_0^\infty e^{-xs}Q_p(\varepsilon(1-E))K_p(s)\,ds
		\nonumber\\
		&\quad=-\frac{\nu\varepsilon}{2}
		\frac1{2\pi i}\int_{c-i\infty}^{c+i\infty}
		\frac{\pi Q_p(\varepsilon z)Z(z)x^z}{\sin\pi z}\,dz
		\nonumber\\
		&\quad=-DQ_p(D)F_p.
		\label{pos:laplace-proof}
	\end{align}
	The last equality follows by differentiating the Mellin representation
	of the logarithm. Equation~\eqref{pos:laplace-proof} is the regularized
	form of the elementary identity above, with the subtraction applied. Two
	points remain to be checked: the fate of the linear term and the
	logarithms at resonant dimensions.
	
	The linear term in $-DF_p$ is $-(\nu\varepsilon/2)\,x\,Z(1)$, and
	it does not survive. Away from coincident poles, with $b=(p-\varepsilon)/2$, the beta integral and
	$\sum_\ell d_\ell t^\ell=(1+t)/(1-t)^p$ give the meromorphic finite-part
	identity
	\begin{equation}
		Z(1)=\frac{\Gamma(\varepsilon-p)}{\Gamma(\varepsilon)}
		\left[\frac{\Gamma(b)}{\Gamma(-b)}
		+\frac{\Gamma(b+1)}{\Gamma(1-b)}\right]=0,
		\label{pos:tadpole}
	\end{equation}
	where the bracket vanishes because $\Gamma(b+1)=b\Gamma(b)$ and
	$\Gamma(1-b)=-b\Gamma(-b)$. When poles coincide, the limit is taken after combining Eq.~\eqref{pos:tadpole} with the
	subtraction polynomial. At
	$\varepsilon=1$ in odd $p$, the value $Z(1)$ need not vanish, but then
	$Q_p(1)=0$. For even $p$ at $\varepsilon=1$, the summand is an even
	polynomial on the half-integer lattice, and $\zeta(-2j,1/2)=0$ gives the
	same cancellation.
	
	At a resonant dimension an integer $k$ coincides with a pole,
	$k\varepsilon=p-2j$, and the free energy contains a logarithm. If
	$r_k=\operatorname{Res}_{z=k}Z(z)$, Eq.~\eqref{pos:physical-determinant}
	gives
	\begin{equation}
		Z_x(0)-Z_0(0)=\sum_{k\geq1}\frac{(-1)^kr_k}{k}x^k.
		\label{pos:resonance}
	\end{equation}
	Only the integer poles with $k\varepsilon=p-2j$ occur in
	Eq.~\eqref{pos:resonance}. At
	fixed $\lambda$, their scale dependence contributes
	$-\nu\sum_k(-1)^kr_kx^k\log x/(2k)$, up to local powers annihilated by
	$Q_p$. Although $Q_p(k\varepsilon)=0$, the logarithm survives the
	subtraction, since
	\begin{equation}
		Q_p(D)[x^k\log x]=\varepsilon Q_p'(k\varepsilon)x^k,
	\end{equation}
	where the prime differentiates the polynomial's argument.
	Correspondingly, the residue of $\pi Q_p(\varepsilon z)Z(z)/\sin\pi z$
	at $z=k$ is $(-1)^k\varepsilon r_kQ_p'(k\varepsilon)$, and the local term
	generated in moving the contour is canceled by applying $-DQ_p(D)$ to the
	logarithm. A determinant built only from the dimensionless eigenvalues
	would therefore miss required terms. Finally, the unperturbed even-dimensional anomaly contributes a constant
	to $C_p=\mathcal P_p(D)F_p$, and hence nothing to $DC_p$.
	
	It remains to express the result through the rescaled function
	$\Phi_p$ of Eq.~\eqref{pos:heat}. Writing
	$K_p=s^{1-p/\varepsilon}\Phi_p(v)$, the identity
	$\prod_{j=0}^{n_p-1}(v\partial_v-j)=v^{n_p}\partial_v^{n_p}$ gives
	\begin{align}
		Q_{2q-1}(\varepsilon(1-E))K_p
		&=\frac{2^q}{(2q-1)!!}s^{1+1/\varepsilon}\Phi_p^{(q)}(v),
		\nonumber\\
		Q_{2q}(\varepsilon(1-E))K_p
		&=-\frac2{q!}s^{1+2/\varepsilon}\Phi_p^{(q+1)}(v),
		\label{pos:heat-algebra}
	\end{align}
	so that the subtraction becomes a single derivative of $\Phi_p$, of
	order $n_p=\lfloor p/2\rfloor+1$. Combining this with
	Eq.~\eqref{pos:laplace-proof} proves Eq.~\eqref{pos:response}. A
	nonnegative density gives complete monotonicity by differentiation under
	the integral. Conversely, the Laplace representation theorem for
	completely monotone functions gives a nonnegative measure, which the
	uniqueness of the Laplace transform identifies with the displayed
	continuous density. This proves the equivalence stated in
	Sec.~\ref{pos:heat-theorem}. For the ordinary monotonicity of $C_p$, the
	nonnegativity of the density is a sufficient condition.
	
	\subsection{The Robin product and normalization}
	
	At $\varepsilon=1$ the eigenvalues are $a_\ell=\ell+(p-1)/2$, and direct
	division of the multiplicity gives
	\begin{equation}
		\frac{d_\ell}{a_\ell}=\frac2{p-1}
		\binom{\ell+p-2}{p-2},
	\end{equation}
	and the generating function
	$\sum_\ell\binom{\ell+p-2}{p-2}t^\ell=(1-t)^{1-p}$, at $t=e^{-s}$,
	gives $K_p(s)=\frac2{p-1}[2\sinh(s/2)]^{1-p}$. Euler's product for
	$\sinh$ then turns $\Phi_p$ into the reciprocal product
	\eqref{pos:robin-product}.
	
	The alternating signs of its derivatives follow from a Laplace
	representation. Each factor of Eq.~\eqref{pos:robin-product} is the
	Laplace transform of a positive normalized function,
	\begin{equation}
		\Bigl(1+\frac{v}{4\pi^2k^2}\Bigr)^{1-p}
		=\int_0^\infty dt\,e^{-vt}g_k(t),
	\end{equation}
	with $g_k(t)=(4\pi^2k^2)^{p-1}t^{p-2}e^{-4\pi^2k^2t}/\Gamma(p-1)$.
	The product is therefore the Laplace transform of the convolution
	$\rho_p=g_1*g_2*\cdots$, which converges because the first moments
	$\int_0^\infty dt\,t\,g_k(t)=(p-1)/(4\pi^2k^2)$ are summable. Since
	$\rho_p\geq0$,
	\begin{align}
		\Phi_p(v)&=\frac2{p-1}\int_0^\infty dt\,\rho_p(t)\,e^{-vt},
		\nonumber\\
		(-1)^j\Phi_p^{(j)}(v)&=\frac2{p-1}\int_0^\infty dt\,\rho_p(t)\,
		t^je^{-vt}>0.
		\label{pos:gamma-measure}
	\end{align}
	Evaluating the convergent product at $v=-u$ gives
	$\int_0^\infty dt\,\rho_p(t)e^{ut}=\prod_k(1-u/4\pi^2k^2)^{1-p}<\infty$
	for $0<u<4\pi^2$, so that all the moments entering the endpoint
	differences below are finite. Substitution in Eq.~\eqref{pos:response}
	gives the strict decrease stated in Sec.~\ref{pos:robin}. Integrating
	Eq.~\eqref{pos:response} with $dx/x$ and using Tonelli's theorem gives
	\begin{align}
		\Delta C_{2q}&=\frac{\nu}{q!(p-1)}\int_0^\infty dt\,\rho_p(t)\,t^q,
		\nonumber\\
		\Delta C_{2q-1}&=\frac{\nu2^{q-1}\sqrt\pi}{(p-1)(2q-1)!!}
		\int_0^\infty dt\,\rho_p(t)\,t^{q-1/2}.
		\label{pos:robin-endpoints}
	\end{align}
	
	\subsection{A differential identity for the Dirac flow}
	\label{pos:dirac-proof}
	
	In this subsection we prove the identity used in Sec.~\ref{pos:dirac}
	and then apply it to the Dirac determinant. Let
	$Q(w)=\prod_{i=1}^r(w+c_i)$, with $c_i>0$, and assume that $B=Q^2f'$ is
	completely monotone on $w>0$. We show that
	\begin{equation}
		k_r(w)=(-1)^r(Qf)^{(r+1)}(w)
		\quad\hbox{is completely monotone}.
		\label{pos:preservation}
	\end{equation}
	
	The mechanism is already visible for $r=1$. With $Q=w+c_1$, one has
	$f'=B/Q^2$ and $Qf''=B'/Q-2B/Q^2$, so that
	\begin{equation}
		-(Qf)''=-2f'-Qf''=-\frac{B'}{Q},
	\end{equation}
	which is completely monotone because $-B'$ and $1/Q$ are. The terms that
	are not manifestly signed cancel. For general $r$, the same cancellation
	is organized by multi-indices.
	
	We put $u_i=(w+c_i)^{-1}$. For a multi-index $\alpha$ with
	$|\alpha|\leq r$, let $h$ be its number of nonzero entries, and define
	\begin{equation}
		I_\alpha=\int_0^1t^{r-h}
		\prod_{\alpha_i>0}[(\alpha_i+1)t-\alpha_i]\,dt.
		\label{pos:Ialpha}
	\end{equation}
	A finite Taylor expansion of
	$f(w-z)=f(w)-\int_0^zB(w-v)/Q(w-v)^2\,dv$, followed by multiplication by
	$Q(w-z)$, gives
	\begin{align}
		(-1)^r(Qf)^{(r+1)}
		={}&\frac{(r+1)!}{Q}
		\sum_{|\alpha|\leq r}\frac{I_\alpha u^\alpha}{(r-|\alpha|)!}
		\nonumber\\[-2pt]
		&\hspace{4mm}\times(-1)^{r-|\alpha|}B^{(r-|\alpha|)}.
		\label{pos:finite-identity}
	\end{align}
	Since only the coefficient of $z^{r+1}$ is used, analyticity is not
	required. Every factor on the right is completely monotone, except
	possibly the numbers $I_\alpha$, so it suffices to show that
	$I_\alpha\geq0$.
	
	Writing each factor with $\alpha_i>0$ in Eq.~\eqref{pos:Ialpha} as
	$t-\alpha_i(1-t)$ and expanding, we obtain
	\begin{equation}
		I_\alpha=\frac1{r+1}\sum_{k=0}^h
		(-1)^k\frac{e_k(\alpha)}{\binom rk},
		\label{pos:coefficient-sign}
	\end{equation}
	where $e_k$ is the elementary symmetric polynomial in the positive
	entries of $\alpha$. The terms of this alternating sum decrease in
	magnitude. Indeed, divide $r$ labeled objects into groups of sizes
	$\alpha_i>0$ and a set of $r-|\alpha|$ excluded objects. Then
	$e_k(\alpha)/\binom rk$ is the probability that a uniformly chosen
	$k$-element subset contains no excluded object and at most one member of
	each group. If the objects are drawn one at a time without replacement,
	the first $k$ draws form a uniform $k$-element subset, and they satisfy
	the condition whenever the first $k+1$ draws do. Hence the probability
	decreases with $k$, and pairing consecutive terms gives
	$I_\alpha\geq0$. Since products and finite sums of completely monotone
	functions are completely monotone, this proves
	Eq.~\eqref{pos:preservation}. The term $\alpha=0$ gives
	$k_r\geq(-1)^rB^{(r)}/Q$, which is strictly positive when $B$ is a
	nonconstant completely monotone function.
	
	For the Dirac field, the eigenvalues of either sign have
	multiplicity
	\begin{equation}
		g(a)=\frac{2^r}{(2r)!}
		\prod_{j=0}^{r-1}[a^2-(j+1/2)^2],
		\qquad a=n+r+\tfrac12,
	\end{equation}
	with $n\geq0$. Pairing the two signs gives the real free energy
	$F_d=-\log|Z|=-\sum_ag(a)\log(a^2+x^2)$, and extending the sum to all positive
	half-integers adds only zero multiplicities. The derivative then follows
	in three steps. Since $g$ is a polynomial in $a^2$, its division by
	$a^2+x^2$ leaves the remainder $g(ix)=(-1)^r2^rQ_r(x^2)/(2r)!$, with $Q_r$ from
	Eq.~\eqref{pos:dirac-polynomial}. The
	polynomial quotient sums to zero, because $\zeta(-2j,1/2)=0$. Finally,
	the remaining sum is
	\begin{equation}
		\sum_{n=0}^\infty\frac1{(n+1/2)^2+x^2}
		=\frac\pi{2x}\tanh(\pi x)
	\end{equation}
	which proves Eq.~\eqref{pos:dirac-resolvent}.
	
	The hypothesis of Eq.~\eqref{pos:preservation} holds for
	$f(w)=\tanh(\pi\sqrt w)$. Writing $a_j=j+1/2$, the product formula for
	$\cosh$ gives
	\begin{align}
		&Q_r(w)^2\frac{d}{dw}\tanh(\pi\sqrt w)\nonumber\\
		&\quad=\frac\pi2\left(\prod_{j=0}^{r-1}a_j^4\right)w^{-1/2}
		\prod_{j=r}^\infty(1+w/a_j^2)^{-2},
		\label{pos:dirac-product}
	\end{align}
	since $Q_r^2$ cancels the first $r$ factors. 
	The remaining product
	converges because $\sum_ja_j^{-2}<\infty$. As for the Robin product,
	each factor, including $w^{-1/2}$, is the Laplace transform of a positive
	function, so that the right-hand side is the Laplace transform of a
	positive density, and hence a nonconstant completely monotone function. Finally, the
	elementary identity
	\begin{equation}
		\prod_{j=0}^r\left(1-\frac D{2j+1}\right)[xh(w)]
		=\frac{(-2)^{r+1}}{(2r+1)!!}w^{r+3/2}h^{(r+1)}(w)
	\end{equation}
	implies, with the constant $A_r=2^r\pi/(2r)!$ of
	Eq.~\eqref{pos:dirac-resolvent},
	\begin{equation}
		-DC_d=\frac{2^{r+1}A_r}{(2r+1)!!}w^{r+3/2}k_r(w)>0,
		\label{pos:dirac-loss}
	\end{equation}
	and Eqs.~\eqref{pos:preservation} and \eqref{pos:dirac-loss} prove
	Eq.~\eqref{pos:dirac-normalized}.
	
	It remains to fix the endpoints. The integration constant is fixed by
	the massless zeta determinant, and the resulting primitive is
	\begin{align}
		(-1)^{r+1}F_d(x)
		={}&A_r\int_0^xQ_r(t^2)\,dt
		\nonumber\\*
		&+2A_r\int_x^\infty\frac{Q_r(t^2)}{e^{2\pi t}+1}\,dt.
		\label{pos:dirac-primitive}
	\end{align}
	To check the constant, we expand $Q_r(w)=\sum_jq_jw^j$ and use
	$\zeta(z,1/2)=(2^z-1)\zeta(z)$. Zeta differentiation gives
	\begin{align}
		(-1)^{r+1}F_d(0)=\frac{2^r}{(2r)!}\bigg[
		&q_0\log2+\sum_{j=1}^r q_j(1-2^{-2j})
		\nonumber\\[-2pt]
		&\hspace{8mm}\times\frac{(2j)!\zeta(2j+1)}{(2\pi)^{2j}}
		\bigg],
	\end{align}
	which is the Fermi integral in Eq.~\eqref{pos:dirac-primitive} at $x=0$.
	The first term of that equation is a polynomial annihilated by
	$\mathcal P_d$, while the second term and all its fixed Euler derivatives
	decay exponentially in the infrared. Hence $C_d(0)>0$ and
	$C_d(\infty)=0$. The proof concerns the absolute determinant, so that a
	parity-anomaly phase does not enter $-\log|Z|$.
	\section{The surface-defect classification}
	\label{dt:defect-core}
	
	In this appendix we prove the results of Sec.~\ref{dt:defect}.
	Appendix~\ref{dt:defect-trace-proof} derives the trace identity
	\eqref{dt:defect-identity} and the endpoint values \eqref{dt:b-endpoints}.
	Appendix~\ref{dt:defect-monotonicity-proof} proves strict monotonicity
	for $0<\varepsilon\le1$, and Appendix~\ref{dt:defect-ir-proof} proves the
	infrared expansion \eqref{dt:defect-IR} for $1<\varepsilon<2$.
	Appendix~\ref{bmin:onset} works out the minimum of
	${\cal S}_b$ near the threshold.
	
	\subsection{The defect trace identity and endpoints}
	\label{dt:defect-trace-proof}
	
	We derive \eqref{dt:defect-identity}, in which
	$F_{\rm def}=-\log(Z_{\rm defect}/Z_{\rm bulk})$ and $b_{\rm UV}$ is the
	UV defect anomaly coefficient, including the contact term, in three
	steps: we compute the trace correlator of the Gaussian defect, we
	relate it to ${\cal S}_b$ through the spherical Ward identity, and we reduce the resulting angular integral to $J_\varepsilon^{(2)}$ of
	Eq.~\eqref{dt:defect-identity}. We then derive the endpoint values.
	
	Let $f>0$ multiply $\tfrac12\sum_{A=1}^{\nu}\widehat{\mathcal O}_A^2$ in
	the defect action, and let $G_0$ be the undeformed two-point function,
	viewed as an integral operator on $S_R^2$. As in Eq.~\eqref{dt:defect-spectrum}, its eigenvalues are
	$\lambda_n=\lambda_0/p_n$, with $x=f\lambda_0$ and the mode variables
	$u_n$ defined there, and $D=\varepsilon x\partial_x$ at fixed $f$. Gaussian integration gives
	the free energy and the deformed two-point function $G_f$,
	\begin{align}
		F_{\rm def}-F_{\rm def,UV}
		&=\frac{\nu}{2}\operatorname{Tr}\log(1+fG_0),\nonumber\\
		G_f&=G_0(1+fG_0)^{-1},
		\label{dt:defect-resummation}
	\end{align}
	with the local terms in $\operatorname{Tr}\log(1+fG_0)$ defined by the
	ultraviolet prescription of Sec.~\ref{ob:section}. Expanding $G_f$ in harmonics orthonormal with
	respect to the physical area measure, the addition theorem gives the
	Legendre series
	\begin{align}
		fG_f(z)&=\frac{\mathcal U(z)}{4\pi R^2},\nonumber\\
		\mathcal U(z)&=\sum_{n=0}^{\infty}(2n+1)u_nP_n(z),
		\qquad z=\Omega\cdot\Omega',
		\label{dt:defect-angular-kernel}
	\end{align}
	where the $P_n$ are Legendre polynomials. At separated points, the
	trace of the defect stress tensor and its connected two-point function,
	obtained by Wick contraction, are
	\begin{align}
		\widehat T&=-\frac{\varepsilon f}{2}
		\sum_{A=1}^{\nu}\widehat{\mathcal O}_A^2,\nonumber\\
		\langle\widehat T(\sigma)\widehat T(0)\rangle_c
		&=\frac{\nu\varepsilon^2 f^2}{2}G_f(\sigma,0)^2 .
		\label{dt:defect-trace-correlator}
	\end{align}
	These formulas are exact in the Gaussian realization. In a parent CFT
	with large-$N$ factorization, they give the leading contribution to the
	defect free-energy difference and to the trace correlator.
	
	Next we relate the trace correlator to the running function. Two
	objects enter: the separated-point correlator of
	Eq.~\eqref{dt:defect-trace-correlator}, which excludes contact terms,
	and the full second variation of the free energy, which includes them.
	The spherical trace Ward identity \cite{ShacharSinhaSmolkin2023} identifies the running function ${\cal S}_b$ of
	Eq.~\eqref{dt:defect-identity} with
	\begin{equation}
		{\cal S}_b=\frac{b_{\rm UV}}3
		-\pi\int_{S_R^2}dA\,s^2(\sigma,0)
		\langle\widehat T(\sigma)\widehat T(0)\rangle_c,
		\label{dt:defect-trace-integral}
	\end{equation}
	where $s^2=2R^2(1-z)$ is the squared chordal distance, and the
	correlator excludes the ultraviolet anomaly contact. To fix the signs,
	we use the dilaton convention
	$\delta\log(Z_{\rm defect}/Z_{\rm bulk})
	=\int dA\,\langle\widehat T\rangle\delta\tau$, in which a constant $\tau$
	implements a logarithmic change of radius. At $\tau=0$, its first two
	derivatives give
	\begin{align*}
		DF_{\rm def}&=-\int dA\,\langle\widehat T\rangle,\\
		D^2F_{\rm def}&=-\int dA\,dA'\,
		\langle\widehat T\widehat T'\rangle_c,
	\end{align*}
	where the second variation includes local contacts from varying the
	density $dA\,\widehat T$. Since
	${\cal S}_b=-DF_{\rm def}+D^2F_{\rm def}/2$, the constant and
	first-harmonic Ward identities take the form
	\begin{align}
		{\cal S}_b
		&=\int dA\,\langle\widehat T\rangle
		-\frac12\int dA\,dA'\,
		\langle\widehat T\widehat T'\rangle_c,\nonumber\\*
		\int dA\,\langle\widehat T\rangle
		&=\frac12\int dA\,dA'\,
		(\Omega\cdot\Omega')\langle\widehat T\widehat T'\rangle_c .
		\label{dt:defect-integrated-ward}
	\end{align}
	For the second identity we choose the first harmonics
	$\tau=\Omega_a$. On the unit sphere the conformal Killing fields
	$\xi_a=\nabla\Omega_a$ obey $\nabla\cdot\xi_a=-2\Omega_a$ and
	$\sum_a\xi_a\cdot\nabla\Omega_a=2$, so that summing the corresponding
	dilaton Ward identities gives the factor $1/2$ and the weight
	$\Omega\cdot\Omega'$ displayed above. Combining the two identities then
	gives the weight $1-z=s^2/(2R^2)$.
	
	It remains to evaluate the anomaly contact, which fixes the constant
	term $b_{\rm UV}/3$. At the ultraviolet fixed point,
	$\langle\widehat T\rangle=b_{\rm UV}\mathcal R/(24\pi)$, and varying the
	trace itself gives the contact
	\begin{equation}
		-\frac{b_{\rm UV}}{12\pi}
		(\mathcal R+\nabla^2)\delta_{S_R^2}(\sigma,\sigma'),
		\qquad \mathcal R=\frac2{R^2}.
	\end{equation}
	The variation of $dA$ adds $2\langle\widehat T\rangle\delta_{S_R^2}$ and
	cancels its $\mathcal R\delta_{S_R^2}$ part in the full second variation.
	Since this undifferentiated delta function also vanishes against the
	weight $s^2$, either expression contributes $b_{\rm UV}/3$, because
	$s^2(0,0)=0$ and $\nabla^2s^2(\sigma,0)|_{\sigma=0}=4$. Rotational
	invariance then gives \eqref{dt:defect-trace-integral}.
	
	We now reduce the angular integral. Substituting
	\eqref{dt:defect-angular-kernel} and \eqref{dt:defect-trace-correlator}
	gives
	\begin{equation}
		{\cal S}_b-\frac{b_{\rm UV}}3
		=-\frac{\nu\varepsilon^2}{8}
		\int_{-1}^{1}dz\,(1-z)\mathcal U(z)^2,
		\label{dt:defect-angular-integral}
	\end{equation}
	and Legendre orthogonality together with
	\begin{equation}
		zP_n(z)=\frac{n+1}{2n+1}P_{n+1}(z)
		+\frac{n}{2n+1}P_{n-1}(z)
	\end{equation}
	reduces the angular integral to $J_\varepsilon^{(2)}$. The reduction
	requires care, because the unweighted harmonic sum need not be square
	integrable. We therefore replace $u_n$ by $r^nu_n$, with $0<r<1$, before
	applying orthogonality. The result is
	\begin{equation}
		2\sum_{n\ge1}n
		(r^{n-1}u_{n-1}-r^nu_n)^2.
	\end{equation}
	We put $v_n=u_{n-1}-u_n$. At fixed $x$, $u_n=O(n^{-\varepsilon})$ and
	$v_n=O(n^{-\varepsilon-1})$. Hence the part $r^{n-1}v_n$ converges in
	the norm $\sum_{n\ge1}n|v_n|^2$, while the squared norm of the remaining
	part is bounded by
	\begin{equation}
		C_x(1-r)^2\sum_{n\ge1}n^{1-2\varepsilon}r^{2n-2}
		\longrightarrow0.
	\end{equation}
	Consequently, the limit $r\to1$ gives
	\begin{equation}
		\int_{-1}^{1}dz\,(1-z)\mathcal U(z)^2
		=2\sum_{n\ge1}n v_n^2 ,
	\end{equation}
	which proves \eqref{dt:defect-identity} for every $0<\varepsilon<2$.
	All fixed orders of scale differentiation preserve the absolute
	convergence, since they multiply $u_n$ and $v_n$ by bounded polynomials
	in the $u_n$.
	
	Finally, we derive the endpoint values. For this purpose we rewrite
	$J_\varepsilon^{(2)}$ as a convergent sum of cubes, from which both
	limits can be read off. The eigenvalue recurrence gives
	\begin{equation}
		nv_n=\frac{\varepsilon}{2}
		(u_{n-1}+u_n-2u_{n-1}u_n).
		\label{dt:defect-v-recurrence}
	\end{equation}
	For $A=u_{n-1}$ and $B=u_n$, the elementary identity
	\begin{equation}
		(A-B)(A+B-2AB)
		=A^2-\frac23A^3-B^2+\frac23B^3
		+\frac23(A-B)^3
	\end{equation}
	telescopes to
	\begin{equation}
		J_\varepsilon^{(2)}
		=\varepsilon\left(u_0^2-\frac23u_0^3\right)
		+\frac{2\varepsilon}{3}\sum_{n\ge1}v_n^3 ,
		\label{dt:defect-cubic}
	\end{equation}
	where $u_0=x/(1+x)$ and $\sum_{n\ge1}v_n=u_0$. For every fixed $n$,
	$v_n\to0$ as $x\to\infty$. The recurrence \eqref{dt:defect-v-recurrence}
	also gives $v_n\le\varepsilon/(2n)$, which bounds the tail of
	$\sum v_n^3$ uniformly by a convergent series and proves that the cubic
	sum tends to zero. It follows that
	\begin{equation}
		J_\varepsilon^{(2)}(0)=0,\qquad
		J_\varepsilon^{(2)}(\infty)=\frac{\varepsilon}{3},
	\end{equation}
	and Eq.~\eqref{dt:b-endpoints} follows directly.
	
	\subsection{Strict monotonicity of the defect function}
	\label{dt:defect-monotonicity-proof}
	
	We now prove that $(J_\varepsilon^{(2)})'>0$ for $0<\varepsilon\le1$.
	The proof has three steps. We reduce the claim to a bound on the cubic
	sum of the differences $v_n$, we show that this sum is largest at
	$\varepsilon=1$, and we compute it at $\varepsilon=1$ in closed form. We
	write
	\begin{equation}
		\mathcal V_\varepsilon(x)=\sum_{n\ge1}v_n^3,
		\qquad
		\mathcal M(x)=u_0^2(1-u_0)^2=\frac{x^2}{(1+x)^4}.
	\end{equation}
	The identities
	\begin{equation}
		x\partial_xu_n=u_n(1-u_n),\qquad
		x\partial_xv_n=v_n(1-u_{n-1}-u_n)
	\end{equation}
	and \eqref{dt:defect-cubic} imply
	\begin{align}
		x(J_\varepsilon^{(2)})'(x)
		=2\varepsilon\bigg[
		\mathcal M(x)+\sum_{n\ge1}v_n^3(1-u_{n-1}-u_n)
		\bigg].
		\label{dt:defect-cubic-derivative}
	\end{align}
	Since $0<u_n<1$, the sum in brackets is bounded below by
	$\mathcal M-\mathcal V_\varepsilon$. Hence it suffices to show that
	\begin{equation}
		\mathcal V_\varepsilon(x)<\frac45\mathcal M(x),
		\qquad 0<\varepsilon\le1.
		\label{dt:defect-cubic-bound}
	\end{equation}
	
	We first compare different values of $\varepsilon$ at fixed $x$. The
	recurrence gives
	\begin{equation}
		\frac{v_{n+1}}{v_n}
		=\frac{n+\varepsilon/2}{n+1-\varepsilon/2}
		\frac{x+p_{n-1}}{x+p_{n+1}}<1
		\quad(0<\varepsilon\le1),
		\label{dt:defect-v-decreasing}
	\end{equation}
	and the inverse eigenvalue
	\begin{equation}
		p_n^{-1}
		=\prod_{j=1}^{n}\frac{j-\varepsilon/2}{j+\varepsilon/2}
	\end{equation}
	strictly decreases with $\varepsilon$. Thus the sequence
	$a_n(\varepsilon)=v_n(\varepsilon)/u_0$, which is positive,
	nonincreasing, and sums to one, has partial sums
	\begin{equation}
		\sum_{n=1}^{m}a_n(\varepsilon)
		=1-\frac{u_m(\varepsilon)}{u_0}
	\end{equation}
	that increase with $\varepsilon$. These ordered partial sums imply
	\begin{equation}
		\sum_{n\ge1}(a_n(\varepsilon)-s)_+
		\le\sum_{n\ge1}(a_n(1)-s)_+,\qquad s\ge0,
	\end{equation}
	since either side is the supremum over $m\ge0$ of the corresponding
	partial sum minus $ms$. In other words, the sequence at $\varepsilon=1$
	is the more concentrated one, and a sum of cubes grows with
	concentration. Multiplication by $6s$, integration, and the identity
	$z^3=6\int_0^\infty ds\,s(z-s)_+$ prove
	\begin{equation}
		\mathcal V_\varepsilon(x)\le\mathcal V_1(x),
		\label{dt:defect-majorization}
	\end{equation}
	where all interchanges are justified by nonnegativity.
	
	Finally, we compute the cubic sum $\mathcal V_1$ at $\varepsilon=1$,
	where $p_n=2n+1$. With
	$c=1+x$,
	\begin{equation}
		v_n=\frac{2x}{(c+2n-2)(c+2n)},
	\end{equation}
	and the partial-fraction identity
	\begin{equation}
		\frac1{t^3(t+1)^3}
		=\frac6t-\frac3{t^2}+\frac1{t^3}
		-\frac6{t+1}-\frac3{(t+1)^2}
		-\frac1{(t+1)^3}
	\end{equation}
	therefore gives
	\begin{equation}
		\mathcal V_1(x)=x^3\left[
		\frac3{2c}+\frac3{2c^2}+\frac1{c^3}
		-\frac34\psi_1\left(\frac c2\right)
		\right],
		\label{dt:defect-cubic-trigamma}
	\end{equation}
	where $\psi_1$ is the trigamma function. Its required lower bound follows
	from the convergent Mittag--Leffler expansion
	\begin{equation}
		y\coth y
		=1+2y^2\sum_{k\ge1}\frac1{y^2+\pi^2k^2}
		\ge1+\frac{y^2}{3}-\frac{y^4}{45},
	\end{equation}
	term by term, using $(a+b)^{-1}\ge a^{-1}-ba^{-2}$. Setting $y=s/2$ and
	using the integral representation of $\psi_1$, we obtain
	\begin{align*}
		\psi_1(z)
		&=\int_0^\infty ds\,e^{-zs}\frac{s}{1-e^{-s}}\\
		&\ge\frac1z+\frac1{2z^2}+\frac1{6z^3}-\frac1{30z^5},
		\qquad z>0.
	\end{align*}
	Substitution into \eqref{dt:defect-cubic-trigamma} gives
	\begin{equation}
		\mathcal V_1(x)\le\frac{4x^3}{5(1+x)^5}
		<\frac45\mathcal M(x),
	\end{equation}
	which, together with \eqref{dt:defect-majorization}, proves
	\eqref{dt:defect-cubic-bound}. Hence \eqref{dt:defect-cubic-derivative}
	implies
	\begin{align}
		x(J_\varepsilon^{(2)})'(x)
		&\ge\frac{2\varepsilon}{5}\frac{x^2}{(1+x)^4}>0,\nonumber\\
		D{\cal S}_b
		&\le-\frac{\nu\varepsilon^4}{20}
		\frac{x^2}{(1+x)^4}<0
		\quad(0<\varepsilon\le1).
		\label{dt:defect-global-bound}
	\end{align}
	This is an analytic proof for every positive coupling, including the
	threshold $\varepsilon=1$. It uses the specific eigenvalue recurrence, in
	addition to the positivity of the trace two-point function.
	
	\subsection{The defect infrared tail}
	\label{dt:defect-ir-proof}
	
	For $1<\varepsilon<2$, the leading infrared term in \eqref{dt:defect-IR}
	follows from the convergent identity \eqref{dt:defect-cubic}, because at
	large $x$ the cubic sum becomes a Riemann sum. The inverse eigenvalues satisfy
	$p_n^{-1}=c_\varepsilon n^{-\varepsilon}[1+O(n^{-1})]$, with
	$c_\varepsilon$ defined below Eq.~\eqref{dt:defect-IR}, and we put $L=(c_\varepsilon x)^{1/\varepsilon}$. For fixed
	$1<\varepsilon<2$, the rescaled differences tend to
	\begin{equation}
		Lv_n\longrightarrow
		\frac{\varepsilon z^{\varepsilon-1}}
		{(1+z^\varepsilon)^2},
		\qquad z=n/L .
	\end{equation}
	Since the recurrence supplies integrable bounds at small and large $z$,
	the Riemann sum gives
	\begin{align}
		\sum_{n\ge1}v_n^3
		&=L^{-2}\varepsilon^2
		\mathrm B(3-2/\varepsilon,3+2/\varepsilon)
		+o(L^{-2}).
	\end{align}
	The same argument applies after any fixed number of scale derivatives,
	which insert bounded polynomials in $u_n$. Meanwhile,
	$u_0^2-2u_0^3/3=1/3-x^{-2}+O(x^{-3})$. Since $2/\varepsilon<2$, the cubic sum dominates, and substitution in
	\eqref{dt:defect-cubic} proves \eqref{dt:defect-IR}. The Riemann-sum
	argument requires only that $V^3$, with
	$V(z)=\varepsilon z^{\varepsilon-1}/(1+z^\varepsilon)^2$, be integrable
	at small $z$, and the low modes $n<\sqrt L$ obey the first inequality in
	Eq.~\eqref{bmin:low-modes}, whose right-hand side vanishes for
	$\varepsilon>2/3$. The argument therefore holds for every
	$2/3<\varepsilon<2$. For $2/3<\varepsilon<1$ the cubic sum is then
	$O(x^{-2/\varepsilon})=o(x^{-2})$, and
	$J_\varepsilon^{(2)}-\varepsilon/3=-\varepsilon x^{-2}+o(x^{-2})$. At
	$\varepsilon=1$ the two exponents coincide, and
	$J_1^{(2)}-1/3=(A_2(1)-1)x^{-2}+o(x^{-2})$, with $A_2(1)=8/15$. Together with the endpoint ordering in
	\eqref{dt:b-endpoints}, this establishes the obstruction for
	$1<\varepsilon<2$ through the infrared criterion of Sec.~\ref{ob:ir}.
	
	\subsection{Onset of the minimum near the threshold}
	\label{bmin:onset}
	
	We now work out the location and the depth of the minimum of
	${\cal S}_b$ near the threshold. Set $\delta=\varepsilon-1>0$. At a
	stationary point, the two leading IR terms in
	$x\partial_xJ_\varepsilon^{(2)}$, given in Eq.~\eqref{bmin:uniform}
	below, balance as
	$A_2(\varepsilon)x_{\min}^{-2/\varepsilon}\sim\varepsilon^2x_{\min}^{-2}$,
	with $A_2(\varepsilon)$ from Eq.~\eqref{dt:defect-IR}, which defines the
	scale
	\begin{equation}
		K_\varepsilon=\frac{\varepsilon^2}{A_2(\varepsilon)},
		\qquad X_\varepsilon=K_\varepsilon^{\varepsilon/(2\delta)}.
		\label{bmin:scale}
	\end{equation}
	For sufficiently small $\delta$, ${\cal S}_b$ has a unique stationary
	point, its global minimum, with
	\begin{equation}
		x_{\min}\sim X_\varepsilon,\qquad
		d_\varepsilon\equiv{\cal S}_{b,\rm IR}-{\cal S}_b(x_{\min})
		\sim\frac{\nu\varepsilon^3\delta}{8X_\varepsilon^2}.
		\label{bmin:location-depth}
	\end{equation}
	Here ${\cal S}_{b,\rm IR}=b_{\rm IR}/3$ denotes the infrared value. The
	scale diverges exponentially,
	\begin{align}
		X_\varepsilon&\sim X_* e^{\log(15/8)/(2\delta)},\nonumber\\
		X_*&=\sqrt{\frac{15}{32}}\,e^{31/12-\gamma}.
		\label{bmin:essential}
	\end{align}
	In this formula $\gamma$ is Euler's constant.
	
	\label{bmin:proof}
	We now prove these statements, together with the limiting profile of ${\cal S}_b$ near its minimum: uniformly on compact subsets of
	$z>0$,
	\begin{equation}
		\frac{{\cal S}_{b,\rm IR}-{\cal S}_b(zX_\varepsilon)}
		{d_\varepsilon}\longrightarrow\frac{1+2\log z}{z^2}.
		\label{bmin:profile}
	\end{equation}
	We work in the normalization $p_0=1$ and write $J=J_\varepsilon^{(2)}$,
	$E=x\partial_x$, $v_n=u_{n-1}-u_n$, $w_n=1-u_{n-1}-u_n$, and
	$a_n=u_n(1-u_n)$. The proof has three steps: uniform estimates of the
	exact identities for $J$, $EJ$, and $E^2J$, the localization of all
	stationary points, and the evaluation of the depth.
	
	For the first step, the exact cubic identity \eqref{dt:defect-cubic} and
	its first derivative \eqref{dt:defect-cubic-derivative} give
	\begin{align}
		E^2J=2\varepsilon\bigg[
		&\frac{2x^2(1-x)}{(1+x)^5}
		\nonumber\\
		&+\sum_{n\ge1}v_n^3
		\big(3w_n^2-a_{n-1}-a_n\big)\bigg],
		\label{bmin:exact-second}
	\end{align}
	where the series and their differentiated versions converge locally
	uniformly in $x>0$ and $\varepsilon$ near one. We restrict to
	$1\le\varepsilon\le9/8$ and define
	\begin{align*}
		L&=(c_\varepsilon x)^{1/\varepsilon},
		&U(z)&=\frac1{1+z^\varepsilon},\\
		V(z)&=\frac{\varepsilon z^{\varepsilon-1}}
		{(1+z^\varepsilon)^2},
		&W(z)&=\frac{z^\varepsilon-1}{z^\varepsilon+1}.
	\end{align*}
	The exact gamma ratio gives
	$p_n^{-1}=c_\varepsilon n^{-\varepsilon}[1+O(n^{-1})]$, uniformly on this
	parameter interval. For $n\ge N=\lfloor\sqrt L\rfloor$, its recurrence
	implies
	\begin{align*}
		v_n&=L^{-1}V(n/L)[1+O(n^{-1})],\\
		u_n&=U(n/L)+O(n^{-1}),
		\qquad w_n=W(n/L)+O(n^{-1}),
	\end{align*}
	while for $n<N$ the bound $v_n\le C L^{-\varepsilon}n^{\varepsilon-1}$
	gives instead
	\begin{equation}
		L^2\sum_{n<N}v_n^3
		\le C L^{2-3\varepsilon}N^{3\varepsilon-2}
		\le C L^{-1/2}.
		\label{bmin:low-modes}
	\end{equation}
	The integral over the omitted interval $0<z<N/L$ obeys the same bound.
	The function $V$ is uniformly bounded and unimodal, while $U$ and $W$ are
	bounded and monotone. Thus $V^3$, times any of the bounded polynomial
	multipliers in the three exact identities, has uniformly bounded
	variation, and its Riemann-sum error is $O(L^{-1})$. The errors in the
	discrete approximations for $n\ge N$ are $O(N^{-1})$ after
	multiplication by $L^2$.
	
	We put $B_\varepsilon=\mathrm B(3-2/\varepsilon,3+2/\varepsilon)$. The
	required integrals are
	\begin{align*}
		\int_0^\infty V^3\,dz&=\varepsilon^2B_\varepsilon,\\
		\int_0^\infty V^3W\,dz&=-\frac{2\varepsilon}{3}B_\varepsilon,\\
		\int_0^\infty V^3\big[3W^2-2U(1-U)\big]\,dz
		&=\frac43 B_\varepsilon.
	\end{align*}
	For example, $t=z^\varepsilon$ reduces the first integral to a beta
	integral with parameters $3-2/\varepsilon$ and $3+2/\varepsilon$, and the
	remaining two follow by inserting $W=(t-1)/(t+1)$ and $U=1/(1+t)$. The
	preceding estimates therefore give an error $O(L^{-5/2})$ in each cubic
	sum, and expanding the explicit $u_0=x/(1+x)$ terms proves
	\begin{align}
		J-\frac\varepsilon3
		&=A_2(\varepsilon)x^{-2/\varepsilon}
		-\varepsilon x^{-2}+O(x^{-2-\kappa}),
		\nonumber\\
		EJ&=2\varepsilon x^{-2}
		-\frac2\varepsilon A_2(\varepsilon)x^{-2/\varepsilon}
		+O(x^{-2-\kappa}),
		\nonumber\\
		E^2J&=-4\varepsilon x^{-2}
		+\frac4{\varepsilon^2}A_2(\varepsilon)x^{-2/\varepsilon}
		+O(x^{-2-\kappa}),
		\label{bmin:uniform}
	\end{align}
	uniformly for $x\ge1$, with $\kappa=2/9$. Indeed, $5/(2\varepsilon)\ge
	20/9$ on the chosen interval. The three estimates were obtained
	separately from exact identities, without differentiating an unspecified
	remainder.
	
	For the second step, we locate all stationary points as
	$\delta=\varepsilon-1\downarrow0$. At $\varepsilon=1$, $EJ>0$ for every
	$x>0$ by \eqref{dt:defect-global-bound}. This sign persists on each
	compact positive $x$ interval, and it also holds uniformly near zero. To
	see the latter statement, we set $q_n=p_n^{-1}$ and expand the convergent
	series to obtain
	\begin{equation}
		EJ=4x^2\sum_{n\ge1}n(q_{n-1}-q_n)^2+O(x^3),
	\end{equation}
	where the coefficient of $x^2$ is continuous and strictly positive near
	$\varepsilon=1$, and the remainder is uniform there. Consequently, every
	sequence of stationary points approaching the threshold must have
	$x\to\infty$. At such a point the second line of \eqref{bmin:uniform}
	implies
	\begin{equation}
		x^{2\delta/\varepsilon}
		=K_\varepsilon+O(x^{-\kappa}).
		\label{bmin:root-equation}
	\end{equation}
	Since $K_1=15/8>1$, every root satisfies
	\begin{equation}
		\delta\log x\longrightarrow\tfrac12\log(15/8),
		\qquad
		\log\frac{x}{X_\varepsilon}
		=O(x^{-\kappa}/\delta)=o(1),
	\end{equation}
	where the last limit follows because the first relation makes
	$x^{-\kappa}$ exponentially small in $1/\delta$. This localizes all
	possible roots to $x/X_\varepsilon\to1$.
	
	To show that there is exactly one root, we rescale $x=zX_\varepsilon$.
	The leading balance $A_2X_\varepsilon^{-2/\varepsilon}
	=\varepsilon^2X_\varepsilon^{-2}$ then gives
	\begin{align}
		\frac{EJ(zX_\varepsilon)}{\delta X_\varepsilon^{-2}}
		&\longrightarrow-4z^{-2}\log z,
		\nonumber\\
		\frac{E^2J(zX_\varepsilon)}{\delta X_\varepsilon^{-2}}
		&\longrightarrow4z^{-2}(2\log z-1),
		\label{bmin:derivative-limits}
	\end{align}
	and both limits are uniform on compact positive $z$ intervals. The first
	limit has opposite signs on either side of $z=1$, so that a stationary
	point exists. We choose a fixed $0<c<1/2$. The second limit is strictly
	negative on $|\log z|\le c$, and all stationary points eventually lie in
	this interval, where $EJ$ is strictly decreasing. Hence there is exactly
	one stationary point, which is the global maximum of $J$ and therefore
	the global minimum of ${\cal S}_b$. This proves uniqueness sufficiently
	near the threshold, which is all that the onset law requires.
	
	For the third step, we evaluate the depth. At the minimum, the root
	equation controls the location error exponentially in $1/\delta$.
	Substitution in the first line of \eqref{bmin:uniform} therefore
	controls the cancellation at order $\delta$ and gives
	\begin{equation}
		J(x_{\min})-\frac\varepsilon3
		\sim\varepsilon\delta X_\varepsilon^{-2}.
	\end{equation}
	Multiplication by $\nu\varepsilon^2/8$ proves
	\eqref{bmin:location-depth}. For fixed positive $z$, the same expansion
	yields
	\begin{align*}
		\frac{J(zX_\varepsilon)-\varepsilon/3}
		{\varepsilon\delta X_\varepsilon^{-2}}
		&=\frac{\varepsilon z^{-2/\varepsilon}-z^{-2}}\delta+o(1)\\
		&\longrightarrow z^{-2}(1+2\log z),
	\end{align*}
	which proves \eqref{bmin:profile}. The first limit in
	\eqref{bmin:derivative-limits} also gives
	\begin{equation}
		\frac{D{\cal S}_b(zX_\varepsilon)}
		{\varepsilon d_\varepsilon}
		\longrightarrow\frac{4\log z}{z^2}.
		\label{bmin:derivative-profile}
	\end{equation}
	Finally, direct differentiation of the gamma and beta functions gives
	\begin{equation}
		\left.\partial_\varepsilon\log K_\varepsilon\right|_{\varepsilon=1}
		=\frac{31}{6}-2\gamma-2\log2,
	\end{equation}
	and the expansion of
	$\log X_\varepsilon=\varepsilon\log K_\varepsilon/(2\delta)$ proves
	\eqref{bmin:essential}.
	
	\section{Three-sphere double-trace classification}
	\label{dt:appendix}
	
	In this appendix we prove the sufficient direction of
	Eq.~\eqref{dt:sharp} by discrete summation identities, and we derive the
	infrared expansion, which implies the converse and locates the minimum
	near the threshold. The sign proof uses exact polynomial identities and
	rational coefficient inequalities, and it does not require the
	positivity of a Laplace-transform density. The second summation identity
	proves monotonicity for $0<\varepsilon\le13/10$, while a further
	identity controls the remaining interval up to $\sqrt{37/13}$, with
	separate estimates for the first mode and for the higher modes.
	Appendix~\ref{dt:convergence} defines the summation variables and derives the convergent sum
	\eqref{dt:convergent}, Appendices~\ref{dt:polynomial-proof} and
	\ref{dt:polynomial-estimate} prove monotonicity on the two parameter
	intervals, and the last two subsections derive the infrared expansion
	from the spectral zeta function.
	
	\subsection{Convergence and the first summation identity}
	\label{dt:convergence}
	
	We first state the convergent sum. Define
	\begin{align}
		u_n&=\frac{x}{x+p_\varepsilon(n)},\qquad
		v_n=u_n-u_{n+1},\nonumber\\
		w_n&=1-u_n-u_{n+1},
		\label{dt:variables}\\
		{\cal A}_\varepsilon(w,v)
		&=(1-\varepsilon w)
		(\varepsilon^2w^2-\varepsilon w+1-\varepsilon^2)
		+\varepsilon^3wv^2.
		\label{dt:Apoly}
	\end{align}
	Here $\mathcal A_\varepsilon$ collects the terms produced by the
	recurrence
	$p_\varepsilon(n+1)/p_\varepsilon(n)=(2n+1+\varepsilon)/(2n+1-\varepsilon)$,
	which gives the absolutely convergent sum
	\begin{align}
		-\frac{12DC_\varepsilon}{\nu\varepsilon^2}
		={}&\varepsilon^2u_1^2(1-u_1)^2\nonumber\\
		&+\frac14\sum_{n\ge1}v_n^2
		{\cal A}_\varepsilon(w_n,v_n).
		\label{dt:convergent}
	\end{align}
	We derive it in three steps: a finite-cutoff summation identity relates
	the scale derivative to a positive sum, a telescoping identity rewrites
	that sum, and analytic continuation extends the result to the whole
	interval. For $\varepsilon>1/2$, the sum
	\begin{equation}
		J_\varepsilon=\sum_{n\ge1}n(n+1)v_n^2
		\label{dt:J}
	\end{equation}
	converges. At cutoff $N$, expanding the square, shifting the index, and
	using the recurrence
	$u_nu_{n+1}=[(2n+1+\varepsilon)u_{n+1}-(2n+1-\varepsilon)u_n]/(2\varepsilon)$
	give the finite identity
	\begin{align}
		\sum_{n=1}^Nn(n+1)v_n^2
		={}&\sum_{n=1}^Nn^2\left(2u_n^2+\frac{2(3-\varepsilon)}{\varepsilon}u_n\right)
		+\mathcal T_N,\nonumber\\
		\mathcal T_N={}&N(N+1)u_{N+1}^2\nonumber\\
		&-\frac2\varepsilon N(N+1)\left(N+\frac{1+\varepsilon}{2}\right)u_{N+1}.
		\label{dt:cutoff-identity}
	\end{align}
	Since $Du_n=\varepsilon u_n(1-u_n)$, each mode of the determinant obeys,
	with $P_3$ from Eq.~\eqref{st:filter},
	\begin{align}
		&\frac{\nu\varepsilon}{2}n^2P_3(D)u_n\nonumber\\
		&\quad=-\frac{\nu\varepsilon^2}{12}(D-1)\left[n^2\left(2u_n^2
		+\frac{2(3-\varepsilon)}{\varepsilon}u_n\right)\right],
		\label{dt:mode-identity}
	\end{align}
	and the sum of the left-hand side over $n$ is the scale derivative of
	the subtracted determinant $C_\varepsilon$ of
	Eq.~\eqref{dt:filtered}. For $\varepsilon>3/2$, subtracting the term
	linear in $x$ makes all the sums converge, and the nonlinear part of
	$\mathcal T_N$ tends to zero. Hence, in this range,
	\begin{equation}
		DC_\varepsilon
		=-\frac{\nu\varepsilon^2}{12}(D-1)J_\varepsilon.
		\label{dt:ward}
	\end{equation}
	The positivity of $J_\varepsilon$ does not determine the sign of
	$(D-1)J_\varepsilon$.
	
	To rewrite the sum $J_\varepsilon$, we set $z_n=1-2u_n$. The eigenvalue recurrence
	gives
	\begin{equation}
		z_{n+1}=\frac{(2n+1)z_n+\varepsilon}{2n+1+\varepsilon z_n},
		\label{dt:mobius}
	\end{equation}
	and it also implies
	\begin{equation}
		Dv_n=\varepsilon w_nv_n,\qquad
		(4n+2)v_n=\varepsilon(1-w_n^2+v_n^2).
		\label{dt:recurrence}
	\end{equation}
	For $T_n=-\varepsilon^2n u_n^2(1-u_n)^2$, these relations give
	\begin{align}
		n(n+1)v_n^2(2\varepsilon w_n-1)
		={}&T_{n+1}-T_n\nonumber\\
		&+\frac{v_n^2}{4}{\cal A}_\varepsilon(w_n,v_n),
		\label{dt:telescoping}
	\end{align}
	which is a finite algebraic identity. Since $T_n\to0$ for
	$\varepsilon>1/2$, summing it and using \eqref{dt:ward} proves
	\eqref{dt:convergent} for $\varepsilon>3/2$.
	
	To extend \eqref{dt:convergent} to the whole interval, we continue the convergent
	final identity, rather than the separate divergent terms in
	\eqref{dt:telescoping}. Locally in the complex $\varepsilon$ plane,
	\begin{equation}
		p_\varepsilon(n)=n^\varepsilon[1+O(n^{-2})],
		\qquad v_n=O(n^{-\varepsilon-1}),
	\end{equation}
	so that the sum in \eqref{dt:convergent} converges absolutely and
	locally uniformly near every real $0<\varepsilon<2$. The same
	analyticity holds for the subtracted free energy. Indeed, we choose $K$
	so that $(K+1)\mathop{\rm Re}\varepsilon>3$, subtract the first $K$
	terms in the large-eigenvalue expansion of the logarithm, and restore them with the spectral zeta function
	$\mathcal Z_\varepsilon(k)$ of Sec.~\ref{dt:section}. The subtracted sum also converges
	absolutely and locally uniformly, while the restored traces can have
	positive-$\varepsilon$ poles only at $k\varepsilon=3$ and
	$k\varepsilon=1$. The factor
	\begin{equation}
		k\varepsilon(1-k\varepsilon)(1-k\varepsilon/3)
	\end{equation}
	from the scale derivative and the subtraction polynomial removes both
	poles, and at a resonance its limit agrees with the retained physical
	logarithmic term. Therefore, the real-analytic identity theorem proves
	\eqref{dt:convergent} throughout $0<\varepsilon<2$.
	
	\subsection{The second summation identity}
	\label{dt:polynomial-proof}
	
	The individual summands in Eq.~\eqref{dt:convergent} need not be
	positive. We therefore move part of each summand into a telescoping
	boundary term, chosen so that the remainder has a definite sign below
	the threshold. This summation by parts rests on
	\begin{align}
		G_\varepsilon(w)
		&=\frac{22-5\varepsilon^2-14\varepsilon w+5\varepsilon^2w^2}{15},
		\label{dt:Gpoly}\\
		{\cal A}_\varepsilon(w,0)
		&=\frac{\varepsilon}{2}(1-w^2)G_\varepsilon'(w)
		-(1+2\varepsilon w)G_\varepsilon(w)\nonumber\\
		&\hspace{9mm}+\frac{37-13\varepsilon^2}{15},
		\label{dt:threshold-identity}
	\end{align}
	where the factor $37-13\varepsilon^2$ that fixes the threshold appears
	explicitly in Eq.~\eqref{dt:threshold-identity}. The polynomial
	$G_\varepsilon$ is fixed by requiring a constant remainder in that
	identity: matching powers of $w$ with a quadratic ansatz gives
	$G_\varepsilon$ and the constant. We set
	\begin{align}
		\bar c_\varepsilon&=\frac{37-13\varepsilon^2}{15},\\
		L_n&=\frac{\varepsilon^2u_n^2(1-u_n)^2}{4n}G_\varepsilon(z_n).
	\end{align}
	The following identity is exact:
	\begin{align}
		\frac14v_n^2\mathcal A_\varepsilon(w_n,v_n)
		={}&L_{n+1}-L_n\nonumber\\
		&+\frac{v_n^2}{4}
		\left[\bar c_\varepsilon+\frac{R_\varepsilon(w_n,v_n)}{60n(n+1)}\right],
		\label{dt:second-summation}
	\end{align}
	where
	\begin{align}
		R_\varepsilon(w,v)={}&5\varepsilon^5w(1-w^2)(3-4w^2)\nonumber\\
		&+\varepsilon^4(1-w^2)(1+5w^2)
		-\varepsilon^3w(5w^2+9)\nonumber\\
		&+\varepsilon^2(14w^2-20)+30\varepsilon w+22\nonumber\\
		&+v^2\big[\varepsilon^5w(15-20w^2)
		+\varepsilon^4(42w^2-6)\nonumber\\
		&\hspace{35mm}-51\varepsilon^3w\big]+3\varepsilon^4v^4 .
		\label{dt:Rpoly}
	\end{align}
	One verifies Eq.~\eqref{dt:second-summation} by substituting
	Eq.~\eqref{dt:mobius} and clearing the positive denominators, and its
	leading part at $v=0$ is Eq.~\eqref{dt:threshold-identity}. Thus the
	infrared factor $37-13\varepsilon^2$ also appears in an exact discrete
	summation identity.
	
	For $0<\varepsilon\le13/10$, Eq.~\eqref{dt:recurrence} implies
	\begin{equation}
		6v_n\le\varepsilon(1+v_n),\qquad
		v_n\le\frac\varepsilon{6-\varepsilon}\le\frac{13}{47}<\frac7{25}.
	\end{equation}
	The elementary bound $|4w^3-3w|\le1$ on $[-1,1]$ controls the two terms
	of order $\varepsilon^5$, and discarding the nonnegative terms in
	Eq.~\eqref{dt:Rpoly} gives
	\begin{equation}
		\begin{aligned}
			R_\varepsilon(w_n,v_n)\ge{}&22-5\varepsilon^5-14\varepsilon^3-20\varepsilon^2-30\varepsilon\\
			&-\frac{49}{625}(5\varepsilon^5+6\varepsilon^4+51\varepsilon^3)\\
			\ge{}&-\frac{139632641}{1250000}>-112 .
		\end{aligned}
	\end{equation}
	Since, moreover, $\bar c_\varepsilon\ge501/500$ and $n(n+1)\ge2$, we obtain
	\begin{equation}
		\bar c_\varepsilon+\frac{R_\varepsilon(w_n,v_n)}{60n(n+1)}
		>\frac{501}{500}-\frac{112}{120}
		=\frac{103}{1500}>0.
		\label{dt:low-bound}
	\end{equation}
	
	The endpoint contribution is positive as well. Indeed,
	\begin{equation}
		G_\varepsilon(z)-\frac{22+14\varepsilon}{15}
		=\frac{\varepsilon(z+1)}{15}[5\varepsilon(z-1)-14]\le0,
	\end{equation}
	so that $G_\varepsilon(z)\le67/25<4$. Since $L_n=O(n^{-1-2\varepsilon})$,
	summing Eq.~\eqref{dt:second-summation} in Eq.~\eqref{dt:convergent}
	gives
	\begin{align}
		-\frac{12DC_\varepsilon}{\nu\varepsilon^2}
		={}&\varepsilon^2u_1^2(1-u_1)^2
		\left[1-\frac{G_\varepsilon(z_1)}4\right]\nonumber\\
		&+\frac14\sum_{n\ge1}v_n^2
		\left[\bar c_\varepsilon+\frac{R_\varepsilon(w_n,v_n)}{60n(n+1)}\right]>0 .
	\end{align}
	This proves strict monotonicity for all $x>0$ and
	$0<\varepsilon\le13/10$ by elementary inequalities alone.
	
	\subsection{The remaining parameter interval}
	\label{dt:polynomial-estimate}
	
	The estimate above uses a positive lower bound on $\bar c_\varepsilon$, which
	vanishes at $\varepsilon=\sqrt{37/13}$. Hence, to reach this endpoint, we
	must control the remainder independently of that constant. For a
	boundary function $F_n$ with $F_n(z_n)\to0$, we define
	\begin{align}
		\mathcal E_n={}&\frac{v_n^2}{4}
		[\mathcal A_\varepsilon(w_n,v_n)-\bar c_\varepsilon]\nonumber\\
		&-F_{n+1}(z_{n+1})+F_n(z_n),
		\label{dt:Rn}
	\end{align}
	and we combine the first mode with the lower boundary term,
	\begin{align}
		\mathcal B={}&\varepsilon^2u_1^2(1-u_1)^2
		+\frac14v_1^2\mathcal A_\varepsilon(w_1,v_1)\nonumber\\
		&-F_2(z_2).
		\label{dt:boundary-definition}
	\end{align}
	Telescoping then gives the decomposition
	\begin{equation}
		-\frac{12DC_\varepsilon}{\nu\varepsilon^2}
		=\mathcal B+\sum_{n\ge2}
		\left(\mathcal E_n+\frac{\bar c_\varepsilon v_n^2}{4}\right).
		\label{dt:high-bound}
	\end{equation}
	Since $\bar c_\varepsilon\ge0$ in the desired range, it suffices to prove
	$\mathcal E_n>0$ for $n\ge2$ and $\mathcal B>0$.
	
	To this end we add a term proportional to $n^{-3}$ to $L_n$. The quartic
	$H_\varepsilon$ below is chosen to make the resulting remainder positive,
	as verified by the polynomial inequalities that follow. We define
	\begin{align}
		H_\varepsilon(z)=\frac1{105}\big[&-525\varepsilon^4z^4+600\varepsilon^3z^3\nonumber\\
		&+(875\varepsilon^4-425\varepsilon^2)z^2\nonumber\\
		&-(858\varepsilon^3+78\varepsilon)z\nonumber\\
		&-350\varepsilon^4+1547\varepsilon^2-1458\big],
		\label{dt:Hpoly}\\
		F_n(z)={}&\varepsilon^2(1-z^2)^2
		\left[\frac{G_\varepsilon(z)}{64n}
		+\frac{H_\varepsilon(z)}{3840n^3}\right],
		\label{dt:Ltilde}
	\end{align}
	so that $F_n(z_n)=L_n+
	\varepsilon^2(1-z_n^2)^2H_\varepsilon(z_n)/(3840n^3)$. We prove the two
	required inequalities on $13/10\le\varepsilon\le17/10$, using a rational
	upper bound for $\sqrt{37/13}$. For the higher modes, we expand the
	remainder numerator in $s=n-2\ge0$ and prove the positivity of each
	coefficient on the compact $(\varepsilon,z)$ domain. Splitting
	$-1\le z\le1$ into two intervals allows these signs to be established in
	the Bernstein basis.
	
	The coefficient formula used in both inequalities is stated once. Let
	$B_i^{(\ell)}(a)=\binom\ell i a^i(1-a)^{\ell-i}$. For a polynomial
	$p(a,b)$ of degree at most $\ell$ in each variable, we set
	\begin{equation}
		\beta_{ij}^{(\ell)}[p]=
		\sum_{\alpha=0}^{i}\sum_{\gamma=0}^{j}
		\frac{\binom i\alpha\binom j\gamma}
		{\binom\ell\alpha\binom\ell\gamma}
		[a^\alpha b^\gamma]p ,
		\label{dt:bernstein-coefficients}
	\end{equation}
	where square brackets extract a polynomial coefficient. The binomial
	theorem gives
	\begin{equation}
		p(a,b)=\sum_{i,j=0}^{\ell}\beta_{ij}^{(\ell)}[p]
		B_i^{(\ell)}(a)B_j^{(\ell)}(b).
		\label{dt:bernstein-expansion}
	\end{equation}
	On the unit square these basis functions are nonnegative and their sum
	is one, so that a lower bound on all coefficients bounds $p$.
	
	To write the higher-mode numerator without a long monomial expansion, we
	put $z=z_n$ and introduce
	\begin{align}
		m&=2n+1, & d&=m+\varepsilon z, & r&=1-z^2,\nonumber\\
		Z&=mz+\varepsilon, &
		W&=2mz+\varepsilon(1+z^2).
	\end{align}
	We set $g(z)=15G_\varepsilon(z)$, $h(z)=105H_\varepsilon(z)$, and
	\begin{equation}
		g_d=d^2g(Z/d),\qquad h_d=d^4h(Z/d),
	\end{equation}
	which are polynomials. With
	\begin{align}
		T={}&-15\varepsilon^3W^3+60\varepsilon^2W^2d\nonumber\\
		&+60\varepsilon(\varepsilon^2-2)Wd^2\nonumber\\
		&-16(\varepsilon^2+11)d^3+15\varepsilon^5Wr^2,
	\end{align}
	we define the polynomial $\mathcal P$ by
	\begin{align}
		132096\mathcal P={}&210d^3n^3(n+1)^3T\nonumber\\
		&+d^8(n+1)^3[420n^2g(z)+h(z)]\nonumber\\
		&-n^3(m^2-\varepsilon^2)^2\nonumber\\
		&\quad{}\times[420(n+1)^2d^2g_d+h_d].
		\label{dt:direct-polynomial}
	\end{align}
	The recurrence \eqref{dt:mobius} gives the exact identity
	\begin{equation}
		\mathcal E_n=
		\frac{172\varepsilon^2(1-z^2)^2}
		{525n^3(n+1)^3d^8}\mathcal P(\varepsilon,n,z).
		\label{dt:direct-remainder}
	\end{equation}
	The degrees of $\mathcal P$ in $(\varepsilon,n,z)$ are $(12,10,12)$, and
	the affine substitutions below map each half of the compact domain to
	the unit square. For $\sigma=-1,0$, we define
	\begin{equation}
		p_k^\sigma(a,b)=[s^k]\mathcal P\left(
		\frac{13}{10}+\frac25a,\,2+s,\,\sigma+b\right).
	\end{equation}
	Expansion of the polynomial \eqref{dt:direct-polynomial} then gives
	\begin{equation}
		\beta_{ij}^{(12)}[p_k^\sigma]\ge
		\begin{cases}
			33,&0\le k\le9,\\
			3/2,&k=10,
		\end{cases}
		\qquad 0\le i,j\le12.
		\label{dt:direct-coefficient-bound}
	\end{equation}
	This is a finite calculation in exact rational arithmetic: one expands
	Eq.~\eqref{dt:direct-polynomial} after the indicated substitutions and
	applies Eq.~\eqref{dt:bernstein-coefficients} to each coefficient of
	$s$. For $k\le9$, the smallest coefficients occur at $(i,j,k)=(0,0,9)$,
	and they are $3363129/100000$ for $\sigma=-1$ and $15939/400$ for
	$\sigma=0$. For $k=10$,
	$p_{10}^\sigma=(4-\varepsilon^2)(\varepsilon^2-1)$, with
	$\varepsilon=13/10+2a/5$, and its minimum Bernstein coefficient is
	$15939/10000$, attained at $i=0$ for every $j$ on both halves. Since
	$s=n-2\ge0$, Eq.~\eqref{dt:bernstein-expansion} proves $\mathcal P>0$
	throughout the required domain. For the physical modes, $|z_n|<1$ and
	$d>0$, so that Eq.~\eqref{dt:direct-remainder} gives $\mathcal E_n>0$.
	
	\subsubsection{The first mode}
	\label{dt:boundary}
	
	The first mode is treated in the same way. Writing $t=z_2$, the
	recurrence gives
	\begin{equation}
		\frac{\mathcal B}{\varepsilon^2u_1^2(1-u_1)^2}
		=\frac{3-\varepsilon t}{(9-\varepsilon^2)^2}Q_\varepsilon(t),
		\label{dt:boundary-compact}
	\end{equation}
	where
	\begin{align}
		Q_\varepsilon(t)={}&2(15+\varepsilon t-4\varepsilon^2
		+5\varepsilon^2t^2-\varepsilon^3t^3)\nonumber\\
		&-(3-\varepsilon t)^3
		\left[\frac{G_\varepsilon(t)}8+
		\frac{H_\varepsilon(t)}{1920}\right].
		\label{dt:boundary-polynomial}
	\end{align}
	This polynomial has degree seven in each variable, and the same
	coefficient formula gives, for $0\le i,j\le7$,
	\begin{equation}
		\beta_{ij}^{(7)}\left[
		Q_{13/10+2a/5}(2b-1)\right]
		\ge\frac{4512477379}{4800000000}>\frac9{10},
		\label{dt:boundary-coefficient-bound}
	\end{equation}
	where the minimum coefficient occurs at $(i,j)=(7,0)$. Thus
	$Q_\varepsilon(t)>0$ on the full parameter interval and for
	$-1\le t\le1$. Since the prefactor in Eq.~\eqref{dt:boundary-compact} is
	positive, $\mathcal B>0$.
	
	Finally, $F_n(z_n)=O(n^{-1-2\varepsilon})$, so that its upper boundary
	term vanishes and the telescoping leading to Eq.~\eqref{dt:high-bound} is
	justified. The bounds on $\mathcal B$ and $\mathcal E_n$ make that sum
	strictly positive, including at $\bar c_\varepsilon=0$. Together with
	Eq.~\eqref{dt:low-bound}, this proves the sufficient direction of
	Eq.~\eqref{dt:sharp}.
	
	\subsection{Spectral zeta function and endpoints}
	\label{dt:zeta-proof}
	
	The converse rests on the large-eigenvalue
	expansion of the gamma ratio,
	\begin{align}
		\log p_\varepsilon(n)
		&=\varepsilon\log n+\frac{a_2}{n^2}
		+\frac{a_4}{n^4}+\frac{a_6}{n^6}+O_K(n^{-8}),\nonumber\\
		a_2&=\frac{\varepsilon(1-\varepsilon^2)}{24},\nonumber\\
		a_4&=\frac{\varepsilon(-7+10\varepsilon^2-3\varepsilon^4)}{960},\nonumber\\
		a_6&=\frac{\varepsilon(31-49\varepsilon^2+
			21\varepsilon^4-3\varepsilon^6)}{8064}.
		\label{dt:symbol}
	\end{align}
	The remainder is uniform for $\varepsilon$ in any compact subset
	$K\subset(0,2)$. To justify it directly, we write
	$\delta_n=\log p_\varepsilon(n)-\varepsilon\log n$ and use
	\begin{equation}
		\delta_n=-\int_0^\infty e^{-nt}
		\left[\frac{\sinh(\varepsilon t/2)}{\sinh(t/2)}
		-\varepsilon\right]\frac{dt}{t}.
		\label{dt:log-integral}
	\end{equation}
	Taylor expansion at $t=0$ gives Eq.~\eqref{dt:symbol}. After the first
	three terms are removed, the integrand is $O_K(t^7)$ near zero, while the
	remainder of the real integration contour is exponentially bounded
	uniformly on $K$. Splitting the integral at a fixed positive $t$
	therefore gives the stated $O_K(n^{-8})$ remainder, and the same argument
	applies after any fixed number of parameter derivatives.
	
	We expand
	$p_\varepsilon(n)^{-w}\sim n^{-\varepsilon w}\sum_{j\ge0}d_jn^{-2j}$ at
	large $n$,
	whose first coefficients are
	\begin{align}
		d_0&=1,\qquad d_1=-wa_2,\nonumber\\
		d_2&=-wa_4+\frac{w^2a_2^2}{2},\nonumber\\
		d_3&=-wa_6+w^2a_2a_4-\frac{w^3a_2^3}{6}.
	\end{align}
	The analytically continued spectral zeta function is
	\begin{align}
		\mathcal Z_\varepsilon(w)
		={}&\sum_{j=0}^3d_j\zeta(\varepsilon w+2j-2)\nonumber\\
		&+\sum_{n\ge1}n^{2-\varepsilon w}
		\left[e^{-w\delta_n}-\sum_{j=0}^3d_jn^{-2j}\right],
		\label{dt:zeta-subtraction}
	\end{align}
	where the second sum converges absolutely for
	$\varepsilon\mathop{\rm Re}w>-5$, locally uniformly away from the poles in
	the first sum. On each fixed vertical contour in this region, Taylor's
	integral remainder bounds the second sum by a polynomial in $|w|$. Since
	the Riemann zeta functions have polynomial growth on vertical strips, the
	same property holds for $\mathcal Z_\varepsilon$, and these bounds are
	uniform when $\varepsilon$ ranges over a compact interval that stays
	within the convergence region.
	
	The pole at $w=-1/\varepsilon$ comes from the $d_2$ term, and its residue
	is
	\begin{equation}
		\mathop{\rm Res}_{w=-1/\varepsilon}\mathcal Z_\varepsilon(w)
		=\frac{d_2(-1/\varepsilon)}{\varepsilon}
		=\frac{(\varepsilon^2-1)(37-13\varepsilon^2)}
		{5760\varepsilon},
		\label{dt:residue}
	\end{equation}
	which gives Eq.~\eqref{dt:IR}. The reciprocal
	and direct traces needed at the other poles follow from finite sums. In
	particular, with $\gamma=(1+\varepsilon)/2$,
	\begin{equation}
		\sum_{n=1}^N n^2p_\varepsilon(n)
		=\frac{N(N+1)(N+1-\gamma)}{\varepsilon+3}
		p_\varepsilon(N+1),
		\label{dt:shadow-sum}
	\end{equation}
	while the reciprocal trace is obtained by summing the first difference
	of $n(n-1)(n+\gamma-1)/[(3-\varepsilon)p_\varepsilon(n)]$. At generic
	$\varepsilon$, these expansions have no constant term and imply
	$\mathcal Z_\varepsilon(-1)=\mathcal Z_\varepsilon(1)=0$. At a collision
	of poles, instead, the residues are combined before taking the parameter
	limit: for example, at $\varepsilon=1$ one has
	$\mathcal Z_1(-1)=\zeta(-3)=1/120$.
	
	Finally, the endpoint difference is
	\begin{align}
		C_{\rm UV}-C_{\rm IR}
		=\frac{\nu}{6}\int_0^{\varepsilon/2}
		&u\sin(\pi u)\nonumber\\[-2pt]
		&\times\Gamma(3/2+u)\Gamma(3/2-u)\,du>0,
		\label{dt:endpoints}
	\end{align}
	which is the usual double-trace endpoint formula
	\cite{GiombiKlebanov2014}.
	
	\subsection{Infrared expansion and the first minimum}
	\label{dt:onset-proof}
	
	Subtracting the large-eigenvalue expansion also justifies the infrared
	expansion used in the main text. In a fixed neighborhood of
	$\varepsilon_*=\sqrt{37/13}$,
	\begin{align}
		C_\varepsilon-C_{\rm IR}
		=\nu\left[
		\frac{A(\varepsilon)}y+\frac{B(\varepsilon)}{y^3}
		+\frac{L(\varepsilon)}{y^{2\varepsilon}}
		+O(y^{-7/2})\right],
		\label{dt:three-term}
	\end{align}
	where $A$ is given in \eqref{dt:IR} and
	\begin{align}
		B(\varepsilon)
		&=-\frac{\pi(\varepsilon^2-1)(\varepsilon^2-9)
			(29\varepsilon^2-109)}
		{80640\sin(3\pi/\varepsilon)},\nonumber\\
		L(\varepsilon)
		&=-\frac{\sqrt\pi}{12}
		\frac{\Gamma((3+\varepsilon)/2)^2\Gamma(1/2-\varepsilon)}
		{\Gamma(-\varepsilon)\Gamma(-\varepsilon/2)^2}.
		\label{dt:BL}
	\end{align}
	The remainder and each fixed number of $D$ derivatives are uniform for
	$5/3\le\varepsilon\le17/10$ and $y\ge1$. To see this, we start from Eq.~\eqref{pos:laplace-proof} with $p=3$,
	\begin{equation}
		DC_\varepsilon=\frac{\nu\varepsilon}{2}\,\frac1{2\pi i}
		\int_{c-i\infty}^{c+i\infty}
		\frac{\pi P_3(\varepsilon z)\mathcal Z_\varepsilon(z)x^z}{\sin\pi z}\,dz,
		\label{dt:mellin-S3}
	\end{equation}
	with $1<c<2$, use \eqref{dt:zeta-subtraction} for $\mathcal Z_\varepsilon$, and shift
	the contour to the left until the remaining integral is
	$O(x^{-21/10})$. In particular, the pole of $\mathcal Z_\varepsilon$ at
	$z=-1/\varepsilon$, with the residue \eqref{dt:residue}, gives
	$-\nu A/y$ in $DC_\varepsilon$. Integration with respect to $\log y$
	then gives $\nu A/y$ in $C_\varepsilon-C_{\rm IR}$. The terms crossed
	in this shift decay as $x^{-1/\varepsilon}$, $x^{-3/\varepsilon}$, and
	$x^{-2}$. After integration they give the three displayed terms, while the term in
	$x^{-1}$ cancels by \eqref{dt:shadow-sum}.
	
	The term $L$ comes from the pole of $1/\sin\pi z$ at $z=-2$, which gives
	$(\nu\varepsilon/2)P_3(-2\varepsilon)\mathcal Z_\varepsilon(-2)x^{-2}$ in
	$DC_\varepsilon$. Since $x^{-2}=y^{-2\varepsilon}$, integration gives
	$L(\varepsilon)=-\frac14P_3(-2\varepsilon)\mathcal Z_\varepsilon(-2)$. For
	$\varepsilon<-3/2$ the trace
	$\mathcal Z_\varepsilon(-2)=\sum_{n\ge1}n^2p_\varepsilon(n)^2$ converges.
	With $k=n-1$, it equals $[\Gamma((3+\varepsilon)/2)/\Gamma((3-\varepsilon)/2)]^2$
	times a very-well-poised ${}_5F_4(1)$ series with parameters $a=2$,
	$b=c=(3+\varepsilon)/2$, and $d=3/2$, because
	$(2)_k(2)_k/[k!\,(1)_k]=n^2$ and $(3/2)_k/(3/2)_k=1$. Dougall's
	summation theorem \cite[Sec.~2.3.4]{Slater1966} then gives
	\begin{equation}
		\mathcal Z_\varepsilon(-2)=\frac{\sqrt\pi}{4}\,
		\frac{\Gamma((3+\varepsilon)/2)^2\,\Gamma(-3/2-\varepsilon)}
		{\Gamma(-\varepsilon)\,\Gamma(-\varepsilon/2)^2},
		\label{dt:Zminus2}
	\end{equation}
	which continues meromorphically in $\varepsilon$ to the range considered
	here. With $P_3(-2\varepsilon)=(1+2\varepsilon)(1+2\varepsilon/3)$ and
	$\Gamma(1/2-\varepsilon)=(1/2+\varepsilon)(3/2+\varepsilon)
	\Gamma(-3/2-\varepsilon)$, it reproduces $L$ in Eq.~\eqref{dt:BL}. The contour remainder is
	$O(x^{-21/10})=O(y^{-7/2})$ on this parameter interval. After the
	subtraction, the spectral zeta function grows at most polynomially along
	vertical contours, while the sine factor in the Mellin integrand supplies
	exponential decay, and each scale derivative adds only a polynomial
	factor. This also supplies the differentiated tails required by the
	infrared criterion of Sec.~\ref{ob:ir}.
	
	The above expansion determines the position and the depth of the minimum
	near the threshold. Let
	\begin{equation}
		\delta=\varepsilon-\varepsilon_*,
		\qquad a_*=-A'(\varepsilon_*)>0,\qquad
		b_*=B(\varepsilon_*)>0 .
	\end{equation}
	Substitution into Eqs.~\eqref{dt:IR} and \eqref{dt:BL} gives
	\begin{align}
		a_*&=\frac{\pi\varepsilon_*}{90\sin(\pi/\varepsilon_*)},
		\nonumber\\
		b_*&=-\frac{172\pi}{46137\sin(3\pi/\varepsilon_*)}.
		\label{dt:onset-coefficients}
	\end{align}
	Their signs follow from $3/2<\varepsilon_*<2$. For sufficiently small
	$\delta>0$, there is exactly one stationary point on the positive
	coupling axis, which is the global minimum and obeys
	\begin{align}
		y_{\min}
		&\sim\sqrt{\frac{3b_*}{a_*}}\,\delta^{-1/2},
		\nonumber\\
		C_{\rm IR}-C_\varepsilon(y_{\min})
		&\sim\nu\frac{2a_*^{3/2}}{3\sqrt{3b_*}}\,
		\delta^{3/2}.
		\label{dt:onset}
	\end{align}
	For the infrared uniqueness proof, \eqref{dt:three-term} gives
	\begin{equation}
		\frac{y^3}{\nu}DC_\varepsilon
		=-A(\varepsilon)y^2-3B(\varepsilon)+E_\varepsilon(y),
	\end{equation}
	where $|E_\varepsilon|+|DE_\varepsilon|\le K y^{-1/4}$ uniformly. After
	choosing a fixed large $Y$, the expression is negative below a scale of
	order $\delta^{-1/2}$, strictly increasing through the region in which it
	can vanish, and positive afterward. This proves a unique infrared zero.
	In the ultraviolet, the vanishing of $\mathcal Z_\varepsilon(1)$ and the
	convergent quadratic trace give
	\begin{align}
		DC_\varepsilon={}&-\frac{\nu\varepsilon}{2}(1-2\varepsilon)
		\left(1-\frac{2\varepsilon}{3}\right)
		\mathcal Z_\varepsilon(2)x^2+O(x^3),
		\label{dt:onset-uv}
	\end{align}
	where $\mathcal Z_\varepsilon(2)=\sum_{n\ge1}n^2/p_\varepsilon(n)^2>0$,
	and the sum and remainder are uniform near $\varepsilon_*>3/2$. Since both linear factors in the coefficient of
	Eq.~\eqref{dt:onset-uv} are negative, the scale
	derivative is uniformly negative for sufficiently small $x>0$. On the
	intervening compact scale interval, the strict monotonicity at
	$\varepsilon_*$, proved above, persists by continuity. Thus no other
	stationary point occurs, and the two balances in \eqref{dt:onset} follow
	from \eqref{dt:three-term}. 
	
	\section{The vector models}
	\label{on:appendix}
	\label{lean:ondetails}
	
	In this appendix we analyze the leading saddle and the normalized
	auxiliary-field integral, and we derive the order-$N^0$ coefficient in
	Eqs.~\eqref{on:subleadingIR} and \eqref{on:b} from the curvature
	expansion. The spherical spectral sum controls the remainder and its
	scale derivatives at this order, while an exact flat-space calculation
	verifies the stress-tensor relation in the same prescription.
	The curvature calculation also gives the order-$N^0$ maximum over
	scalar improvements at finite quartic coupling. The final subsections
	derive the leading finite-coupling coefficient, the Gross--Neveu
	moments, and the improvement identity and weak-coupling expansion.
	The mass parameter is tuned to give the flat-space physical mass $m$
	and held fixed during curvature variations.
	
	\subsection{The leading saddle}
	\label{on:saddle-proof}
	\label{on:leading}
	
	We first prove that the leading subtracted free energy
	$C_0=P_3(D)f_0$ of Eq.~\eqref{on:expansion} is positive and strictly
	decreasing, with no algebraic infrared terms. The critical action per component, in which the auxiliary field
	$\sigma$ acts as a mass squared for the scalars and couples to the
	source $J$, is
	\begin{equation}
		S_0[g,\sigma,J]
		=\frac12\operatorname{Tr}\log
		\left(-\nabla_g^2+\frac{\mathcal R}{8}+\sigma\right)
		+J\int d^3x\sqrt g\,\sigma ,
		\label{on:action}
	\end{equation}
	where local volume and curvature terms are understood to be renormalized
	in the fixed ultraviolet prescription. At leading order the physical
	mass is $m=8\pi J$. On the round sphere, we set
	$z=R^2\sigma$ and $a^2=z-1/4$. The known massive saddle
	\cite{AnninosDenefHarlow2012} is
	\begin{align}
		f_0(y)&=F_s(z)+\frac{\pi}{4}yz ,
		&y&=a\coth(\pi a), \label{on:saddle}\\
		F_s(z)&=\frac12\operatorname{FP}
		\sum_{n=1}^{\infty}n^2\log(n^2-1/4+z),
		\label{on:scalar}
	\end{align}
	where $\operatorname{FP}$ denotes the finite part with the specified local
	subtractions, and the additive constant is fixed by the conformal scalar
	determinant. For $0<y<1/\pi$, we write $a=i\theta/\pi$, with
	$0<\theta<\pi/2$. This parametrization is regular, and all scalar
	eigenvalues remain positive. For $y>1/\pi$, instead, $a$ is real and
	positive.
	
	For the gapped saddle \eqref{on:saddle}, $C_0(y)>0$ and $DC_0(y)<0$ for
	every $y>0$, and the endpoints are
	\begin{equation}
		C_0(0)=\frac{\log2}{8}
		-\frac{3\zeta(3)}{16\pi^2},
		\qquad C_0(\infty)=0 .
		\label{on:leadingendpoints}
	\end{equation}
	The complete algebraic infrared expansion vanishes after the
	application of $P_3(D)$. These properties follow directly from the saddle
	equation, as we now show.
	
	Stationarity gives $f_0'=\pi z/4$, where primes on $f_0$ and $z$ denote
	derivatives with respect to $y$. Consequently,
	\begin{equation}
		DC_0=\frac{\pi y^2}{12}(yz''-z').
		\label{on:derivative}
	\end{equation}
	We put $t=\pi a$ and $h(t)=t\coth t$. Implicit differentiation of
	$y=h/\pi$, together with the identity
	\begin{equation}
		hh'-thh''-t(h')^2
		=t^2\operatorname{csch}^2t\,(3h'-2t),
		\label{on:hidentity}
	\end{equation}
	gives
	\begin{equation}
		DC_0=\frac{y^2}{6}
		\frac{t^2\operatorname{csch}^2t}{h'(t)^3}
		\bigl(3h'(t)-2t\bigr).
		\label{on:leadingderivative}
	\end{equation}
	For real $t>0$, the convergent partial-fraction expansion of $\coth t$
	implies
	\begin{equation}
		0<\frac{h'(t)}{t}
		=4\sum_{n=1}^{\infty}
		\frac{\pi^2n^2}{(\pi^2n^2+t^2)^2}<\frac23,
		\label{on:hbound}
	\end{equation}
	so that the derivative in \eqref{on:leadingderivative} is negative. For
	$t=i\theta$, we define $k(\theta)=\theta\csc^2\theta-\cot\theta$.
	Analytic continuation of the same convergent series gives
	$k/\theta>2/3$, and
	\begin{equation}
		DC_0=-\frac{y^2}{6}
		\frac{\theta^2\csc^2\theta}{k(\theta)^3}
		\bigl(3k(\theta)-2\theta\bigr)<0.
		\label{on:imaginaryderivative}
	\end{equation}
	The removable limit at $t=0$ is $DC_0=-3/(20\pi^2)$. This covers the
	entire gapped solution for $y>0$.
	
	For real $a$, the same scalar determinant has the integral representation
	\begin{equation}
		F_s(a^2+1/4)
		=-\frac{\pi a^3}{6}
		+\pi\int_a^\infty\frac{u^2\,du}{e^{2\pi u}-1}.
		\label{on:scalarintegral}
	\end{equation}
	Since $a=y-2ye^{-2\pi y}+O(y^2e^{-4\pi y})$, the stationary action obeys
	\begin{align}
		f_0(y)
		={}&\frac{\pi y^3}{12}+\frac{\pi y}{16}
		+e^{-2\pi y}
		\left(\frac{y^2}{2}+\frac{y}{2\pi}
		+\frac{1}{4\pi^2}\right) \nonumber\\*
		&+O(y^3e^{-4\pi y}),
		\label{on:leadingFIR}
	\end{align}
	where all the terms other than the two local powers are exponentially
	small. Applying $P_3(D)$ of Eq.~\eqref{st:filter} gives
	\begin{align}
		C_0(y)
		={}&e^{-2\pi y}
		\left(\frac{2\pi^2}{3}y^4+\frac{\pi}{3}y^3
		+\frac{y^2}{2}+\frac{y}{2\pi}+\frac{1}{4\pi^2}\right)
		\nonumber\\
		&+O(y^5e^{-4\pi y}),
		\label{on:leadingCIR}
	\end{align}
	which yields the infrared endpoint. The ultraviolet endpoint follows from
	$z=0$ at $y=0$, and the strict decrease to zero implies $C_0>0$.
	
	\subsection{The regulated measure and contour}
	\label{on:measure}
	
	We now fix the normalization of the auxiliary-field integral, which
	determines the sign of the determinant in \eqref{on:determinant}. The
	determinant follows from a normalized identity before the continuum
	limit. We use a finite local regulator with positive integration weights
	$v_x$ and the action
	\begin{equation}
		S[\phi]=\frac12\sum_{a=1}^N
		(\phi^a,(K_g+r_0)\phi^a)_g
		+\sum_xv_x\frac{u}{4N}(\phi_x^2)^2 ,
		\label{on:regulatedaction}
	\end{equation}
	where $K_g=-\nabla_g^2+\mathcal R/8$ denotes the regulated operator and
	$u>0$. The exact identity at each site is
	\begin{align}
		&e^{-v_xu(\phi_x^2)^2/(4N)}
		=\left(\frac{Nv_x}{4\pi u}\right)^{1/2}
		\int_{\mathbb R}d\eta_x \nonumber\\
		&\hspace{3em}\times
		\exp\left[-\frac{Nv_x\eta_x^2}{4u}
		+\frac{i v_x\eta_x\phi_x^2}{2}\right].
		\label{on:HS}
	\end{align}
	The linear change of variables $\sigma=r_0-i\eta$ has a constant
	Jacobian, and no bilocal collective field is needed.
	
	For negative critically tuned $r_0$, we first translate the regulated
	contour so that $K_g+\operatorname{Re}\sigma$ is positive. For each fixed
	scalar configuration the integrand is entire, and the Gaussian decay
	permits this translation, so that the scalar integration can be
	performed on that line. In a neighborhood of the gapped real saddle, the
	auxiliary contour can then be moved through the saddle without
	encountering a zero of the scalar determinant. Integrating the scalars
	gives
	\begin{equation}
		S_{0,u}[\sigma]
		=\frac12\operatorname{Tr}\log(K_g+\sigma)
		-\frac{1}{4u}\int d^3x\sqrt g\,(\sigma-r_0)^2 ,
		\label{on:finiteuaction}
	\end{equation}
	whose Hessian on real $\sigma$ variations is
	\begin{equation}
		S_{0,u}''=-\frac12(B+u^{-1}\mathbb I),
		\qquad B(x,x')=G(x,x')^2 .
		\label{on:hessian}
	\end{equation}
	For a real multiplication operator $f$,
	\begin{equation}
		(f,Bf)_g
		=\operatorname{Tr}\bigl[(G^{1/2}fG^{1/2})^2\bigr]>0
		\label{on:bpositive}
	\end{equation}
	unless $f=0$. Hence, on the physical fluctuation contour
	$\sigma=\sigma_0+i\chi/\sqrt N$, the Gaussian action is
	$\frac14(\chi,(B+u^{-1})\chi)_g$.
	
	Let $L$ be the number of auxiliary variables, and use orthonormal
	coordinates $\chi'_x=\sqrt{v_x}\chi_x$. The $N^{L/2}$ in \eqref{on:HS}
	cancels the fluctuation rescaling, and including its Gaussian
	normalization gives exactly
	\begin{align}
		&(4\pi u)^{-L/2}
		\int d^L\chi'\,
		e^{-\chi'(B+u^{-1})\chi'/4} \nonumber\\
		&\hspace{4em}=\det(\mathbb I+uB)^{-1/2}.
		\label{on:Gaussianmeasure}
	\end{align}
	Thus
	\begin{equation}
		F_{1,u}=\frac12\operatorname{Tr}\log(\mathbb I+uB).
		\label{on:finiteudet}
	\end{equation}
	In the critical limit, the term $L\log u/2$ is independent of the
	relevant source, and the remaining dynamical contribution is
	$\operatorname{Tr}\log B/2$. The critical source and local counterterms
	must be tuned before removing the regulator. Indeed, expanding the last
	term in \eqref{on:finiteuaction} gives
	$-\sigma^2/(4u)+(r_0/(2u))\sigma-r_0^2/(4u)$. After critical mass tuning,
	the finite linear coefficient is the renormalized $J$ in
	\eqref{on:action}, while the quadratic term vanishes in the strict critical limit. Retaining it describes the theory at finite quartic
	coupling.
	
	After the scalar integrations there are no other dynamical fields. An
	auxiliary cubic vertex has order $N^{-1/2}$ and a quartic vertex has
	order $N^{-1}$, so that non-Gaussian auxiliary vacuum graphs begin at
	$N^{-1}$. The displacement $\sigma_0\mapsto\sigma_0+\sigma_1/N$ does not
	contribute at first subleading order, because
	\begin{equation}
		NS_0(\sigma_0+\sigma_1/N)
		=NS_0(\sigma_0)+\sigma_1\cdot S_0'(\sigma_0)+O(N^{-1})
		\label{on:saddledisplacement}
	\end{equation}
	and $S_0'(\sigma_0)=0$. The remaining terms at this order are explicit
	source and local counterterms.
	
	Redundant field normalizations do not change this. A
	wavefunction insertion is proportional to
	$\operatorname{Tr}(GK_g)=\operatorname{Tr}\mathbb I-
	\operatorname{Tr}(G\sigma)$, and the stationary equation reduces its
	second term to a local source contribution. In the critical theory,
	an improvement insertion gives a local $J\mathcal R$ term, while a change of the
	auxiliary normalization must include both its constant Jacobian and its
	compensating source change. None of these generates an additional
	independent mass-dependent determinant.
	
	\subsection{The local curvature coefficient}
	\label{lean:curvaturedetails}
	
	We compute the order-$N^0$ curvature coefficient at criticality and
	at finite quartic coupling. On hyperbolic three-space of sectional curvature
	$\kappa=-k^2$, we define
	\begin{equation}
		S_\kappa(r)=\frac{\sinh kr}{k}.
		\label{on:hyperbolicradius}
	\end{equation}
	The scalar Green function and its renormalized coincidence limit are
	\begin{align}
		G_\kappa(r)&=\frac{e^{-\mu r}}{4\pi S_\kappa(r)},
		&\mu^2&=\sigma-\frac{\kappa}{4}, \nonumber\\
		G_{\kappa,\mathrm{ren}}(x,x)&=-\frac{\mu}{4\pi}.
		\label{on:hyperbolicG}
	\end{align}
	The gap equation from \eqref{on:action} fixes $\mu=m$, hence
	$\sigma_0=m^2+\kappa/4$. Holding $\sigma$ fixed instead
	would omit the response of the saddle.
	
	The radial harmonic is $\sin(pr)/(pS_\kappa(r))$, with measure
	$p^2dp/(2\pi^2)$ per unit physical volume, and it gives the eigenvalue
	\eqref{on:hyperbolicB}. Expanding $S_\kappa^{-1}$ gives
	\begin{align}
		B_\kappa&=B_0+\kappa B_{[1]}+\kappa^2B_{[2]}+\cdots, \nonumber\\*
		B_0(p)&=\frac{\arctan(p/(2m))}{4\pi p}, \nonumber\\*
		B_{[1]}(p)&=\frac{m}{6\pi(p^2+4m^2)^2}, \nonumber\\*
		B_{[2]}(p)&=\frac{7m(4m^2-p^2)}
		{30\pi(p^2+4m^2)^4},
		\label{on:curvaturevertices}
	\end{align}
	and the coefficient of $\kappa^2$ in $f_1/\operatorname{Vol}$, with $f_1$
	from Eq.~\eqref{on:determinant}, is
	\begin{equation}
		Q=\frac{1}{4\pi^2}\int_0^\infty dp\,p^2
		\left[
		\frac{B_{[2]}}{B_0}
		-\frac12\left(\frac{B_{[1]}}{B_0}\right)^2
		\right].
		\label{on:Qintegral}
	\end{equation}
	Its integrand is $O(p^2)$ at zero and $O(p^{-3})$ at infinity, so that no
	four-derivative subtraction is needed. The massive Green function
	justifies the curvature derivatives at $\kappa=0^-$ by exponential decay
	at large $r$. The resulting coefficient is local and covariant, and it
	therefore also determines the analytic curvature term on a large sphere.
	
	We put $u=p/(2m)$ and $A(u)=\arctan u$. Equation \eqref{on:Qintegral}
	then becomes $Q=c_{-1}/(2\pi^2m)$, where
	\begin{equation}
		c_{-1}=\int_0^\infty du
		\left[
		\frac{7u^3(1-u^2)}{60(1+u^2)^4A(u)}
		-\frac{u^4}{72(1+u^2)^4A(u)^2}
		\right].
		\label{on:cintegral}
	\end{equation}
	Setting $u=\tan\theta$ and integrating by parts, with vanishing
	boundary terms, gives
	\begin{equation}
		c_{-1}=\frac1{180}\int_0^{\pi/2}
		\frac{\sin^3\theta\cos\theta(27\cos^2\theta-16)}{\theta}\,d\theta.
		\label{on:cangular}
	\end{equation}
	The identity
	\begin{align}
		&32\sin^3\theta\cos\theta(27\cos^2\theta-16) \nonumber\\
		&\hspace{1em}=-47\sin(2\theta)+64\sin(4\theta)
		-27\sin(6\theta)
		\label{on:trigonometric}
	\end{align}
	proves the sine-integral formula in Eq.~\eqref{on:b}. Splitting
	the sine integrals into intervals of length $\pi$ yields
	\begin{equation}
		b_1=\frac1{2160}\int_0^\pi\sin t
		\left[\frac{10}{t}+\frac{37}{t+\pi}
		+\frac{27}{t+2\pi}\right]dt>0.
		\label{on:bpositiveintegral}
	\end{equation}
	The integrand is positive for $0<t<\pi$.
	
	At finite quartic coupling, the same calculation gives the first
	subleading coefficient at $\xi=1/6$. Let $M$ be the mass in the
	hyperbolic Green function. With the flat-space physical mass held at
	$m$, the leading gap equation is
	\begin{equation}
		M^2+\frac{\lambda M}{8\pi}
		=m^2+\frac{\lambda m}{8\pi}-(1-6\xi)\kappa.
		\label{fin:curvature-gap}
	\end{equation}
	Thus $M=m$ at $\xi=1/6$, for every $\lambda\geq0$, and the leading
	free-energy density has no curvature-squared term. Using the coupling
	normalization of Eq.~\eqref{on:finiteudet}, the first subleading
	determinant is $\tfrac12\Tr\log(B_\kappa+2/\lambda)$. At this saddle
	the mass-source and allowed local terms are at most linear in
	$\kappa$, so its curvature-squared coefficient fixes the moment
	difference through Eq.~\eqref{st:AB}.
	
	Set $\ell=\lambda/m=8\pi\bar\lambda$. In the following integrals,
	$p$ is momentum in units of $m$, and $B_0$, $B_{[1]}$ and $B_{[2]}$
	are the functions of Eq.~\eqref{on:curvaturevertices} evaluated at
	$m=1$. The order-$N^0$ moment difference at $\xi=1/6$ is $g_1(\ell)/m$,
	where, for $\ell>0$,
	\begin{equation}
		g_1(\ell)=\frac1{3\pi^2}\int_0^\infty dp\,p^2
		\left[\frac{B_{[2]}}{B_0+2/\ell}
		-\frac{B_{[1]}^2}{2(B_0+2/\ell)^2}\right].
		\label{fin:g1}
	\end{equation}
	The integral converges at both endpoints. Its sign follows from an
	integration by parts. Set $U(p)=p^3/(p^2+4)^3$. Then
	\begin{align}
		p^2B_{[2]}(p)&=\frac7{90\pi}U'(p),\nonumber\\
		B_0'(p)&=\frac{2p/(p^2+4)-\arctan(p/2)}{4\pi p^2}<0.
		\label{fin:bubble-derivative}
	\end{align}
	The last inequality follows from $\arctan z>z/(1+z^2)$ for $z>0$.
	The boundary term vanishes, giving
	\begin{align}
		g_1(\ell)&=-\frac1{3\pi^2}\int_0^\infty
		\frac{W(p)}{[B_0(p)+2/\ell]^2}\,dp,\nonumber\\
		W(p)&=\frac{p^2B_{[1]}(p)^2}{2}
		-\frac7{90\pi}U(p)B_0'(p)>0.
		\label{fin:g1positive}
	\end{align}
	Hence $g_1(\ell)<0$ for every positive coupling. Differentiating under
	the integral also gives
	\begin{equation}
		g_1'(\ell)=-\frac4{3\pi^2\ell^2}\int_0^\infty
		\frac{W(p)}{[B_0(p)+2/\ell]^3}\,dp<0.
		\label{fin:g1derivative}
	\end{equation}
	
	Both coupling limits can be taken analytically. Substituting the bubble
	vertices into $W$ gives
	\begin{equation}
		W(p)=\frac{7p\arctan(p/2)}{360\pi^2(p^2+4)^3}
		-\frac{p^2}{40\pi^2(p^2+4)^4}.
	\end{equation}
	Integrating its first term by parts reduces $\int W\,dp$ to rational
	integrals, with the result
	\begin{equation}
		\int_0^\infty W(p)\,dp=\frac1{30720\pi}.
	\end{equation}
	Since $0<B_0(p)\leq1/(8\pi)$, expansion in $\ell$ is uniform near
	zero. At strong coupling the integrand in Eq.~\eqref{fin:g1positive}
	is bounded by $W/B_0^2$, which is $O(p^2)$ at zero and $O(p^{-3})$ at
	infinity. Dominated convergence therefore gives
	\begin{align}
		g_1(\ell)&=-\frac{\ell^2}{368640\pi^3}+O(\ell^3)
		&& (\ell\to0),\nonumber\\
		g_1(\infty)&=\frac{2c_{-1}}{3\pi^2}
		=-\frac{b_1}{4\pi^2}.
		\label{fin:g1limits}
	\end{align}
	The second line also follows directly by setting $2/\ell=0$ in
	Eq.~\eqref{fin:g1} and comparing with Eq.~\eqref{on:Qintegral}.
	In particular, $-b_1/(4\pi^2)<g_1(\ell)<0$ at every finite positive
	$\ell$.

	\subsection{The spherical bubble and its uniform expansion}
	\label{on:symbol}
	
	We now evaluate the critical coefficient $c_{-1}$ from the discrete
	spectrum on the sphere, which also controls the remainder. On the unit sphere the scalar
	Green function is
	\begin{equation}
		G_a(\theta)=
		\frac{\sinh[a(\pi-\theta)]}
		{4\pi\sin\theta\,\sinh(\pi a)} ,
		\label{on:sphereG}
	\end{equation}
	and the addition theorem for scalar harmonics gives the bubble eigenvalue
	\begin{equation}
		B_n(a)=\frac{1}{4\pi n\sinh^2(\pi a)}
		\int_0^\pi d\theta\,
		\frac{\sin(n\theta)}{\sin\theta}
		\sinh^2[a(\pi-\theta)] ,
		\label{on:sphereB}
	\end{equation}
	with multiplicity $n^2$. The exact massive bubble was obtained in
	Ref.~\cite{DiPietroGorbenkoKomatsu2023}. For real $a>0$, it is given in terms of the digamma function $\psi$ by
	\begin{equation}
		nB_n(a)=\frac18-\frac{\coth(\pi a)}{4\pi}
		\operatorname{Im}\psi\left(\frac{n+1}{2}+ia\right).
		\label{on:digamma}
	\end{equation}
	On the imaginary-$a$ branch, the imaginary part in this expression is
	continued as
	\begin{equation}
		\frac{1}{2i}\left[
		\psi\left(\frac{n+1}{2}+ia\right)
		-\psi\left(\frac{n+1}{2}-ia\right)\right].
		\label{on:digammacontinuation}
	\end{equation}
	This difference agrees with the imaginary part for real $a$ and has an
	analytic continuation along the physical saddle. The bubble normalization
	can also be checked directly from Eq.~\eqref{on:sphereB}, since taking
	the difference at $n+2$ and $n$ gives
	\begin{equation}
		(n+2)B_{n+2}-nB_n
		=\frac{y}{\pi[4a^2+(n+1)^2]} ,
		\label{on:recurrence}
	\end{equation}
	with the initial values
	\begin{align}
		B_1&=\frac{\coth(\pi a)}{8\pi a}
		-\frac{\operatorname{csch}^2(\pi a)}8, \nonumber\\
		2B_2&=\frac{y}{\pi(4a^2+1)} .
		\label{on:seeds}
	\end{align}
	The apparent singularities at $a=0$ and at the ultraviolet endpoint
	$a=i/2$ have removable limits, and at the latter endpoint
	$B_n=1/(8n)$.
	
	For the infrared expansion we take $a$ real and large, and we put
	$u=n/(2a)$ and $A(u)=\arctan u$. The large-$w$ asymptotic expansion
	\begin{equation}
		\psi(w+1/2)=\log w+\frac{1}{24w^2}
		-\frac{7}{960w^4}+O(|w|^{-6}),
		\label{on:digammaasymptotic}
	\end{equation}
	with $w=n/2+ia$, is uniform in the closed first quadrant. After retaining
	the ultraviolet local subtractions, the factor $\coth(\pi a)-1$ gives
	only exponentially small terms, and the algebraic expansion is
	\begin{equation}
		B_n(a)=\frac{A(u)}{4\pi n}
		\left[1+\frac{q_2(u)}{a^2}
		+\frac{q_4(u)}{a^4}+\cdots\right],
		\label{on:bubblesymbol}
	\end{equation}
	where
	\begin{align}
		q_2(u)&=\frac{u}{12(1+u^2)^2A(u)}, \nonumber\\
		q_4(u)&=\frac{7u(1-u^2)}
		{240(1+u^2)^4A(u)} .
		\label{on:q}
	\end{align}
	Taking the logarithm and using $\frac12n^2dn=4a^3u^2du$, we obtain
	\begin{equation}
		c_{-1}=4\int_0^\infty du\,u^2
		\left(q_4(u)-\frac12q_2(u)^2\right),
		\label{on:symbolintegral}
	\end{equation}
	which is the same convergent integral as \eqref{on:cintegral}, now
	obtained from the discrete spectrum on the sphere.
	
	The Euler--Maclaurin endpoints produce no additional inverse-radius
	term. Indeed, the scaled summands are even and analytic at $n=0$, their
	$n^2$ factor makes the value there vanish, and every odd derivative at
	that endpoint also vanishes. For the first two terms, we apply
	Euler--Maclaurin after retaining their ultraviolet finite parts: six
	derivatives of the leading term, four of the second, and two of the
	third are sufficient to bound the respective remainders by $O(a^{-3})$.
	
	The uniform remainder in \eqref{on:digammaasymptotic} gives a second
	trace estimate. We write $H=nB_n$ and $H_0=A(n/(2a))/(4\pi)$. The
	absolute algebraic error in $H$ is bounded by $K(n^2+a^2)^{-3}$, while
	$H_0\geq k n/a$ for $n\leq a$ and $H_0\geq k>0$ for $n>a$. It follows
	that
	\begin{align}
		\sum_{n\leq a}n^2\frac{|\delta H|}{H_0}
		&\leq K a^{-5}\sum_{n\leq a}n=O(a^{-3}), \nonumber\\
		\sum_{n>a}n^2\frac{|\delta H|}{H_0}
		&\leq K\sum_{n>a}n^{-4}=O(a^{-3}).
		\label{on:tracebound}
	\end{align}
	The omitted logarithmic products have the same bound, because $q_2$ and
	$q_4$ are bounded and their relevant products are integrable with the
	weight $u^2$, and differentiating the uniform expansion gives the
	corresponding scale-derivative estimates. The exponentially small factors must be summed only after their ultraviolet
	local terms have been subtracted, since their unrenormalized trace is not
	convergent.
	
	\subsection{An exact flat-space stress-tensor calculation}
	\label{flat:check}
	
	The metric response of the critical flat-space determinant gives an
	independent analytic derivation of Eq.~\eqref{on:moment}. The calculation has two
	parts: the integrand at zero regulator, which reduces to a total
	derivative plus the curvature integrand, and the finite contribution of
	the regulator, which cancels the resulting boundary term. Two
	regularizations enter, with different roles. The inner scalar loops are
	dimensionally regularized, so that integrals such as
	Eq.~\eqref{flat:saddletensorintegral} below are regularized identities,
	while their bare forms diverge. The auxiliary-field determinant
	$\frac12\operatorname{Tr}\log B$ requires a separate analytic regulator,
	introduced in Eq.~\eqref{flat:regulator}, whose finite contribution is
	computed at the end. We use
	$G_k=(k^2+m^2)^{-1}$, $\int_k=\int d^3k/(2\pi)^3$, and
	$B_p=\arctan[p/(2m)]/(4\pi p)$. For external momentum $q\hat e_3$, let
	$H_\tau$ and $H_\eta$ be the second derivatives of $f_1$ along the
	leading saddle for $g_{ij}=e^{2\tau}\delta_{ij}$ and
	$g_{ij}=\operatorname{diag}(e^\eta,e^{-\eta},1)$, respectively. The
	normalization of Sec.~\ref{st:sec} gives
	\begin{equation}
		(M_2-3M_0)^{(1)}
		=-2[q^4]H_\eta(q)+\frac32[q^4]H_\tau(q),
		\label{flat:moment}
	\end{equation}
	where $[q^4]$ denotes a Taylor coefficient, and both derivatives include
	the metric seagulls and the response of the saddle.
	
	We first set $m=1$ and combine the integrands with the regulator of the determinant
	$\frac12\operatorname{Tr}\log B$ removed. We define the scalar triangles and boxes by
	\begin{align}
		T(p,q)&=\int_kG_kG_{k+p}G_{k+p+q},\nonumber\\
		X(p,q)&=\int_kG_kG_{k+q}G_{k+p}G_{k+p+q},\nonumber\\
		Y(p,q)&=\int_kG_k^2G_{k+q}G_{k+p},
		\label{flat:scalarvertices}
	\end{align}
	whose contribution to the Hessian with respect to the potential is
	\begin{equation}
		K(q)=\int_p\left[
		\frac{X+2Y}{B_p}-\frac{2T^2}{B_pB_{p+q}}\right],
		\qquad K_j=[q^j]K(q).
		\label{flat:K}
	\end{equation}
	For the transverse traceless metric perturbation, we write
	$V_k=k_2^2-k_1^2$ and
	\begin{align}
		T_\eta&=-2\int_kG_kG_{k+p}G_{k+p+q}V_{k+p},\nonumber\\
		Y_\eta&=\int_kG_kG_{k+p}^2G_{k+p+q}V_{k+p}^2,\nonumber\\
		X_\eta&=\int_kG_kG_{k+q}G_{k+p}G_{k+p+q}V_kV_{k+p},\nonumber\\
		\mathcal H_\eta(q)&=\frac12\int_p\left[
		\frac{4Y_\eta+2X_\eta}{B_p}
		-\frac{T_\eta^2}{B_pB_{p+q}}\right].
		\label{flat:tensorvertices}
	\end{align}
	Only its fourth-order coefficient is used. The explicit metric seagull
	has degree at most two in $q$, while the induced saddle term must also be
	included.
	
	For the Weyl perturbation, we set $v=e^{2\tau}\sigma_0[g]$.
	Differentiating the leading gap equation, with $J=1/(8\pi)$, gives
	\begin{align}
		L(q):=v_{,\tau}(q)&=\frac{2J}{B_q}
		=2+\frac{q^2}{6}-\frac{q^4}{90}+O(q^6),\nonumber\\
		v_{,\tau\tau}(q,-q)&=4-\frac{q^2}{6}+\frac{q^4}{30}+O(q^6),\nonumber\\
		[q^4]\sigma_{,\eta\eta}&=\frac1{120},
		\label{flat:saddleresponse}
	\end{align}
	where the last identity follows from
	\begin{equation}
		\int_kG_k^2G_{k+q}V_k^2
		=-\frac1{8\pi}\int_0^1\sqrt{1+x(1-x)q^2}\,dx.
		\label{flat:saddletensorintegral}
	\end{equation}
	The derivative of the determinant $\frac12\operatorname{Tr}\log B$ with
	respect to the saddle value $\sigma$ is
	\begin{equation}
		f_{1,\sigma}=-\frac1{8\pi}\int_p\frac1{(p^2+4)B_p}.
		\label{flat:tadpole}
	\end{equation}
	At zero regulator, the Weyl multiplication factors do not contribute. On
	the flat Hilbert space the Weyl-transformed bubble is
	$\widehat B_g=e^{\tau/2}B_ve^{\tau/2}$, where $B_v$ is the bubble of
	$-\partial^2+v$, and the determinant of a product is the product of the
	determinants, so that
	\[
	\frac12\operatorname{Tr}\log\widehat B_g
	=\frac12\operatorname{Tr}\log B_v+\frac12\operatorname{Tr}\tau .
	\]
	The last term is linear in $\tau$ and drops out of the second variation
	$H_\tau$. In momentum space this cancellation is implemented by the
	shift $p\mapsto-p-q$ of the outer loop momentum. The remaining terms
	follow from the chain rule along the saddle. The Weyl response contains
	$L(q)^2K(q)+f_{1,\sigma}v_{,\tau\tau}$, and the transverse response
	contains $\mathcal H_\eta+f_{1,\sigma}\sigma_{,\eta\eta}$, up to terms of
	degree at most two in $q$. Since $L(q)^2=4+2q^2/3-q^4/60+O(q^6)$, the
	Weyl contribution to Eq.~\eqref{flat:moment} is
	$6K_4+K_2-K_0/40+f_{1,\sigma}/20$, while the transverse saddle term
	contributes $-f_{1,\sigma}/60$. Their sum is
	\begin{equation}
		I_0=-2[q^4]\mathcal H_\eta+6K_4+K_2
		-\frac{K_0}{40}+\frac{f_{1,\sigma}}{30}.
		\label{flat:rawintegral}
	\end{equation}
	These terms are combined before the outer integration: their sum is
	convergent, although the individual terms need not be.
	
	All inner integrals can be evaluated exactly. The expansion
	\begin{equation}
		[q^d]G_{k+q}=\sum_{r=\lceil d/2\rceil}^{d}
		(-1)^r\binom r{d-r}2^{2r-d}k_3^{2r-d}G_k^{r+1}
		\label{flat:propagatorexpansion}
	\end{equation}
	leaves two distinct denominators in each loop. After Feynman
	parametrization and tensor integration, we put $u=p/2$, and every
	remaining parameter integral reduces to
	\begin{align}
		&\int_0^1\frac{t^{2j}\,dt}
		{(1+u^2-u^2t^2)^{n+1/2}}\nonumber\\*
		&\qquad=\frac{(1+u^2)^{j-n}}{u^{2j+1}}
		\int_0^u z^{2j}(1+z^2)^{n-j-1}\,dz ,
		\label{flat:masterintegral}
	\end{align}
	where $j$ is a nonnegative integer and $n$ is an integer, so that the
	last integral is elementary. We use the normalized angular average
	$\langle f\rangle_{\widehat p}=(4\pi)^{-1}\int d\Omega_{\widehat p}\,f$,
	so that the outer loop measure becomes
	$p^2dp\,\langle\cdot\rangle_{\widehat p}/(2\pi^2)$. Thus the angularly
	averaged radial integrand $R_{\mathrm{flat}}$, defined by
	$I_0=\int_0^\infty R_{\mathrm{flat}}(p)\,dp$, is rational in $p$ and
	$A=\arctan(p/2)$.
	
	The exact result takes the form of a total derivative plus
	the curvature integrand,
	\begin{align}
		R_{\mathrm{flat}}(p)&=H'(p)+R_c(p),\nonumber\\
		R_c(p)&=\frac{14p^3(4-p^2)}{45\pi^2(p^2+4)^4A}
		-\frac{2p^4}{27\pi^2(p^2+4)^4A^2},\nonumber\\
		H(p)&=\frac1{\pi^2(1+u^2)^5}
		\sum_{j=1}^5\frac{h_j(u)}{A^j},
		\qquad u=\frac p2,
		\label{flat:totalderivative}
	\end{align}
	where
	\begin{align}
		151200h_1={}&4032u^{10}+14787u^8+24560u^6\nonumber\\*
		&+20720u^4+1300u^2+305,\nonumber\\
		90720h_2={}&-u(1035u^6+2841u^4-1879u^2+15),\nonumber\\
		181440h_3={}&-u^2(135u^4+8351u^2-294),\nonumber\\
		20160h_4={}&u^3(337u^2-218),\nonumber\\
		5040h_5={}&37u^4.
		\label{flat:primitivecoefficients}
	\end{align}
	Equations~\eqref{flat:scalarvertices}--\eqref{flat:masterintegral}
	reduce the flat propagator expression to a rational radial integrand,
	and comparing that integrand with $H'+R_c$, using $A'=2/(p^2+4)$ and
	clearing denominators, gives Eq.~\eqref{flat:totalderivative}. No
	curvature coefficient enters this reduction. The apparent singularities
	of $H$ at the origin cancel, with
	\begin{equation}
		H(p)=\frac{161p^3}{12960\pi^2}+O(p^5),\qquad
		H(\infty)=\frac4{75\pi^3},
		\label{flat:boundary}
	\end{equation}
	while the remaining integral is precisely
	$\int_0^\infty R_c(p)\,dp=2c_{-1}/(3\pi^2)$, by the elementary integration
	by parts in Appendix~\ref{lean:curvaturedetails}. Consequently,
	\begin{equation}
		I_0=\frac4{75\pi^3}
		+\frac{-47\Si(\pi)+64\Si(2\pi)-27\Si(3\pi)}{8640\pi^2}.
		\label{flat:exactraw}
	\end{equation}
	The positive representation \eqref{on:bpositiveintegral} implies
	\begin{equation}
		b_1<\frac{10\pi^2+101}{2160\pi}<\frac{16}{75\pi},
	\end{equation}
	using $\sin t/t\le1$ and $\pi<4$. Hence $I_0>0$: the naive flat-space
	integral has the opposite sign to the sphere coefficient.
	
	We now show that the boundary term in Eq.~\eqref{flat:exactraw} is
	cancelled by the finite contribution of the determinant regulator. With the Weyl-transformed bubble $\widehat B_g$ introduced above, we
	define
	\begin{equation}
		\Gamma_\omega[g]=-\frac1{2\omega}
		\operatorname{Tr}(\mu_{\mathrm r}\widehat B_g)^{-\omega},
		\label{flat:regulator}
	\end{equation}
	with reference inverse length $\mu_{\mathrm r}$, and we take the metric
	derivatives before the finite part at $\omega=0$. For $\beta=B_p$,
	$\gamma=B_{p+q}$, and $r=\gamma/\beta$, the divided difference in the
	term with two first variations is
	\begin{equation}
		\frac{\beta^{-1-\omega}-\gamma^{-1-\omega}}{2(\beta-\gamma)}
		=-\frac{\beta^{-\omega}}{2\beta\gamma}
		\left[1+\omega\frac{\log r}{r-1}+O(\omega^2)\right].
		\label{flat:divided}
	\end{equation}
	The common weight has a regular limit on the convergent combined
	integrand, while the term linear in $\omega$ has a logarithmic
	ultraviolet divergence. To compute its residue, we define the first Weyl
	vertex
	\begin{equation}
		X_\tau(p,q)=\frac{\beta+\gamma}{2}-2L(q)T(p,q),
		\label{flat:weylvertex}
	\end{equation}
	whose fourth-order radial contribution to Eq.~\eqref{flat:moment} is
	\begin{align}
		E(p)=\frac{p^2}{2\pi^2}
		\left\langle[q^4]\left\{
		\frac{T_\eta^2-3X_\tau^2/4}{\beta\gamma}
		\frac{\log r}{r-1}\right\}\right\rangle_{\widehat p}.
		\label{flat:Eradial}
	\end{align}
	Thus the residue is fixed independently by the flat propagators.
	
	Restoring $m$, we put $L(q)=m/(4\pi B_q)$ and
	$J_{\mathrm{tad}}=-m/(4\pi)$. In the large-$p$ expansion, the hard region
	has inner loop momentum of order $p$, while the soft regions keep an
	internal momentum of order $m$ and $q$, after shifting the loop variable
	to the corresponding propagators. Let $T_{[4]}$ denote the coefficient of
	$p^{-4}$ in the soft-region expansion of the scalar triangle at large
	$p$. Its angular average obeys
	\begin{align}
		\langle T_{[4]}\rangle
		&=\frac43J_{\mathrm{tad}}-\frac43m^2B_q,\nonumber\\
		[q^4]\{L(q)\langle T_{[4]}\rangle\}
		&=\frac1{270\pi m}.
		\label{flat:scalarresidue}
	\end{align}
	For the transverse triangle, the hard term is $-V_n/(16p)$, with
	$n=\widehat p$, and the relevant soft term is $-16V_n C_4(q)/p^4$, where
	\begin{align}
		C_4(q)&=\frac1{24\pi}\int_0^1
		[m^2+x(1-x)q^2]^{3/2}\,dx,\nonumber\\
		[q^4]C_4(q)&=\frac1{1920\pi m},\qquad
		\langle V_n^2\rangle=\frac4{15}.
		\label{flat:tensorresidue}
	\end{align}
	The tadpole and the lower terms of the soft integrals are dimensionally
	regularized. The logarithmic divergence is the coefficient of $p^{-1}$ in
	$E(p)$, and the displayed hard and soft vertices give cross terms of
	order $p^{-5}$. With the leading Weyl vertex $1/(8p)$, these terms are
	$3L T_{[4]}/(8p^5)$ and $2V_n^2C_4/p^5$ in the numerator of
	Eq.~\eqref{flat:Eradial}. The radial measure, $\beta\gamma\sim1/(64p^2)$,
	and $\log r/(r-1)\to1$ give the factors below. After combining the
	remaining terms, the channel residues and their sum are
	\begin{align}
		\mathcal R_\tau
		&=\frac{12}{\pi^2}[q^4]
		\{L(q)\langle T_{[4]}\rangle\}
		=\frac2{45\pi^3m},\nonumber\\
		\mathcal R_\eta
		&=\frac{64}{\pi^2}\langle V_n^2\rangle[q^4]C_4(q)
		=\frac2{225\pi^3m},\nonumber\\
		[p^{-1}]E(p)=\mathcal R
		&=\mathcal R_\tau+\mathcal R_\eta
		=\frac4{75\pi^3m}.
		\label{flat:totalresidue}
	\end{align}
	As shown above, the additional terms from the Weyl multiplication
	factors cancel at zero regulator, and the remaining differences of
	regulator weights have no logarithmic divergence at order $q^4$. Since
	$B_p^{-\omega}\sim(8p)^\omega$ and
	$\omega\int_{P_0}^\infty p^{\omega-1}dp\to-1$, the finite regulator
	contribution is $-\mathcal R$. The source counterterm contributes zero to
	the moment difference as well: its transverse and Weyl fourth-order
	coefficients are $\delta J/(120m^2)$ and $\delta J/(90m^2)$,
	respectively, whose combination in Eq.~\eqref{flat:moment} vanishes.
	
	We therefore obtain the exact flat-space result
	\begin{align}
		m(M_2-3M_0)^{(1)}
		&=I_0-\frac4{75\pi^3}\nonumber\\*
		&=\frac{1}{8640\pi^2}\bigl[-47\Si(\pi)+64\Si(2\pi)\nonumber\\*
		&\qquad-27\Si(3\pi)\bigr]<0,
		\label{flat:exactfinal}
	\end{align}
	which agrees with Eq.~\eqref{on:moment}. The reduction to
	Eqs.~\eqref{flat:totalderivative} and \eqref{flat:totalresidue} uses only
	the displayed parameter integrals and polynomial identities. The result holds in the stated covariant prescription: an
	independent four-derivative background coupling would change the
	observable.
	
	\subsection{Physical source and local terms}
	\label{on:source}
	
	For the critical model, the allowed source counterterms and the
	finite-$N$ scaling of the source leave $c_{-1}$ unchanged. The ultraviolet expansion
	of the exact spectrum determines the source subtraction. We define
	$u_{\mathrm{UV}}=4y/\pi$ and $v_{\mathrm{UV}}=16yz/(3\pi)$. Then
	\begin{align}
		n^2\log(8nB_n)
		={}&-u_{\mathrm{UV}}n-\frac{u_{\mathrm{UV}}^2}{2}
		+\frac{v_{\mathrm{UV}}-u_{\mathrm{UV}}^3/3}{n}
		\nonumber\\
		&+O(n^{-2}),
		\label{on:UVsymbol}
	\end{align}
	and the logarithmic divergence in the determinant
	$\frac12\operatorname{Tr}\log B$ is proportional to
	\begin{equation}
		\frac{8yz}{3\pi}-\frac{32y^3}{3\pi^3}.
		\label{on:logdivergence}
	\end{equation}
	The $yz$ term renormalizes the critical source, while the $y^3$ term is a
	local source contribution. The angular cutoff is
	$n_{\max}\simeq\Lambda_{\mathrm{phys}}R$, with the physical cutoff held fixed. Holding the dimensionless angular cutoff fixed while
	differentiating with respect to $R$ would compute a different
	derivative. At the leading saddle, $z=y^2+1/4+O(y^2e^{-2\pi y})$, so that
	the allowed source terms $J\sigma$, $J^3$, and $J\mathcal R$ contribute
	only volume and curvature terms, together with exponentially small
	terms, and they do not change $c_{-1}$. With a fixed physical
	renormalization scale, the local effective action depends on the radius
	as $R^3$ or $R$, also when its coefficients contain logarithms of the
	physical mass and of the renormalization scale.
	
	At finite $N$, the critical source has dimension
	$d_N=1+\gamma/N+O(N^{-2})$, and we write $x_J=JR^{d_N}$. If
	$E_J=x_J\partial_{x_J}$, the physical derivative is $D=d_NE_J$, and
	expanding the subtraction $P=P_3$ of Eq.~\eqref{st:filter} gives
	\begin{align}
		P(d_NE_J)F_N
		={}&NP(E_J)f_0 \nonumber\\*
		&+P(E_J)f_1+\gamma E_JP'(E_J)f_0
		+O(N^{-1}).
		\label{on:physicalfilter}
	\end{align}
	The last term is required at first subleading order. Indeed, a local
	contribution proportional to $m^jR^j$, with $j=1,3$, is proportional to
	$x_J^{j/d_N}$, and its expansion contains $-\gamma j x_J^j\log x_J/N$.
	The contribution of this logarithm to the subtracted free energy is
	cancelled by the last term in \eqref{on:physicalfilter}, since $P(j)=0$,
	and using $y=mR$ implements this cancellation directly. Applying
	$P(E_J)$ to an unrenormalized determinant while omitting the source
	correction would instead leave spurious local powers.
	
	A radius-independent change of flat-space mass normalization at order
	$1/N$ changes $f_1$ by a multiple of $yf_0'(y)$. Since $f_0'=\pi z/4$ and
	$z=y^2+1/4+$ exponentially small terms, this cannot produce a first
	subleading $y^{-1}$ term, consistently with Eq.~\eqref{on:leadingFIR}. A
	curvature-dependent redefinition $y\mapsto y+\alpha/(Ny)$, with
	$\alpha$ constant, would produce such a term, but it changes the coupling
	of the source to curvature.
	Likewise, independent terms $\int\sqrt g\,\mathcal R^2/J$ or
	$\int\sqrt g\,\mathcal R^2/\mu$, where $\mu$ is an independent mass
	scale, would change the answer. These are additional background
	couplings, which are excluded by the ultraviolet prescription.
	
	\subsection{Leading saddle at finite quartic coupling}
	\label{on:finite-coupling-proof}
	
	For $\lambda>0$, integrating out the scalars in
	Eq.~\eqref{on:finite-coupling-action} gives the renormalized leading
	action per component
	\begin{equation}
		S_{0,\lambda}[g,\sigma]
		=\frac12\Tr\log(K_g+\sigma)
		-\frac1{2\lambda}\int d^3x\sqrt g\,(\sigma-r)^2,
		\label{on:finite-coupling-effective-action}
	\end{equation}
	up to the local metric terms fixed by the prescription of
	Sec.~\ref{st:prescription}. Here $r$ is the renormalized mass parameter
	in the same finite-part prescription as the scalar determinant.
	The renormalized flat-space coincident propagator is $-m/(4\pi)$, so stationarity
	at $\sigma=m^2$ fixes
	\begin{equation}
		r=m^2+\frac{\lambda m}{8\pi}=m^2(1+\bar\lambda),
		\qquad \bar\lambda=\frac{\lambda}{8\pi m}.
		\label{on:finite-coupling-tuning}
	\end{equation}
	On the sphere, put $a^2=R^2\sigma-1/4$. The renormalized coincident propagator is
	$-a\coth(\pi a)/(4\pi R)$, continued analytically for imaginary $a$.
	Using the scalar determinant $F_s$ of Eq.~\eqref{on:scalar}, the
	stationary free energy and its gap equation are
	\begin{align}
		f_{0,\lambda}(y)
		&=F_s(a^2+1/4)\nonumber\\
		&\quad-\frac{\pi}{8\bar\lambda y}
		\bigl[a^2+1/4-(1+\bar\lambda)y^2\bigr]^2,
		\label{on:finite-coupling-free-energy}
	\end{align}
	\begin{equation}
		a^2+\frac14+\bar\lambda y\,a\coth(\pi a)=(1+\bar\lambda)y^2.
		\label{on:finite-coupling-gap}
	\end{equation}
	The omitted local metric terms do not affect $P_3(D)f_{0,\lambda}$.
	At $\bar\lambda=0$, the gap equation reduces to $a^2=y^2-1/4$ and
	the free energy is the free-scalar determinant.
	For $y>0$, the left side of Eq.~\eqref{on:finite-coupling-gap}
	is strictly increasing in $a^2\geq-1/4$, from zero to infinity,
	so the physical saddle is unique and smooth. In the UV,
	$z=a^2+1/4=(1+\bar\lambda)y^2+O(y^3)$ and
	$a\coth(\pi a)=\pi z/2+O(z^2)$. Since
	$F_s'(z)=-\pi a\coth(\pi a)/4$, the free energy is
	$F_s(0)-\pi^2(1+\bar\lambda)^2y^4/16+O(y^5)$, modulo the local terms.
	This verifies the UV endpoint and derivative decay required by
	Sec.~\ref{ob:ir}.
	
	For fixed finite $\bar\lambda\geq0$ and large $y$, replacing
	$\coth(\pi a)$ by one changes the saddle only exponentially. The
	positive solution of the resulting quadratic equation gives
	\begin{align}
		a&=\frac{\sqrt{(2+\bar\lambda)^2y^2-1}-\bar\lambda y}{2}
		+O(ye^{-2\pi y})\nonumber\\
		&=y-\frac{1}{4(2+\bar\lambda)y}
		-\frac{1}{16(2+\bar\lambda)^3y^3}+O(y^{-5}).
		\label{on:finite-coupling-saddle-ir}
	\end{align}
	Equation~\eqref{on:scalarintegral} gives
	$F_s(a^2+1/4)=-\pi a^3/6+O(a^2e^{-2\pi a})$.
	On shell the second term in
	Eq.~\eqref{on:finite-coupling-free-energy} is
	$-\pi\bar\lambda y a^2\coth^2(\pi a)/8$. Substitution therefore yields
	\begin{equation}
		f_{0,\lambda}(y)
		=f_{0,\lambda,\mathrm{local}}(y)
		-\frac{\pi}{128(2+\bar\lambda)y}+O(y^{-3}),
		\label{on:finite-coupling-free-energy-ir}
	\end{equation}
	where $f_{0,\lambda,\mathrm{local}}$ contains only the $y^3$ and
	$y$ terms. The algebraic expansion of the quadratic root and the
	exponentially small correction from the exact gap equation remain
	valid after any fixed number of scale derivatives.
	Multiplication by $P_3(-1)=8/3$ proves
	Eq.~\eqref{on:finite-coupling-tail}, and curvature matching gives
	Eq.~\eqref{on:finite-coupling-moments}.
	
	\subsection{Flat-space moments of the Gross--Neveu flow}
	\label{on:gn-flat}
	
	For a complex two-component Dirac field with
	$S(p)=(-i\gamma\cdot p+m)/(p^2+m^2)$, dimensional regularization gives
	\begin{equation}
		G_{\bar\psi\psi}(q)=-(q^2+4m^2)B(q)-\frac m{2\pi},
	\end{equation}
	where the constant comes from the tadpole
	$\int d^3k\,(2\pi)^{-3}(k^2+m^2)^{-1}=-m/(4\pi)$. It does not affect the
	spectral density in Eq.~\eqref{st:dirac-densities}, but it enters the
	auxiliary propagator $G_{\sigma\sigma}=1/(NK)$, with $K=-G_{\bar\psi\psi}$.
	The kernel obeys $K(0)=m/\pi=V''(m)$, as it must, and
	\begin{equation}
		K(q)=\frac m\pi\Bigl(1+\frac{q^2}{12m^2}-\frac{q^4}{240m^4}
		+O(q^6)\Bigr).
	\end{equation}
	With the separated trace identity $\Theta=2NJ\sigma$, which holds
	because $\Delta_\sigma=1$ at leading order,
	\begin{align}
		G_{\Theta\Theta}(q)&=\frac{(2NJ)^2}{NK(q)}\nonumber\\
		&=\frac{Nm^3}{\pi}\Bigl(1-\frac{q^2}{12m^2}+\frac{q^4}{90m^4}
		+O(q^6)\Bigr)
	\end{align}
	up to contacts. As below Eq.~\eqref{st:inverse-bubble}, the $q^4$
	coefficient is $2M_0$, which gives $M_0^{\rm GN}=N/(180\pi m)$. The
	auxiliary exchange affects only spin zero, so that
	$M_2^{\rm GN}=NM_2^\psi=N/(160\pi m)$, and the difference is
	Eq.~\eqref{on:gn-moments}.
	
	\subsection{Improvement dependence and weak quartic coupling}
	\label{fin:app}

	We consider the finite-coupling scalar theory of
	Sec.~\ref{on:finite-coupling}, with curvature coupling $\xi$ and
	$\ell=\lambda/m=8\pi\bar\lambda$. The flat-space mass parameter is tuned so that the
	physical mass is $m>0$. Curvature derivatives are then taken at fixed
	source and coupling, without retuning the mass on the curved background.
	Throughout this subsection $f$ is the total free-energy density,
	including all $N$ components. The operator $\phi^2$ is normalized
	by the mass source, $\partial S/\partial r_0=\tfrac12\int dV\,\phi^2$.

	The improvement changes the separated trace by
	$\Theta_\xi=\Theta_0+2\xi\partial^2\phi^2$, while its spin-two projection
	vanishes. By Eq.~\eqref{st:trace-density},
	\begin{equation}
		M_0=\frac12\int\frac{ds}{s^3}\,
		\rho_{\Theta_\xi\Theta_\xi}(s).
	\end{equation}
	The coefficient of $\xi^2$ is
	$2\int ds\,\rho_{\phi^2\phi^2}(s)/s=2\chi$. At finite $\lambda$ the
	ultraviolet fixed point is Gaussian, and the connected $\phi^2$
	two-point function has scaling degree two. The prescription of
	Sec.~\ref{st:prescription} therefore admits no delta-function contact
	in this correlator. Its integral is the positive spectral susceptibility
	$\chi$ in Eq.~\eqref{fin:chi}; the mass gap makes it convergent
	in the infrared. Completing the square in $\xi$ proves
	Eq.~\eqref{fin:quadratic}, with a unique maximum $G_*$.

	On hyperbolic space of sectional curvature $\kappa$,
	$\mathcal R=6\kappa$ and $\mathcal R_{\mu\nu}^2=12\kappa^2$.
	Equation~\eqref{st:AB} gives the coefficient
	$12(3A+B)=3(M_2-3M_0)/4$ of $\kappa^2$ in $f$. Consequently,
	\begin{equation}
		M_2-3M_0=\frac23\left.\frac{\partial^2f}{\partial\kappa^2}
		\right|_{\kappa=0}.
		\label{fin:susceptibility}
	\end{equation}
	The derivatives refer to the local curvature expansion from
	$\kappa<0$. The free propagator follows from
	Eq.~\eqref{on:hyperbolicG}:
	\begin{align}
		G_{\kappa,\xi}(r)&=\frac{e^{-Mr}}{4\pi r}\,h(\kappa r^2),
		\qquad M^2=m^2-(1-6\xi)\kappa,\nonumber\\*
		h(z)&=\frac{\sqrt z}{\sin\sqrt z}
		=1+\frac z6+\frac{7z^2}{360}+O(z^3).
		\label{fin:green}
	\end{align}
	At $\xi=1/6$, $M=m$ and the free determinant density
	$-NM^3/(12\pi)$ is independent of curvature.

	For the weak-coupling calculation we Wick order with respect to the
	flat massive propagator. The difference of the renormalized coincident
	propagators is $-(M-m)/(4\pi)$. Wick contraction gives the first-order
	contribution
	\begin{equation}
		f^{(1)}=\frac{\lambda(N+2)}{128\pi^2}(M-m)^2.
		\label{fin:firstorder}
	\end{equation}
	Its curvature-squared coefficient is proportional to $(1-6\xi)^2$,
	so the maximizing improvement has no shift at order $\ell$.
	The remaining mass counterterm starts at order $\lambda^2$.
	At $\xi=1/6$ its free expectation value is independent of curvature,
	and hence it does not change the coefficient in
	Eq.~\eqref{fin:susceptibility} at this order.

	At second order, the connected contraction is
	\begin{equation}
		\bigl\langle{:}(\phi^2)^2(x){:}\,{:}(\phi^2)^2(0){:}\bigr\rangle
		=8N(N+2)G_{\kappa,1/6}(x,0)^4.
	\end{equation}
	The second cumulant is therefore
	\begin{equation}
		f^{(2)}=-\frac{(N+2)\lambda^2}{16N}\int dV\,G_{\kappa,1/6}^4.
	\end{equation}
	With $G_0(r)=e^{-mr}/(4\pi r)$, the volume element is
	$4\pi r^2h(\kappa r^2)^{-2}dr$. With a common short-distance regulator,
	\begin{equation}
		\int dV\,G_{\kappa,1/6}^4
		=\frac1{64\pi^3}\int_0^\infty
		\frac{e^{-4mr}}{r^2}\,h(\kappa r^2)^2\,dr.
	\end{equation}
	The curvature-squared coefficient is convergent at both endpoints.
	Since $h(z)^2=1+z/3+z^2/15+O(z^3)$, it gives
	\begin{align}
		[\kappa^2]f^{(2)}
		&=-\frac{(N+2)\lambda^2}{15360\pi^3N}
		\int_0^\infty r^2e^{-4mr}\,dr\nonumber\\
		&=-\frac{(N+2)\lambda^2}{491520\pi^3Nm^3}.
		\label{fin:melon}
	\end{align}
	Here $[\kappa^2]$ denotes the coefficient of $\kappa^2$.
	Together with Eq.~\eqref{fin:susceptibility}, this proves
	\begin{equation}
		\left.m(M_2-3M_0)\right|_{\xi=1/6}
		=-\frac{N+2}{368640\pi^3N}\,\ell^2+O(\ell^3).
	\end{equation}
	At fixed $N$, $\chi=N/(4\pi m)+O(\ell/m)$ and
	$\xi_*=1/6+O(\ell^2)$. By Eq.~\eqref{fin:quadratic}, the gain in
	$mG_*$ from this shift is $O(\ell^4)$, which establishes
	Eq.~\eqref{fin:weak}.

	For the large-$N$ expansion, the leading response formula
	\eqref{on:response-formula} gives
	$\chi=4N/(\lambda+16\pi m)+O(N^0)$ and places the leading maximum at
	$\xi=1/6$. At each fixed finite $\ell$, the first subleading
	contribution shifts $\xi_*$ by $O(N^{-1})$ and raises $mG_*$ by
	$O(N^{-1})$. Thus the coefficient $g_1$ computed at $\xi=1/6$ in
	Appendix~\ref{lean:curvaturedetails}, with the positive representation
	\eqref{fin:g1positive}, is also the order-$N^0$ coefficient of $mG_*$.

\end{document}